\documentclass[aps,prb,twocolumn,superscriptaddress,longbibliography,english]{revtex4-2}
\usepackage{graphicx}
\usepackage{dcolumn}
\usepackage{color,soul}
\usepackage{soul}
\usepackage{amsfonts,amsmath,amssymb,bm}
\usepackage{xcolor}
\usepackage{hhline}
\usepackage[utf8]{inputenc}
\usepackage[colorlinks=true,allcolors=blue]{hyperref}
\usepackage{booktabs}
\usepackage{multirow}
\usepackage{ifthen}
\DeclareUnicodeCharacter{2212}{-}

\usepackage{comment}

\def\ket#1{|#1\rangle}

\def\Cc#1{\hat{C}_{#1}}

\graphicspath{{./Figures/}}

\begin{document}
\title{Magneto-optical properties of Group-IV--vacancy centers in diamond upon hydrostatic pressure}
\date{\today}

 \author{Meysam Mohseni}
 \affiliation{HUN-REN Wigner Research Centre for Physics, PO.\ Box 49, H-1525 Budapest, Hungary}
 \affiliation{E\"otv\"os Lor\'and University, P\'azm\'any P\'eter S\'et\'any 1/A, H-1117 Budapest, Hungary}
 
 \author{Lukas Razinkovas}
 \affiliation{Center for Physical Sciences and Technology (FTMC), Vilnius LT-10257, Lithuania}
 
 \author{Vytautas \v{Z}alandauskas} 
 \affiliation{Center for Physical Sciences and Technology (FTMC), Vilnius LT-10257, Lithuania} 

 \author{Gerg\H{o} Thiering}
\affiliation{HUN-REN Wigner Research Centre for Physics, PO.\ Box 49, H-1525 Budapest, Hungary} 
 
 \author{Adam Gali}
\affiliation{HUN-REN Wigner Research Centre for Physics, PO.\ Box 49, H-1525 Budapest, Hungary} 
\affiliation{Budapest University of Technology and Economics, M\H{u}egyetem rakpart 3., H-1111 Budapest, Hungary}
\affiliation{MTA-WFK Lend\"ulet "Momentum" Semiconductor Nanostructures Research Group, PO.\ Box 49, H-1525 Budapest, Hungary}

\begin{abstract}
 In recent years, the negatively charged group-IV–vacancy defects in diamond, labeled as G4V($-$) or G4V centers, have attracted significant attention in quantum information processing. In this study, we investigate the magneto-optical properties of G4V centers under high compressive hydrostatic pressures up to 180~GPa. The spin-orbit splitting of the electronic ground and excited states, as well as the hyperfine tensors, are calculated using plane-wave supercell density functional theory, providing distinctive fingerprints that uniquely characterize these defects.
To this end, we develop a theory for calculating the hyperfine tensors when the electronic states are subject to the Jahn–Teller effect. We find that the zero-phonon-line energy increases with hydrostatic pressure, with the deformation potential increasing from SiV($-$) to PbV($-$). On the other hand, our calculated photoionization threshold energies indicate that PbV($-$)-based quantum sensors can operate only up to 32~GPa, whereas SnV($-$), GeV($-$), and SiV($-$) remain photostable up to 180~GPa.
We also find that the spin-orbit splitting increases in both the electronic ground and excited states with increasing pressure. The optical transitions associated with the hyperfine fine structure of the dopant atoms are interpreted using our theoretical framework, which reproduces existing experimental data at zero strain. We show that the hyperfine levels are weakly dependent on magnetic field, and increasing pressure leads to optical transitions at longer wavelengths. Finally, we estimate the spin coherence times of the G4V centers under increasing hydrostatic pressure across different temperature regimes.
\end{abstract}

\maketitle

\section{Introduction}\label{sec:intro}
\vspace{-0.5em}

\begin{figure*}
  \includegraphics[width=\textwidth]{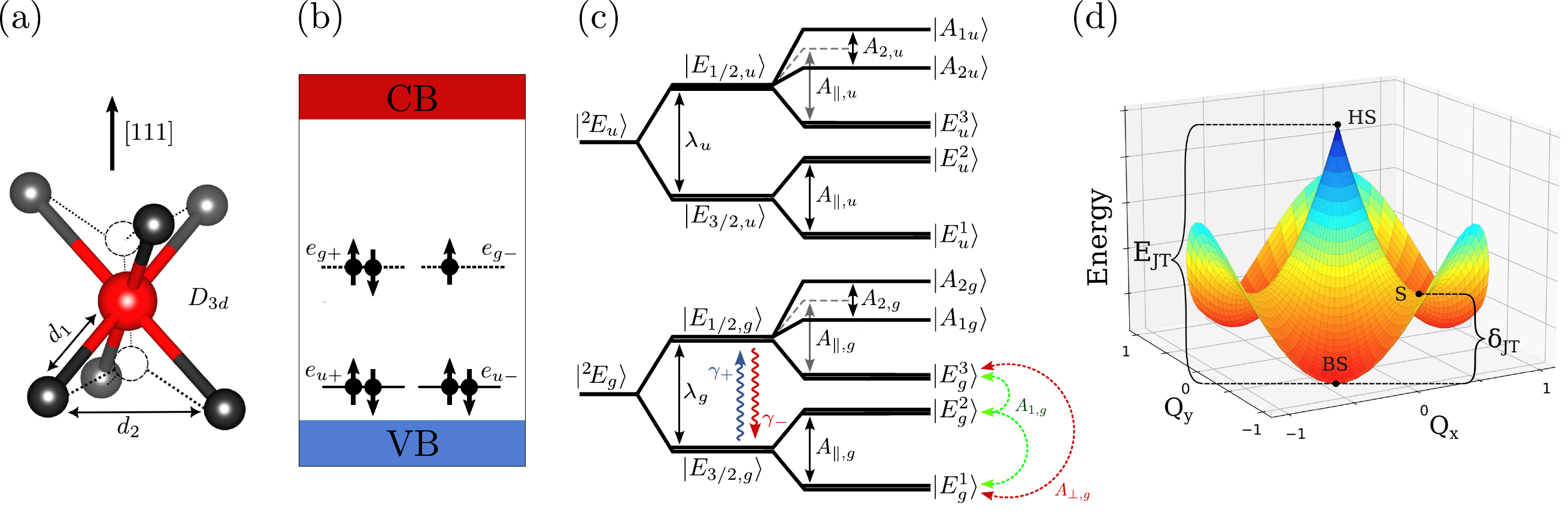}
\caption{\label{fig:JT}%
  Schematic diagrams illustrating the fundamental
    properties of G4V($-$) color centers in diamond.
  (a) Geometry of the defect. Dashed spheres represent the two vacancies, and the interstitial dopant atom (silicon, germanium, tin, or lead) is shown as a red sphere. The critical distances under high $D_{3d}$ symmetry are labeled as $d_1$ and $d_2$. The dopant atom is positioned at the inversion center of the diamond.
  (b) Schematic single-particle defect levels in the electronic ground state \textcolor{black}{(small spin-polarization splittings of \(\lesssim20\)\,meV omitted)}. The positions of the fully
  occupied double degenerate $e_u$ levels and the partially occupied $e_g$ levels
  vary in the various G4V($-$) color centers. In the electronic excited state (not
  shown), the $e_g$ levels are fully occupied, whereas a hole is left on the $e_u$
  level. 
  (c) Electronic structure in the ground and excited states, including fine and hyperfine interactions.
  The hyperfine structure originates from the coupling with the impurity atom positioned at the inversion symmetry point. The fine and hyperfine states are labeled according to the double-group representations of the $D_{3d}$ point group, following the notation in Ref.~\cite{altmann1994}.
  $\lambda_{g,u}$ are the effective spin-orbit splitting in the $g$ (even parity) ground and $u$ (odd parity) excited states. Transitions between two Kramers states occur with orbital relaxation at rates $\gamma_{\pm}$. The hyperfine coupling parameters $A_{\parallel,\{g,u\}}$ and $A_{2,\{g,u\}}$ lift the degeneracy of the fine-structure states, while $A_{\perp,\{g,u\}}$ and $A_{1,\{g,u\}}$ facilitate the mixing of spin-orbit states.
  (d) Adiabatic potential energy surface (APES) of the quadratic
  Jahn-Teller system along the ionic degrees of freedom corresponding to a single degenerate $e_g$-symmetry mode, described by the configuration coordinates $Q_{x,y}$. The points of BS, HS, and S represent the lowest
  energy broken-symmetry, the highest energy of highest symmetry and saddle
  point configurations, respectively. $E_\text{JT}$ is the difference in energy
  between the high symmetry configuration and the distorted configuration. The
  three equivalent global minima are separated by energy barriers of~$\delta_\text{JT}$.\looseness=1}
\smallskip
\end{figure*}

The negatively charged group-IV--vacancy G4V($-$) defects in
diamond, commonly referred to as
  G4V centers, have attracted significant
attention over the past decades due to their
  promising optical properties (see Fig.~\ref{fig:JT} for a schematic
  representation of their geometry and electronic structure). These color
centers, including silicon-vacancy (SiV), germanium-vacancy (GeV), tin-vacancy
(SnV), and lead-vacancy (PbV) defects are promising qubits for quantum
communication and sensor
applications~\cite{Vavilov1980,Hepp2014,Muller2014,Rogers2014, Ekimov2015,
  Iwasaki2015, Iwasaki2017, Meesala2018, Tchernij2018,
  Trusheim2019, Chen2019, Trusheim2020, Gorlitz2020, Krivobok2020, Rugar2021,
  Aghaeimeibodi2021, Wang2021, Vindolet2022, Wang2024, Bhaskar2020, Bersin2024,
  Knaut2024}. The G4V centers have a $S=1/2$ spin state and exhibit $D_{3d}$
inversion symmetry~\cite{Goss1996, Gali2013, Thiering2018}, leading to outstanding optical stability
compared to other color centers, such as the
nitrogen-vacancy center in diamond~\cite{Doherty2013, Gali2019}, which can be
utilized in quantum communication technologies~\cite{Bhaskar2020, Bersin2024}.
The electronic structure of G4V centers provides an advantage in achieving coherent control of spin states using only optical
methods without the need for microwave fields~\cite{Pingault2014, Becker2016,
  Siyushev2017, Becker2018, Weinzetl2019, Debroux2021}. The ability of
microwave-free control of spin states can be practical under extreme conditions
such as high pressures, where reducing the complexity
of measurement is a very important issue~\cite{Vindolet2022}. On the other hand,
the effect of pressure on the magneto-optical properties of the G4V centers has
not yet been explored in detail~\cite{Vindolet2022, Harris2023}, which is an
inevitable step in determining the potential of these
color centers in high-pressure quantum sensor applications.\looseness=1

In this paper, we explore the magneto-optical properties of G4V centers upon
compressive hydrostatic pressure using density functional
theory (DFT) plane wave supercell calculations. In
particular, we focus on the fine-level structure of the electronic
system, i.e., the spin-orbit and hyperfine interaction in
the respective ground and excited states. These properties
are intertwined with the strong electron-phonon coupling that can be described
by the Jahn--Teller theory~\cite{Thiering2018}. We developed a theory to
accurately compute the hyperfine tensors subject to the Jahn--Teller
effect. We implemented it to calculate all the critical
hyperfine tensor elements of the respective group-IV dopant and the proximate
$^{13}$C nuclear spins. The photoionization threshold energies are also
monitored in our study, along with the shift in the zero-phonon-line energy as a
function of compressive hydrostatic pressure. We provide the list of
pressure-dependent magneto-optical parameters that indicate the
coupling strength of pressure to zero-phonon-line (ZPL) energies, spin-orbit
gaps, and hyperfine levels. These parameters can be used to calibrate the actual
pressure acting on the color centers. We find that the coupling strength
increases when going from lighter to heavier group-IV elements in the G4V
centers. Our calculations also reveal that the operation of the PbV($-$) quantum
sensor is limited up to 32~GPa.

\section{Methods}\label{sec:method}
We performed density functional theory (DFT)
calculations~\cite{PhysRev.140.A1133,PhysRev.136.B864} using plane-wave
supercell and the projector-augmented-wave (PAW)
approach~\cite{PhysRevB.50.17953,PhysRevB.59.1758} as implemented in the Vienna
Ab-initio Simulation Package (VASP)~\cite{PhysRevB.54.11169,Kresse1996}.
For the
  exchange-correlation functional, we employed the strongly constrained and
  appropriately normed (SCAN) meta-GGA functional~\cite{Sun2015,Isaacs2018SCAN,Rodriguez2018SCAN}, which has
  demonstrated accurate performance in describing structural and electronic
  properties of color centers in diamond~\cite{scandiamond}. For the hyperfine
tensor calculations, we additionally provide
  results obtained using the hybrid HSE06 functional~\cite{Heyd2023}. The
defect structure was simulated within a $4\times 4\times 4$
supercell consisting of 512 atoms, with lattice
constants of 14.21~\AA\ and 14.18~\AA\ obtained using the SCAN and
hybrid functionals, respectively. $\Gamma$-point was used to sample the Brillouin zone. The structures
were optimized using a convergence
  criterion of 10$^{-2}$~eV/\AA\ per atom for the Hellman-Feynman
forces, with a kinetic energy cutoff of~600~eV. 

The G4V($-$) defects feature double-degenerate $|e_u\rangle$ and $|e_g\rangle$ localized
single-particle levels, corresponding to irreducible representations of the
$D_{3d}$ symmetry group [see~Fig.~\ref{fig:JT}(b)]. These levels lie within
the bandgap and are occupied by seven electrons. The ground-state
configuration can be described as a single-hole occupation of the $|e_g\rangle$
orbitals, giving rise to the spin-doublet $^{2}\!E_{g}$ state [see
Fig.~\ref{fig:JT}(c)]. The excited state is formed by promoting an electron
from an orbital with $e_u$ symmetry to the $|e_g\rangle$ orbital, resulting in the
$^{2}\!E_{u}$ state. Geometry optimization of the electronic excited state was
conducted using the $\Delta$SCF method~\cite{Gali2009}, with the Kohn–Sham
orbital occupation constrained to achieve the $^{2}\!E_{u}$ configuration. ZPL
energies were computed as the total energy difference between the respective
electronic states at their global minima on the adiabatic potential energy
surfaces. The orbital degeneracy in both the ground and excited states induces
Jahn–Teller coupling, which in adiabatic DFT calculations manifests as
symmetry-breaking distortions along the nuclear degrees of freedom associated
with the $|e_g\rangle$ symmetry~\cite{Thiering2018}.

Next, we address the calculation of spin-orbit
coupling (SOC) and hyperfine tensors, which provide insights into the fine electronic structure of
G4V($-$) defects. We determined the SOC using the SCAN functional
within a non-collinear framework of DFT, with the quantization axis of the spin aligned along the symmetry axis of the defect. The SOC has a
small perturbation effect on electronic states; therefore, it was calculated using the optimized
high-symmetry $D_{3d}$ configuration.
After applying
  SOC to the system, the associated $z$-component of the SOC, $\lambda_0$,
  splits the $^{2}\!E_{g}$ and $^{2}\!E_{u}$ states into two degenerate
  doublets. Within the hole notation, these doublets are defined as
  $|E_{3/2,\{g,u\}}\rangle = \{|e_{\{g,u\}+}^{\uparrow}\rangle, |e_{\{g,u\}-}^{\downarrow}\rangle\}$
  and
  $|E_{1/2,\{g,u\}}\rangle = \{|e_{\{g,u\}-}^{\uparrow}\rangle, |e_{\{g,u\}+}^{\downarrow}\rangle\}$
  [see~Fig.~\ref{fig:JT}(c)], where, for example,
  $\ket{e_{g+}^{\uparrow}} \equiv \ket{e_{g+}}\ket{{\uparrow}}$ denotes the
  spinor orbitals. Under the operations of the $D_{3d}$ symmetry group, the
  $|e_{u\pm}\rangle$ orbitals are chosen to transform as the spherical harmonics 
  $Y_{1,\pm1} \sim $"$ \mp (x \pm i y$)" of the Condon-Shortley phase convention~\cite{CondonShortley1935,altmann1994},
  while the $|e_{g\pm}\rangle$ orbitals transform equivalently to angular
  momentum raising and lowering operators (axial vectors)
  $\hat{L}_{\pm} = \hat{L}_{x}\pm i \hat{L}_{y}$.
  The Hamiltonian for SOC is given by
\begin{align}
  \notag
  {\hat{H}_\text{SOC}}
  & =
   -\lambda_{\{g,u\}}\hat{\sigma}_{z}\hat{S}_z
  \\\notag
  & =
  \frac{\lambda_{\{g,u\}}}{2}
    \left[
    \bigl|e^{\uparrow}_{\{g,u\}+} \bigr\rangle \bigr\langle e^{\uparrow}_{\{g,u\}+}\bigl|
    +
    \bigl|e^{\downarrow}_{\{g,u\}-} \bigr\rangle \bigl\langle e^{\downarrow}_{\{g,u\}-}\bigr|
    \right.
  \\
  & \quad
    \left.
    -
    \bigl|e^{\uparrow}_{\{g,u\}-} \bigr\rangle \bigl\langle e^{\uparrow}_{\{g,u\}-}\bigr|
    -
    \bigl|e^{\downarrow}_{\{g,u\}+} \bigr\rangle \bigl\langle e^{\downarrow}_{\{g,u\}+}\bigr|
    \right]
    \text{,}
    \label{eq:SOC-hammiltonian}
\end{align}
where the $\hat{\sigma}_{z}=\hat{L}_z$ represents the orbital
angular momentum operator: $\hat{\sigma}_{z}|e_{\pm}\rangle=\pm|e_{\pm}\rangle$, and $\hat{S}_{z}$ is the $z$
component of the electronic spin. The negative sign of $\lambda_{\{g,u\}}$ occurs
because of the hole quasiparticle. \textcolor{black}{Equation~\ref{eq:SOC-hammiltonian} is obtained by projecting the atomic SOC operator into the $E_g$ manifold of G4V centers under $D_{3d}$ symmetry, yielding the single invariant $\lambda_{\rm SO}\,L_zS_z\,$.}

The hyperfine (HF) interaction arises from the coupling between nuclear and
electron spins, contributing additional splitting in the electron spin resonance
spectrum. This interaction is particularly significant when the electron spin
density overlaps with the nuclear positions, providing a characteristic
spectroscopic fingerprint of the atomic structure. In systems with orbital
degeneracy, the HF interaction also couples to degenerate orbital components,
making interaction dependent on the vibronic states of the Jahn–Teller system.

For the hyperfine interaction with the central \mbox{Group-IV} ion, the HF
Hamiltonian is invariant under the $D_{3d}$ double group symmetry operations. Ensuring
that the HF Hamiltonian transforms as a scalar under this symmetry results in
the canonical form (for a group theoretical derivation, see Appendix~\ref{app:HFHam}):
\begin{align}
  \hat{H}_{\text{HF}}
&=
  \left[
    \frac{1}{2}A_{\perp\{g,u\}}(\hat{S}_{+}\hat{I}_{-} + \hat{S}_{-}\hat{I}_{+}) + 2A_{\parallel\{g,u\}}\hat{S}_z \hat{I}_z
    \right]
  \hat{\mathbb{I}}
  \nonumber
  \\
  &
    - A_{1\{g,u\}}
    \left[
    \left(\hat{S}_z \hat{I}_{+} + \hat{S}_{+}\hat{I}_z\right)
    \!\hat{\sigma}_{+}
    \!+\!
    \left(\hat{S}_z \hat{I}_{-} + \hat{S}_{-}\hat{I}_z\right)
    \!\hat{\sigma}_{-}
    \right]
    \nonumber
  \\
&
    + \frac{A_{2\{g,u\}}}{2}
    \left[
        \hat{S}_{-}\hat{I}_{-}\hat{\sigma}_{+} + \hat{S}_{+}\hat{I}_{+}\hat{\sigma}_{-}
    \right] 
    \text{.}
    \label{eq:HFD3d}
\end{align}
Here, $\hat{S}_{i}$ and $\hat{I}_{i}$ denote the spin projection and spin raising or
lowering operators acting on the electronic and nuclear spin degrees of freedom,
respectively. The operators $\mathbb{I}$ and
$\sigma_{\pm} = |e_{\{g,u\}\pm}\rangle\langle e_{\{g,u\}\mp}|$ represent the
identity operator and the orbital angular momentum raising and lowering
operators, respectively, acting within the two-dimensional subspace of
degenerate orbitals. Here, we note that while $\hat{\sigma}_{z}$ is equivalent 
with $\hat{L}_{z}$ the ladder operators $\hat{\sigma}_{\pm}$'s are not the same as $\hat{L}_{\pm}$.
When one looks into Eq.~\eqref{eq:HFD3d} more closely angular momentum conservation appears to be violated.
However, one has to take into consideration that $\hat{\sigma}_{\mp}$ flips the "real" angular momentum 
twice: $\hat{\sigma}_{\mp}|e_{\{g,u\}\pm}\rangle\sim\hat{L}_{\mp}^{2}|e_{\{g,u\}\pm}\rangle\sim\hat{L}_{\mp}|a_{\{1g,2u\}}\rangle\sim|e_{\{g,u\}\mp}\rangle$ 
because one has to flip though $|a_{\{1g,2u\}}\rangle$ orbitals first. 
However, $|a_{\{1g,2u\}}\rangle$ lie deep in the valence band~\cite{Hepp2014_supp, Haubler_2017} 
thus we omit $\hat{L}_\pm$s from discussion from now on. Finally, we note that Eq.~\eqref{eq:HFD3d} neglects the orbital hyperfine interaction, which is justified because the spin density does not build up from the atomic-like orbitals of the dopant atom.

Figure~\ref{fig:JT}(c) illustrates the hyperfine energy level structure arising
from the interaction with the impurity atoms in G4V defects, with the
corresponding states labeled according to their irreducible representations. The
hyperfine interaction splits the $E_{3/2,\{g,u\}}$ levels into two doublets of
$E_{\{g,u\}}$ symmetry. In contrast, the upper $E_{1/2,\{g,u\}}$ branch is divided
into a doublet of $E_{\{g,u\}}$ symmetry and two singlets of $A_{1,\{g,u\}}$ and
$A_{2,\{g,u\}}$ symmetries.

The first term in Eq.~\eqref{eq:HFD3d} is isotropic with respect to the orbital
degrees of freedom and remains unaffected by the ionic motion associated with
the Jahn--Teller system. As such, it can be regarded as a ``static''
contribution to the HF interaction, unaffected by the ionic motion. Within this
term, $A_{\parallel, \{g,u\}}$ induces splitting between spin--orbit-coupled (SOC)
branches, while $A_{\perp, \{g,u\}}$ facilitates coupling between SOC eigenstates, as
illustrated by the red dashed arrow in Fig.~\ref{fig:JT}(c).

In contrast, the last two terms of Eq.~\eqref{eq:HFD3d} describe interactions
that couple degenerate components of the orbital states via the
$\hat{\sigma}_\pm$ operators. These contributions depend explicitly on the
orbital composition and are sensitive to the vibronic state of the defect,
reflecting a \emph{dynamical} HF coupling. The parameter $A_{1,\{g,u\}}$ enables
the mixing of $E_{\{g, u\}}$ states, as indicated by green dashed arrows in
Fig.~\ref{fig:JT}(c), while $A_{2,\{g,u\}}$ induces splitting between the
$A_{1,\{g,u\}}$ and $A_{2,\{g,u\}}$ levels. In this study, we compute the values
of $A_1$ and $A_2$ for the lowest vibronic state using the reduction factors
proposed by Ham~\cite{Ham1965}.

For a selected spin density $\rho_{\alpha}$, the hyperfine (HF) tensor
parameters can be computed using the expression:
\begin{align}
  \notag
  \mathcal{A}_{ij}^{(\alpha)}=
  & \frac{\mu_{0}\hbar^2}{4\pi}
    \frac{\gamma_I\gamma_e}{2\langle \hat{S}_{z} \rangle}\int d^3r \rho_\alpha(r) \\
  &{} \times
    \left[
    \left(
    \frac{8\pi}{3}\delta(r)
    \right)
    +
    \left(
    \frac{3r_ir_j}{|r|^5}
    -
    \frac{\delta_{ij}}{|r|^3}
    \right)
    \right]\text{,}
    \label{eq:HF}
\end{align}
where $\gamma_e$ and $\gamma_I$ are the gyromagnetic ratios of the electron and
nucleus, respectively; $\rho_\alpha(r)$ is the spin density at position $r$,
$\langle \hat{S}_{z}\rangle$ is the expectation value of the $z$ component of
the total electronic spin, and $r_i$ denotes the $i = x, y, z$ components
of the position vector relative to the nucleus. The first term represents the
Fermi-contact interaction, which strongly depends on the spin density localized
at the nucleus, while the second term corresponds to the dipole–dipole
interaction. In VASP, the HF tensor expression in Eq.~\eqref{eq:HF} is modified
within the projector-augmented wave (PAW) formalism~\cite{Blochl2000}, and the
spin polarization of the core orbitals is included to ensure accurate
calculation of the Fermi-contact term~\cite{Szasz2013}.

The HF parameters in Eq.~\eqref{eq:HFD3d} can be calculated using the
real-valued representation of $e$-orbitals.
The parameter values, derived for a coordinate system with the
$z$-axis aligned along the symmetry axis of the $D_{3d}$ group, are as follows
[see Eq.~\eqref{eq:HFtensorD3d} in Appendix~\ref{app:HFHam} and Ref.~\onlinecite{Thiering2024_supp}]:
\begin{align}
  \notag
  & A_{\parallel} = \tfrac{1}{2}\mathcal{A}_{zz}^{x}  , \quad A_{\perp} = \frac{1}{2}(\mathcal{A}_{xx}^{x} + \mathcal{A}_{yy}^{x})
  \\
  & A_{1} = q\mathcal{A}_{xz}^{x},
  \quad
    A_{2} = q\left(\mathcal{A}_{yy}^{x} - \mathcal{A}_{xx}^{x}\right) \text{,}
    \label{eq:HFparameters}
\end{align}
where the subscript $x$ indicates that the HF tensor [Eq.~\eqref{eq:HF}] was
computed for the hole spin density constrained to the $|e_x\rangle$ orbital of $C_{3v}$. The factor $q$
represents vibronic reduction factors~\cite{Ham1965, ham1968}, which account for
the quenching of the HF interaction caused by the dynamic Jahn–Teller (DJT)
effect in the ground vibronic state.

When the HF interaction involves nearby $^{13}$C ions, the overall symmetry is
reduced to $C_{2h}$, complicating the calculation of HF parameters. In
Appendix~\ref{app:hfder} we outline the procedure for estimating the HF
parameters under this reduced symmetry.

The DJT effect suppresses
the orbital moment, at least partially, which is referred to as the
Ham-reduction~\cite{Ham1965, Bersuker2006, Bersuker2012}. The inherent
electronic $\lambda_0$ can be read directly from the scalar-relativistic
spin-orbit splitting~\cite{Steiner2016} of the respective $e_g$ or $e_u$
Kohn-Sham orbitals. The electronic spin-orbit splitting is suppressed due to DJT
by the Ham reduction factor $p$; thus, the observed value in experiments is
$\lambda=p\lambda_0$. The value of $p$ depends on the vibronic states, which we compute by
  explicitly incorporating the DJT effect as
  implemented in this study (see the derivation in
  Ref.~\onlinecite{Thiering2018}).

In our study, we explore the Jahn--Teller effect by calculating parameters for
an effective single-degenerate-mode model. This model represents motion along a
symmetry-breaking direction within the $e_g$ irreducible representation,
adhering to the methodology outlined in previous works~\cite{Thiering2018}.
Within the linear Jahn--Teller theory, the $E_g \otimes e_g$ (electronic ground
state) and $E_u \otimes e_g$ (electronic excited state) APESs have a sombrero
shape. The inclusion of quadratic terms disrupts the axial symmetry, resulting
in three equivalent minima and saddle points~\cite{Bersuker2012}, as illustrated
in Fig.~\ref{fig:JT}(d).

Below, we discuss the \textit{ab initio} DFT analysis of the DJT effect and the
respective magneto-optical properties. To address convergence issues in
degenerate high-symmetry configurations and accurately determine $D_{3d}$
symmetry geometries, we adopt a strategy involving half-half fractional
occupations in half-filled degenerate orbitals within the spin-minority channel.
This method simulates the ensemble of two degenerate states, effectively
suppressing JT interactions and relaxing to the $D_{3d}$ configuration [HS point in Fig.~\ref{fig:JT}(d)]. Subsequently, shifting to integer
occupations enables relaxation along the $e_g$-symmetry
direction towards the lowest energy broken-symmetry configuration [BS point in Fig.~\ref{fig:JT}(d)]. This technique enables the identification of the lowest
energy point, $Q_{e_g;\text{BS}}$.

For a comprehensive examination of the APES, we parameterize
$\Delta Q_{e_g;\text{BS}}$ using a parameter $w$. In scenarios where $w=1$, a
broken symmetry configuration emerges, whereas $w=0$ corresponds to a
high-symmetry configuration. This method allows us to systematically explore the
APES for $\Delta Q_{e_g;x}$ component of an effective mode, adiabatically
transitioning from $w=1.2$ to $w=0$ to identify minima and from $w=-1.2$ to
$w=0$ to locate saddle point, thereby mitigating convergence issues associated
with degeneracy close to the high-symmetry point.

However, this technique sometimes results in divergent energies as $w$
approaches zero from positive and negative directions due to the
inability of the exchange-correlation functional to yield identical
energies for degenerate densities~\cite{Bruyndonckx1997}. We address
this issue by applying a rigid vertical shift to one of the branches,
ensuring identical energies at $w=0$.

Assuming only linear Jahn--Teller interaction, to simplify the estimation of
Jahn--Teller relaxation energy $E_{\mathrm{JT}}$ at different pressures and to
reduce the computational complexity, we develop a novel practical computational
procedure that avoids issues related to the degeneracy of electronic states.
Since the slice cut of sombrero APES is described by
$U(Q_{e_g;x}) = \omega^2Q_{e_g;x}^2/2 - V Q_{e_g;x}$, where $V$ is vibronic
coupling constant, energy minimum is at $\Delta Q_{e_g;\text{BS}} = V/\omega^2$,
yielding Jahn--Teller relaxation energy
$E_{\mathrm{JT}}=V^2/2\omega^2$~\cite{Bersuker2012}. However, using fractional occupations described above, we suppress Jahn--Teller
interactions, and the potential energy surface attains the harmonic form
$U_{\text{ad}}(Q_{gx}) = \omega^2Q_{e_g;x}^2/2$. Plugging in the coordinate of
the Jahn--Teller minimum into the harmonic potential we obtain
$U_{\text{ad}}(\Delta Q_{e_g;\text{BS}}) = V^2/2\omega^2 \equiv E_{\mathrm{JT}}$.
This approach allows us to estimate
$E_{\mathrm{JT}}$ from high-symmetry and broken-symmetry geometries and one
single-point calculation with factional occupations at broken-symmetry geometry.

\section{Results}
\label{sec:results}

Throughout this study, we characterize the magneto-optical properties of the negatively charged G4V($-$) color centers under compressive hydrostatic pressure. While not explicitly stated henceforth, all references to pressure will assume compressive hydrostatic conditions.

Under hydrostatic pressure, the fundamental band gap of the diamond widens, which also shifts the photoionization threshold levels of G4V defects. To account for this effect, we first compute the charge transition levels of G4V defects to identify their photostable excitation energies, assuming single photon absorption. Then, we compute the ZPL energies under increasing hydrostatic pressure up to 180~GPa, which gives information about the fluorescence conditions for each G4V($-$) color center at a given hydrostatic pressure.

After
  exploring charge dynamics under optical illumination, we focus on the effects
  of compressive hydrostatic pressure on the fine-level structure of G4V($-$)
  centers in the electronic ground and excited states. In particular, the spin-orbit splitting is determined with hyperfine tensors for the relevant isotopes of the
dopant and neighbor $^{13}$C nuclear spins and the
quadrupole moments of the relevant dopants in G4V($-$) centers. These data serve as spectroscopic fingerprints of the G4V($-$)
centers under specific hydrostatic pressure conditions. Furthermore, this information is essential in understanding the temperature-dependent
electron spin coherence times of G4V($-$) color centers. These spin interactions
are strongly interwoven with electron-phonon interaction, which we treat with Jahn--Teller theory,
as briefly described in the Method section. Consequently, the Jahn--Teller parameters are calculated at each
considered hydrostatic pressure for each G4V($-$) color center at the electronic
ground and excited states and will be reported before providing the fine-level
structure.

\subsection{Photoionization thresholds and zero-phonon line energies}
\label{ssec:PL}

\begin{figure*}[t]
\includegraphics[scale=0.30]{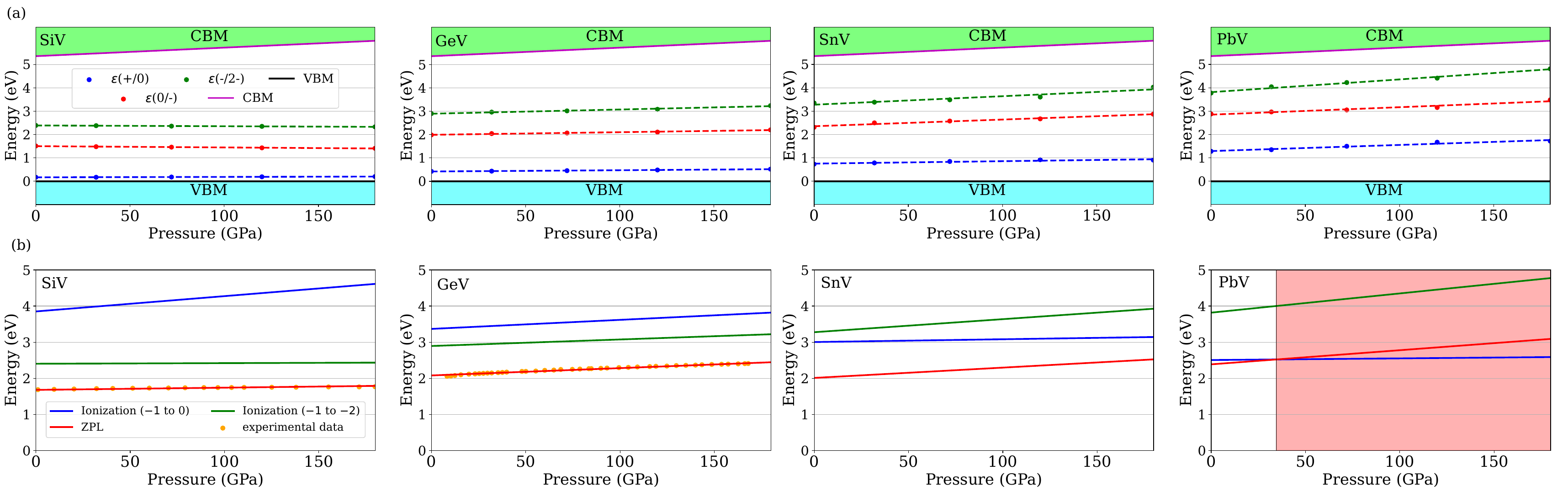}
\caption{\label{fig:CHGtrns}
(a) Calculated charge transition levels of SiV, GeV, SnV and PbV defects under hydrostatic pressure in the range of $0$ to $180$~GPa as obtained by HSE06 functional. (b) The ZPL shift under hydrostatic pressure was calculated by SCAN functional where the zero pressure value was aligned to the experimental data. 
\textcolor{black}{Calculated ZPL curves have been rigidly shifted to their experimental
zero-pressure values (1.68 eV for SiV, 2.06 eV for GeV, and 2.00 eV for
SnV). The slight deviation of the GeV data below 20 GPa is due to a
pressure-calibration bias during diamond
anvil cell loading (see Ref.~\cite{Vindolet2022}).} For SiV and GeV defects the experimental data points for non-zero pressure are also given from Ref.~\onlinecite{Vindolet2022}. We highlight the hydrostatic pressure region by light red color where PbV($-$) is not photostable. We list the charge correction to the total energy and dielectric constant dependency on pressure in Table~\ref {table:eps}. 
}
\end{figure*}

The photoionization threshold energies can be calculated from the charge
transition levels referenced to the appropriate band edge. In particular, the
photostability of G4V($-$) is of interest. In this case, the $(-|0)$ level with
respect to the conduction band minimum (CBM) yields the photoionization
threshold for converting the negatively charged defect to a neutral one, whereas
the $(2-|-)$ level with respect to the valence band maximum (VBM)
yields the photoionization threshold energy for converting the negatively
charged defect to doubly negative charged one. If the ZPL energy of the G4V($-$) center exceeds either of these
  photoionization threshold energies, the G4V($-$) center cannot be considered
  photostable. In this case, excitation energies at or above the ZPL could
  induce photoionization, transferring the defect to an alternative charge
  state.

The charge transition levels of defects in diamond can be accurately calculated using the HSE06
  functional, with an accuracy of approximately $0.1$~eV~\cite{Deak2010}.
  Therefore, we employ this method in our study. The computational
  approach for determining the charge transition levels of G4V defects follows
  the standard procedure outlined in our previous study~\cite{Thiering2018}
and is based on the methodology
  detailed by Freysoldt \textit{et al.}~\cite{Freysoldt2014}, which we do not reiterate here. We note that the reported values are
valid at $T=0$~K, and phonon-assisted photoionization can decrease 
the effective thresholds at elevated temperatures (e.g., Ref.~\cite{csore2022}).

We find that the increase of hydrostatic pressure from $0$ to $180$~GPa leads to
the widening of the fundamental band gap of diamond by $0.66$~eV. The calculated
charge transition levels of G4V($-$) shift upward with increasing pressure. The
general trend is that the heavier the dopant ion, the steeper the upward shift in the charge transition level (Fig.~\ref{fig:CHGtrns}).
The same trend can be observed in the shift of ZPL. We already discussed its
origin in our previous study for SiV($-$) and GeV($-$) color
centers~\cite{Vindolet2022}, which can be qualitatively explained through an
investigation of Kohn–Sham molecular orbital states using a
single-particle picture. In the first order, the ZPL energy scales with the
difference between the unoccupied and occupied states, named $e_g$ and $e_u$.
Thus, the change in ZPL energy upon varying hydrostatic pressures can be
estimated as $E_{\text{ZPL}}(p) \propto \varepsilon_u(p)-\varepsilon_g(p)$,
where $\varepsilon_g(p)$ and $\varepsilon_u(p)$ are the Kohn–Sham energies of
the orbitals $|e_g\rangle$ and $|e_u\rangle$ at pressure $p$, respectively. The crucial
distinction among the response to hydrostatic pressures is the deformation
potential of the $|e_g\rangle$ orbital, which transitions from bonding to antibonding
character as the atomic number of impurity atom increases~\cite{Vindolet2022}.
Although PbV($-$) shows the most sensitive pressure
dependence, this also applies to photoionization
thresholds. We find that the photoionization threshold towards CBM crosses the
ZPL energies around $32$~GPa hydrostatic pressure, which
sets a limit for PbV($-$) in pressure sensor applications.

\subsection{Dynamic Jahn-Teller effect and electron-phonon parameters}
\label{ssec:JT}

The electronic structure of both the ground and
excited states of the G4V($-$) centers is subject to the
dynamic Jahn--Teller effect, as thoroughly characterized in
Ref.~\onlinecite{Thiering2018} and subsequently utilized in further studies~\cite{Krivobok2020, Vindolet2022,Harris2023}. In DFT calculations,
geometry optimization follows the
  sombrero hat potential, resulting in a low-symmetry $C_{2h}$ geometry with
the global energy minimum at zero pressure.
  However, the system can be constrained to retain the high-symmetry
  $D_{3d}$ configuration by employing the fractional occupations described
  above. In $D_{3d}$ symmetry, the distances between the three symmetrically
equivalent neighbor carbon atoms are labeled by $d_{2}$, whereas the distance
between the dopant ion and the nearest neighbor carbon atoms is labeled by $d_1$
in Fig.~\ref{fig:JT}(a), and the values are listed in Table~\ref{table:str}. The
general trend is that $d_1$ and $d_2$ values increase with heavier ions in
G4V($-$) centers at zero pressure. Such behavior may be expected
as a larger ion creates a stronger strain field around the dopant ion. For a
given G4V($-$) center, the $d_1$ and $d_2$ values slowly decrease with
the increase of hydrostatic pressure.
This response may be also expected as the hydrostatic pressure
generally decrease the bond length in the diamond crystal which also holds for
the bond lengths in the center of G4V defects.

\begin{table}[!ht]
    \caption{\label{table:str}
    The structural parameters of G4V centers in $D_{3d}$ symmetry are denoted as $d_1$ and $d_2$ as obtained by SCAN calculations.
    These parameters correspond to the bond length between the dopant atom and their first neighbor carbon atoms and the distances
    between these carbon atoms, respectively, measured in \AA ngstr\"om unit~(\AA).
    The $d_1$ and $d_2$ distances in the ground state at zero pressure as obtained by SCAN/HSE06 are also listed for comparison.}
    \begin{ruledtabular}
    \setlength\extrarowheight{4pt}
    \setlength\tabcolsep{6pt}
    \begin{tabular}{l cc cc}
        \multicolumn{1}{c}{} & \multicolumn{2}{c}{0 GPa} & \multicolumn{2}{c}{180 GPa} \\
        \cline{2-3}\cline{4-5}
        Defect& $d_1$ & $d_2$ & $d_1$ & $d_2$ \\ \hline
         \multicolumn{5}{c}{Ground state} \\ \hline
        SiV($-$) & 1.97$/$1.96\footnotemark[1] & 2.67$/$2.67\footnotemark[1] & 1.79 & 2.40 \\
        GeV($-$) & 2.01$/$2.01\footnotemark[1] & 2.74$/$2.73\footnotemark[1] & 1.84 & 2.48 \\
        SnV($-$) & 2.09$/$2.08\footnotemark[1] & 2.83$/$2.83\footnotemark[1] & 1.92 & 2.58 \\
        PbV($-$) & 2.12$/$2.12\footnotemark[1] & 2.89$/$2.88\footnotemark[1] & 1.96 & 2.64 \\
         \multicolumn{5}{c}{Excited state} \\ \hline
        SiV($-$) & 1.96 & 2.66 & 1.78 & 2.39 \\
        GeV($-$) & 2.02 & 2.74 & 1.85 & 2.48 \\
        SnV($-$) & 2.09 & 2.84 & 1.93 & 2.59 \\
        PbV($-$) & 2.15 & 2.91 & 1.98 & 2.66 \\
    \end{tabular}
    \end{ruledtabular}
    \footnotetext[1]{Ref.~\onlinecite{Thiering2018}}
\end{table}

\begin{table*}[!ht]
\setlength\extrarowheight{8pt}
\caption{\label{table:lambda}
The calculated Jahn-Teller (JT) energy in meV ($E_\text{JT}$), the barrier energy in meV ($\delta_\text{JT}$), the effective mode in meV ($\hbar \omega$), spin-orbit splitting $\lambda_0$ (meV), $p$ reduction factor, $g$ the orbital reduction factor, and $\lambda$ (GHz) which is Ham reduced $\lambda_0$ for the G4V($-$) defects ($\lambda=p\lambda_0$) with no applied pressure. The left/right hand side data are from SCAN/HSE06 calculations.}
\begin{ruledtabular}
    \begin{tabular}{ccccccccc}
            Defects & $E_\text{JT}$ (meV) & $\delta_\text{JT}$ (meV) & $\hbar \omega$ (meV) & $\lambda_\text{0}$ (meV) & $p$ & $g$ &$\lambda$ (GHz) & $\lambda_\text{exp}$ (GHz)\\ \hline
    &&&&Ground state&&\\ \hline
      SiV($-$)   & 40.9 / 42.3\footnotemark[1] & 3.79 / 3.0\footnotemark[1] &89.7 / 85.2\footnotemark[1] & 0.86 / 0.82\footnotemark[1] & 0.34 / 0.31\footnotemark[1] &0.328\footnotemark[1] & 70.26 / 61.0\footnotemark[1]& 50\footnotemark[2]  \\
        GeV($-$) & 30.6 / 30.1\footnotemark[1] & 4.05 / 2.0\footnotemark[1] & 77.0 / 82.2\footnotemark[1] & 2.45 / 2.20\footnotemark[1] & 0.38 / 0.39\footnotemark[1] &0.328\footnotemark[1]&222.8 / 207\footnotemark[1] & 181\footnotemark[3]  \\
        SnV($-$) & 20.8 / 21.6\footnotemark[1] & 1.15 / 1.6\footnotemark[1] & 64.9 / 79.4\footnotemark[1] & 8.69 / 8.28\footnotemark[1] & 0.44 / 0.47\footnotemark[1] &0.328\footnotemark[1]& 915.2 / 946\footnotemark[1] &850\footnotemark[4] \\
        PbV($-$) & 15.0 / 15.6\footnotemark[1] & 3.88 / 0.6\footnotemark[1] & 52.0 / 74.9\footnotemark[1] & 35.0 / 34.6\footnotemark[1] & 0.48 / 0.54\footnotemark[1] &0.328\footnotemark[1]& 4097 / 4514\footnotemark[1] & 3914\footnotemark[5] \\
    &&&&Excited state&&\\ \hline
        SiV($-$) & 62.6 / 78.5\footnotemark[1]& 1.12 / 2.7\footnotemark[1] & 61.0 / 73.5\footnotemark[1] & 8.864 / 6.96\footnotemark[1] & 0.133 / 0.128\footnotemark[1] &0.782\footnotemark[1]& 286.2 / 215\footnotemark[1]  & 260\footnotemark[2] \\
        GeV($-$) & 71.5 / 85.7\footnotemark[1]& 2.31 / 5.4\footnotemark[1] & 70.6 / 73.0\footnotemark[1] & 35.03 / 36.1\footnotemark[1] & 0.136 / 0.113\footnotemark[1] &0.782\footnotemark[1]& 1155 / 987\footnotemark[1]  & 1120\footnotemark[3] \\
        SnV($-$) & 67.7 / 83.1\footnotemark[1]& 4.20 / 6.8\footnotemark[1] & 68.1 / 75.6\footnotemark[1] & 94.77 / 96.8\footnotemark[1] & 0.140 / 0.125\footnotemark[1] &0.782\footnotemark[1]& 3214 / 2897\footnotemark[1]  & 3000\footnotemark[4]  \\
        PbV($-$) & 87.3 / 91.6\footnotemark[1]& 6.69 / 12.3\footnotemark[1]& 77.9 / 78.6\footnotemark[1] & 241.4 / 245\footnotemark[1]  & 0.116 / 0.119\footnotemark[1] &0.782\footnotemark[1]& 6782 / 7051\footnotemark[1]  & - \\
    \end{tabular}
\end{ruledtabular}
\footnotetext[1]{Ref.~\onlinecite{Thiering2018}}
\footnotetext[2]{Ref.~\onlinecite{Hepp2014}}
\footnotetext[3]{Ref.~\onlinecite{Ekimov2015}}
\footnotetext[4]{Ref.~\onlinecite{Iwasaki2017}}
\footnotetext[5]{Ref.~\onlinecite{Wang2021}}
\end{table*}

For each applied hydrostatic pressure, we calculated the APES by SCAN
functional, as illustrated in Fig.~\ref{fig:JT}(d). For the zero-pressure case,
we compared the SCAN data with previous HSE06 data (see
Table~\ref{table:lambda}). We found a good agreement between the
SCAN and HSE06 data for the
critical JT parameters (such as the Jahn--Teller energy $E_\text{JT}$, barrier
energy $\delta_\text{JT}$, and the effective frequency $\hbar \omega$), so we
continued the calculations with SCAN functional for non-zero pressure
cases. 
Table~\ref{table:lambda-HS-pressure} in the
  Appendix lists the respective JT parameters for G4V($-$) centers under pressure.

Having these JT parameters in hand is critical in calculating the effective
spin-orbit coupling as well as the effective hyperfine tensors for the
respective nuclear spins proximate to the G4V defects. These will be discussed
below in the next sections.

\subsection{Effective spin-orbit splitting}
\label{ssec:SOC}

We calculated the spin-orbit coupling using the SCAN
functional, which is an essential parameter in capturing
the fine electronic structure of G4V($-$) centers. We label this electronic
spin-orbit coupling as $\lambda_0$, listed for each G4V($-$)
for the zero-pressure case in Table~\ref{table:lambda}. Our results show strong agreement between the SCAN data and the
previously published HSE06 $\lambda_0$ values
for each electronic state in the G4V($-$) color centers. This
  consistency supports confidence that using SCAN functional, the electronic spin-orbit coupling can be further calculated for the
G4V($-$) color centers under non-zero hydrostatic pressures.

\begin{figure*}[t]
\includegraphics[scale=0.30]{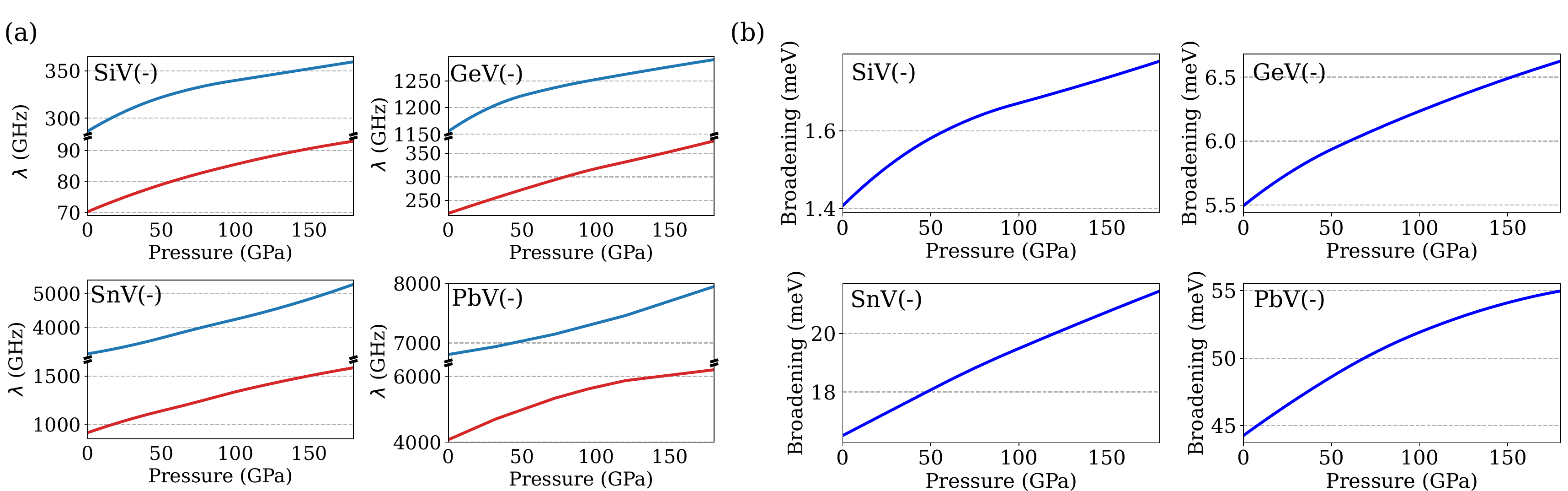}
\caption{\label{fig:Lambda} Pressure dependent spin-orbit splitting as obtained
  with SCAN functional where the spin-orbit interaction added as perturbation to
  the Jahn--Teller-effect. (a) The calculated hydrostatic pressure (GPa)
  dependence of the effective spin-orbit splitting $\lambda$ in GHz unit for
  G4V($-$) color centers in the electronic excited and ground states plotted by
  blue and red lines, respectively. (b) The sum of the electronic excited and
  ground state's $\lambda$ values associated with the zero-phonon line
  broadening at elevated measurement temperatures with off-resonant excitation
  for each G4V($-$) color center as a function of the applied hydrostatic
  pressure.}
\end{figure*}

In Jahn-Teller active systems, the electronic spin-orbit coupling
is reduced due to strong electron-phonon coupling, and this reduction is quantified by the Ham reduction factor that we
label by $p$. In our previous work~\cite{Thiering2017},
  we described the \textit{ab initio} calculation of the Ham reduction factor
  for the $E \otimes e$ Jahn-Teller system, which was later applied to G4V($-$)
  defects in a subsequent study~\cite{Thiering2018}. For
the zero pressure case, we find that the calculated $p$
factors and the resulting $\lambda$ for the electronic ground and excited state
of G4V($-$), as obtained by SCAN and HSE06 functionals, are consistent. The calculated
  $\lambda$ values also show good agreement with experimental data for both
  functionals (see Table~\ref{table:lambda}). This
  result further supports the applicability of the SCAN functional for
  investigating the dependence of effective spin-orbit splitting on hydrostatic
  pressure in G4V($-$) color centers.

Fig.~\ref{fig:Lambda}(a) shows the calculated effective spin-orbit
splitting as a function of hydrostatic pressure,
obtained using SCAN functional. The general trend is that the effective spin-orbit
splitting is monotonously increasing with increasing hydrostatic pressure both
in the electronic ground and excited states. The calculated effective spin-orbit
splitting has two factors: the Ham-reduction parameter $p$ and the calculated
electronic spin-orbit splitting $\lambda_{0}$ (see
Table~\ref{table:lambda-HS-pressure}). We find that the Ham-reduction parameter
only slightly varies as a function of hydrostatic pressure.
Therefore, the primary
  contribution to the overall increase in spin-orbit coupling under rising
  hydrostatic pressure comes from the inherent electronic spin-orbit splitting.
This change can be explained by analyzing the
  dominant contribution to the electronic spin-orbit splitting,
  which arises from the spin-orbit coupling on the dopant
ion~\cite{Thiering2018}. As
  hydrostatic pressure increases, the distance ($d_1$) between the dopant ion
and the neighbor carbon atoms decreases, enhancing
the overlap of the dangling bonds with the dopant ion
(see also Sec.~\ref{ssec:hyperfine}). Consequently,
the wavefunction cloud around the dopant ion becomes denser, contributing significantly to the spin-orbit coupling. The
steepest shift in the effective spin-orbit splitting is
observed for the SnV($-$) color center.

In Fig.~\ref{fig:Lambda}(b), we plot the sum of $\lambda_g$ and
$\lambda_u$ -- defined in Fig.~\ref{fig:JT}(c) -- as a function of the applied
hydrostatic pressure, which we discuss
in the context of the broadening of the ZPL emission. Off-resonant excitation of
G4V($-$) color centers should result in emission in all the possible
combinations of states and levels depicted in Fig.~\ref{fig:JT}(c). By assuming
a homogeneous strain field and neglecting the isotope effects (in the GHz
region), the sum of $\lambda_g$ and $\lambda_u$ yields the width of the
zero-phonon line. In experiments, the broadening of ZPL for SiV($-$) varied from 10 to 30~meV going from zero to $180$~GPa of
hydrostatic pressure, whereas it varied from 15 to
  30~meV for GeV($-$), respectively~\cite{Vindolet2022}. In our calculations,
the spin-orbit splitting related broadening of the ZPL emission varies between
1.4~meV to 1.8~meV for SiV($-$) going from zero to $180$~GPa of hydrostatic
pressure, whereas it varies between 5.5~meV to 6.7~meV for GeV($-$),
respectively. The calculated values are smaller in order of
magnitude both in terms of absolute values and changes upon
hydrostatic pressures. We conclude that before applying pressure to the sample,
non-uniform local strain fields could be already present in the samples, which
causes inhomogeneous broadening in the region of 10~meV. Applying hydrostatic pressure to the samples could
further increase the variety of the local strain fields that may explain the
observed wide broadening of the ZPL emission for SiV($-$) and GeV($-$) optical centers.

\subsection{Hyperfine interaction}
\label{ssec:hyperfine}

We computed the hyperfine tensors for the relevant isotopes of the dopant ion
and proximate $^{13}$C $I=1/2$ nuclear spins that can be optically addressed in
a resonant manner~\cite{Becker2016,Green2017}. The $^{13}$C nuclear spins occur
with a 1.1\% abundance in natural diamonds. The G4V($-$) color centers are often
generated by ion implantation, for which the isotope can be well controlled. The
non-zero nuclear spin isotopes are $^{29}$Si, $^{73}$Ge, $^{117}$Sn, and
$^{207}$Pb with respective spins of $I=1/2$, $9/2$, $1/2$, and $1/2$. For
$^{73}$Ge ($I=9/2$), the nuclear quadrupole moment also contributes to the
fine-level structure of the GeV($-$) color center, which we also consider in
this section.

If these nuclear spins can be well controlled, then they could be applied as
quantum memory of the system~\cite{Metsch2019, Knaut2024} but may also contribute
to extra spin flip-flop processes by reducing the coherence time of the electron
spin of G4V($-$) centers. Therefore, understanding the full hyperfine tensor and
its effects on the system is important in quantum technology applications.
As phonons dynamically distort the
  electronic states via the Jahn-Teller effect, the hyperfine and quadrupole
  matrix elements become modulated by the vibrational motion, which plays a
  vital role in spin relaxation processes~\cite{Monge2023, Thiering2024}.

We note that the orbital angular momentum is significantly quenched but cannot
be entirely neglected~\cite{Thiering2018}. As a consequence, the remaining orbital
angular momentum may contribute to the effective dipole-dipole hyperfine
coupling. On the other hand, we assume that this contribution is small (e.g.,
a few percent), and we do not consider it further in this context.

The strong electron-phonon coupling is responsible for relatively short
coherence times of the electron spin in G4V($-$) defects, even at cryogenic
temperatures~\cite{Norambuena2016}. Therefore, no electron spin resonance
measurements have been reported for G4V($-$) defects. The hyperfine structure
can be probed optically. Resonant excitation is often performed at the
$|{}^2E_{3/2,u}\rangle$ level, resulting in $|{^2\!E_{3/2,u}}\rangle \rightarrow |{^2\!E_{3/2,g}}\rangle$
optical transitions that appear in the zero-phonon line (ZPL) fluorescence.
Because the nuclear spin state is conserved during such optical transitions,
only transitions between states with the same nuclear spin projection $m_I$ are
allowed. This gives rise to transitions such as
$|{^2\!E_u^{1}}\rangle \leftrightarrow |{^2\!E_g^{1}}\rangle$ and
$|{^2\!E_u^{2}}\rangle \leftrightarrow |{^2\!E_g^{2}}\rangle$. [see
Fig.~\ref{fig:JT}(b)]. In the case of $I=1/2$, two hyperfine-resolved lines
appear in the spectrum, separated by
$A_{\text{PLE}}^{\text{exp}}=A_{\|,u} - A_{\|,g}$.

In typical experimental setups, relatively weak magnetic fields in the range of
1.5--40~G are applied to lift the degeneracy of the Kramers doublets that serve
as qubit states. When the magnetic field is aligned along the symmetry axis of
the defect, the $|E^{i}_{\{g,u\}\pm}\rangle$ levels are split to first order in
perturbation theory by $g \mu_B B$ [see
Eqs.~\eqref{eq:HFEnergiesE_1}--\eqref{eq:HFEnergiesA}], where $g \approx 2$ is the
effective $g$-factor in both the ground and excited state manifolds, $\mu_B$ is
the Bohr magneton, and $B$ is the applied magnetic field strength. Nuclear
Zeeman shifts are typically negligible in this regime due to their much smaller
magnitude.

\begin{figure}
\includegraphics[width=0.7\linewidth]{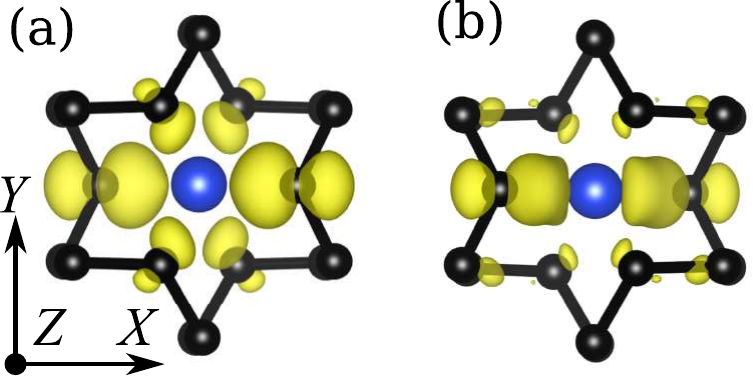}
\caption{\label{fig:spin-hf} The electron spin density of the SiV($-$) center with
  low $C_{2h}$ symmetry in the electronic (a) ground and (b) excited states
  depicted with the isosurface value of $5\times 10^{-3}$~$e/$\AA.}
\end{figure}
%
\begin{table*}[!ht]
  \caption{Hyperfine parameters of the G4V($-$) centers calculated using the
    HSE06 functional in the $C_{2h}$ symmetry configuration. The parameters
    $A_{\|}$, $A_{\perp}$, $A_1$, and $A_2$ are defined in Eqs.~\eqref{eq:HFD3d}
    and~\eqref{eq:HFparameters}, and reported here as reduced values using the
    reduction factors listed in Table~\ref{table:hyperfine-C13}. The calculated
    $A_\text{PLE}$ values correspond to the zero-field splitting observed in
    optical transitions between the $E_{3/2,u}$ and $E_{3/2,g}$ states (see main
    text for details). Experimental values $A_\text{PLE}^\text{exp}$ are taken
    from different studies conducted at various magnetic field strengths. Our
    calculations indicate that fields in the range of 10--50~G affect
    $A_\text{PLE}$ by less than 0.15\%.\label{table:hyperfine}}
    \begin{ruledtabular}
        \setlength\extrarowheight{6pt}
        \setlength\tabcolsep{-2pt}
        \begin{tabular}{cccccccc}
        Nuclear        & Spin & $A_1$~(MHz) & $A_2$~(MHz) & $A_{\|}$~(MHz) & $A_{\bot}$~(MHz) & $A_\text{PLE}$~(MHz) & $A_\text{PLE}^\text{exp}$~(MHz) \\ \hline
           &      &             &             &          Ground state      &                                                                           \\ \hline
         $^{29}$Si     & $1/2$  & $-2.9$      & $-5.9$      & $41.7$        & 88.7             & $-39.3$              & 35\footnotemark[1]              \\
         $^{73}$Ge     & $9/2$  & $0.8$       & $1.7$       & $20.6$         & 44.0             & $-18.4$              & $-12.5(5)$\footnotemark[2]      \\
         $^{117}$Sn    & $1/2$  & $1.1$       & $1.9$       & $488.0$        & 1029.7           & $-473.1$            & $-484(8)$\footnotemark[2]       \\
         $^{207}$Pb    & $1/2$  & $-1.6$      & $1.0$       & $-574.7$      & $-1192.0$        & 565.0                & -                               \\
          &      &             &             &         Excited state       &                                                                           \\ \hline
         $^{29}$Si     & $1/2$  & $-0.03$     & $-0.02$     & $2.4$          & $5.1$              & -                    & -                               \\
         $^{73}$Ge     & $9/2$  & $0.22$      & $0.52$      & $2.3$         & 4.6              & -                    & -                               \\
         $^{117}$Sn    & $1/2$  & $0.1$       & $-0.43$      & $15.0$        & $32.3$           & -                    & -                               \\
         $^{207}$Pb    & $1/2$  & $0.06$      & $0.77$       & $-9.7$         & $-20.76$         & -                    & -                               \\
        \end{tabular}
      \end{ruledtabular}
      \footnotetext[1]{Ref.~\onlinecite{Rogers2014b}}
      \footnotetext[2]{Ref.~\onlinecite{Harris2023}}
\end{table*}

Table~\ref{table:hyperfine} lists the calculated hyperfine parameters for
dopants in G4V($-$) centers. These values were obtained using the HSE06
functional to ensure accurate spin density distributions near the nuclear spins.
In all investigated G4V($-$) color centers, we find that the spin density around
the dopant ion is lower in the electronic excited state compared to the
electronic ground state, as illustrated in Figs.~\ref{fig:spin-hf}(a) and (b).
This difference can be understood from symmetry considerations: the excited
state wavefunction has odd (ungerade) parity, which excludes contributions from
$s$-type orbitals since they do not change signs under inversion. In contrast,
the ground state wavefunction has even (gerade) parity, allowing $s$-orbital
contributions that enhance the Fermi-contact interaction. As a result, the
static hyperfine parameters of dopant ions located at the inversion center are
more significant in the ground state than in the excited state. Consequently,
for dopants with non-zero nuclear spin ($I > 0$), the hyperfine splitting
$A_{\text{PLE}}$ becomes finite, enabling the optical resolution of hyperfine
lines. 

As shown in Table~\ref{table:hyperfine}, the dynamic hyperfine parameters are
small but non-zero in both the $|{}^2E_u\rangle$ and $|{}^2E_g\rangle$ states. The parameter $A_1$
reflects the degree to which the spin density bends out of the $xy$ plane toward
the $z$-axis, while $A_2$ quantifies the rotational intensity of the $\pi$
orbital within the $xy$ plane. Both $A_1$ and $A_2$ remain small in the ground
and excited states, indicating limited contribution from $\pi$ orbitals and
minimal out-of-plane spin density distortion [see Figs.~\ref{fig:spin-hf}(a,b)].
Nevertheless, these components of the hyperfine tensor play a role in shaping
the fine structure of the spin levels in G4V($-$) centers, as discussed below.

In the relevant $|{}^2E_{3/2, \{g,u\}}\rangle$ states split by the external 
magnetic field and hyperfine interaction terms $A_{\perp,\{g,u\}}$ and $A_{1,\{g,u\}}$ see Eqs.~\eqref{eq:HFEnergiesE_1} 
and \eqref{eq:HFEnergiesE_2} in the Appendix for details. However, in optical measurements, 
only the hyperfine splitting ($A_{\text{PLE}}$) of the photoluminescence excitation 
(PLE) peaks is measured that we show below
\begin{equation}
\label{eq:hfcorrection}
A_{\text{PLE}}(B)=\overset{A_{\text{PLE}}}{\overbrace{A_{\parallel,g}-A_{\parallel,u}}}\pm\left(\frac{A_{\perp,g}^{2}}{\lambda_{g}^{2}}-\frac{A_{\perp,u}^{2}}{\lambda_{u}^{2}}\right)\frac{\mu_{B}g_{S}B}{4} \text{,}
\end{equation}
where the magnetic field aligned to the symmetry axis. We note that the magnetic
field independent terms cancel out or just being negligible see Eq.~\eqref{eq:HF_PLE} 
for details. As a consequence, the optical transition between hyperfine lines will be weakly
magnetic field dependent: there will be an extra splitting in the hyperfine peaks at large fields which is very small ($\sim0.02$~MHz) at $B\sim 10\dots50$~Gauss that were typically applied in experimental studies~\cite{Harris2023, Rogers2014b}.

Nevertheless, by comparing the experimental data and our theory about
hyperfine-related splitting in the optical transition we find good agreement. We
conclude that the hyperfine couplings are accurately calculated with our method
at zero pressure. The hyperfine parameters in the ground and excited states
cannot be directly observed in quantum optics measurements so our data can be
used for analyzing G4V($-$) color centers. We provide further hyperfine data
about proximate $^{13}$C nuclear spins. These nuclear spins do not reside in the
symmetry axis therefore the static and dynamic hyperfine tensors have to be
derived for such cases that we present in Appendix~\ref{app:hfder}. The results
are listed in Table~\ref{table:hyperfine-C13}.

We also studied the quadrupole moment of Ge isotope with $I=9/2$ nuclear spin in
GeV($-$) for which the dynamical $Q_{1,2}$ may flip the nuclear spin as
explained in Refs.~\cite{Monge2023, Thiering2024}. The quadrupole moment ($Q$)
can be calculated as
\begin{equation}
\label{eq:quadrupolemoments}
    Q=\frac{\rho e^2 V_{zz}}{h}
\end{equation}
where $\rho$ is the nuclear quadrupole moment, $e$ is the elementary charge,
$V_{zz}$ is the electric field gradient (EFG) at the nucleus site and $h$ is
Planck's constant~\cite{HAOUAS2016}. With using
$\rho=-196\times10^{−31} \mathrm{m}^2$ (Ref.~\onlinecite{Pekka2008}) we obtained
$Q=-13.5$~MHz, $Q_1=-5.6$~MHz and $Q_2=-5.0$~MHz.

\begin{figure*}
\includegraphics[scale=0.23]{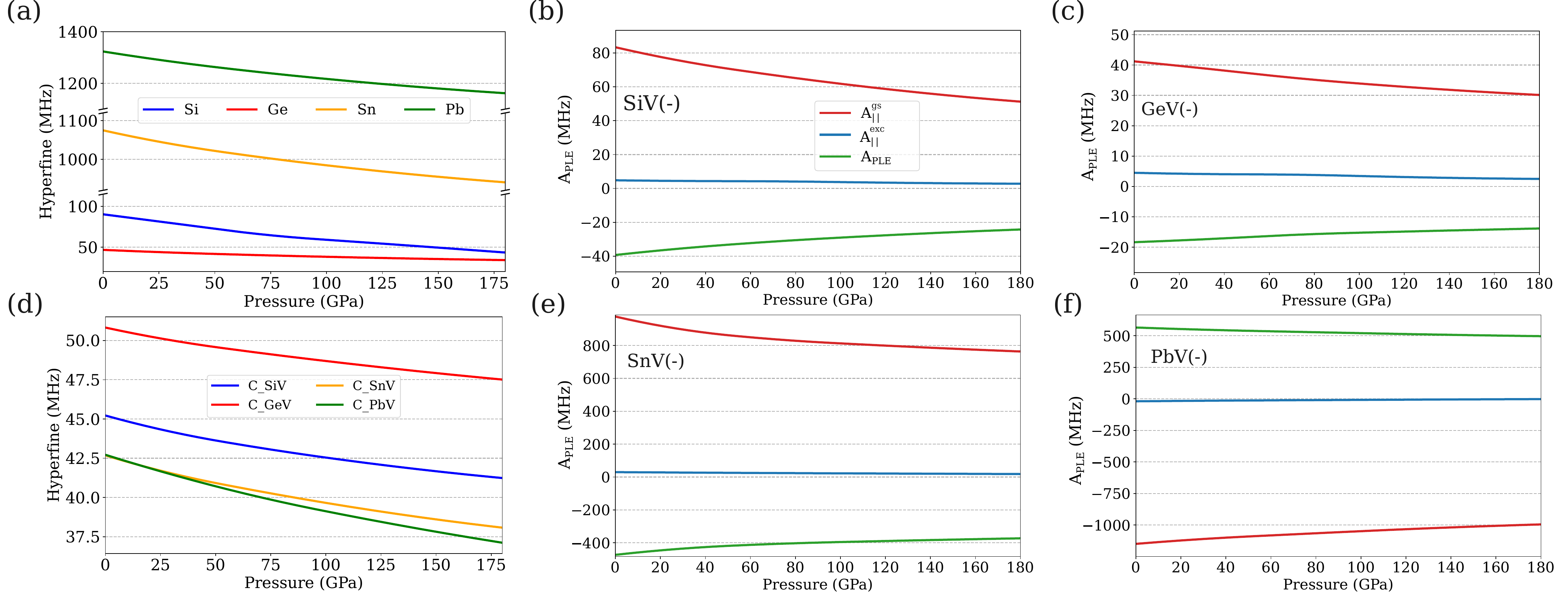}
\caption{\label{fig:hyperfine} Absolute value of the Fermi-contact hyperfine
  constants in G4V($-$) centers as obtained with HSE06 functional as a function
  of hydrostatic pressure for (a) dopant atoms and \textcolor{black}{(d)} first neighbor carbon
  atoms. 
  These calculations were carried out in $D_{3d}$ symmetry configurations.
  The illustration of A$_\text{PLE}$, $A_{\|}^\text{gs}$ and $A_{\|}^\text{exc}$ 
  as a function of hydrostatic pressure for dopant atoms\textcolor{black}{---depicted in green, red and blue colors, respectively---}(see text for details)
  with low $C_{2h}$ point group symmetry \textcolor{black}{for given defect centers of (b) SiV($-$), (c) GeV($-$) , (e) SnV($-$) and (f) PbV($-$)}. The plots were fit to quadratic functions.}
\end{figure*}

\begin{figure*}[t]
\includegraphics[scale=0.15]{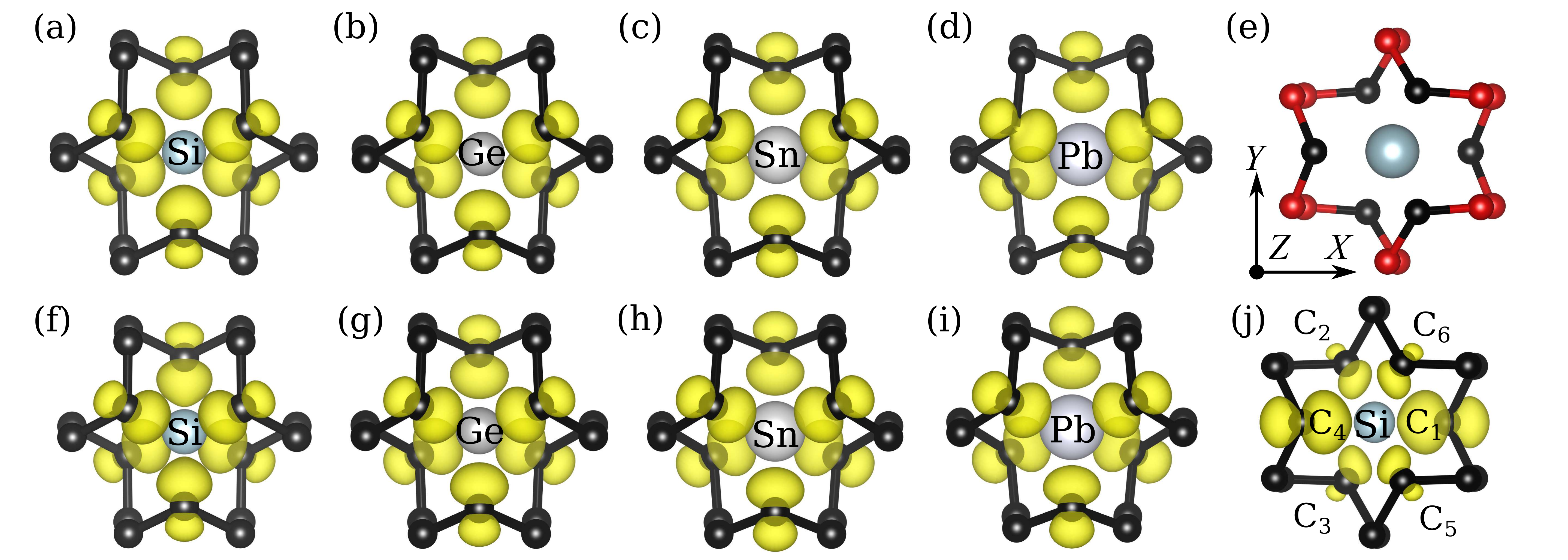}
\caption{\label{fig:spindens}
Illustration of the spin density in $D_{3d}$ symmetry. (a) SiV($-$), (b) GeV($-$), (c) SnV($-$) and (d) PbV($-$) defects with zero pressure, and (f) SiV($-$), (g) GeV($-$), (h) SnV($-$) and (i) PbV($-$) defects under 180~GPa hydrostatic pressure as obtained by HSE06 functional. (e) The positions of the first and second carbon neighbor atoms are depicted in black and red spheres, respectively. (j) The spin density of SiV($-$) representative of all G4V centers in $C_{2h}$ symmetry with no pressure. The electron spin density is depicted with an isosurface value of $5\times 10^{-3}$~$e/$\AA.
}
\end{figure*}

As a next step, we computed the pressure dependence of the hyperfine parameters.
For the sake of simplicity, we plot the absolute value of the Fermi-contact
hyperfine constant for the dopants and the nearest neighbor $^{13}$C nuclear
spins as a function of the applied hydrostatic pressure in
Figs.~\ref{fig:hyperfine}(a,b). We find that the hyperfine constants are
generally decreasing with increasing hydrostatic pressures in the electronic
ground state because the spin density is less localized with increasing
hydrostatic pressures (see Fig.~\ref{fig:spindens}). We also find that the
heavier the dopant the larger the change in the hyperfine constant on the dopant
ion upon hydrostatic pressure. We also plot the computed $A_\text{PLE}$ under
hydrostatic pressure which can be directly observed in quantum optics
measurements. We find that the absolute values of $A_\text{PLE}$ also reduces
with increasing hydrostatic pressure because the absolute value of the hyperfine
constants reduces much faster in the electronic ground state than that in the
electronic excited state.

\subsection{Zeeman shift of the optical emission}

Next, we investigated the effect of the constant external magnetic field on the
optical emission of the G4V centers when the magnetic field is perfectly aligned
with the symmetry axis of the defect. The most general microscopic theory of the
spin Hamiltonian is given in Refs.~\cite{Thiering2018, Thiering2020} that we do
not wish to reiterate here. In this microscopic theory, the magnetic field is in
part coupled to the orbital moment of the electron that is quenched by the Ham
reduction factor (already defined above) and the Stevens orbital reduction
factor~\cite{Stevens542} where the second one is associated with the shape of
the orbital (see Ref.~\onlinecite{Thiering2018}). We found earlier that the
Stevens orbital reduction factor is smaller in the electronic ground state than
that in the electronic excited state. We assume that the Steven reduction
parameters of the respective electronic states at zero pressure as listed in
Appendix~D of Ref.~\onlinecite{Thiering2018} are unaltered under hydrostatic
pressures. In the simulation of Zeeman shift we go beyond the above applied
theory and solve the spin-orbit and electron-phonon Hamitonians simultaneously
with the polaronic basis functions $\tilde{\Psi}_{\Gamma}$ with the spin degrees
of freedom as
\begin{equation}
\left|\tilde{\Psi}_{\Gamma}\right\rangle =\sum_{n,m}\left[c_{nm}^{\chi}\left|e_{g\pm}\right\rangle \left|n,m\right\rangle \left|\chi\right\rangle +d_{nm}^{\chi}\left|e_{g\mp}\right\rangle \left|n,m\right\rangle \left|\chi\right\rangle \right] \text{,}
\label{eq:DJTSOCseries}
\end{equation}
where $\chi$ can be either $\uparrow$ or $\downarrow$ spin state. This solution
represents a coupling between spins and phonons and goes \emph{beyond} the
perturbation theory of spin-orbit coupling acting on the polaronic
wavefunctions. Here, the subindex $\Gamma$ refers to the total angular momentum
of the wavefunction which is either $\frac{3}{2}$ or $\frac{1}{2}$. The
parameters of the electron-phonon coupling parameters via JT-effect and the
electronic spin-orbit parameter are listed in Tables~\ref{table:lambda} and
\ref{table:lambda-HS-pressure}.

In practical point of view, the orbital moment corrections contribute to the
effective $\mathbf{g}$-tensor of the system that may be simply written for the
ground state ($g$) and excited state ($u$) as
\begin{equation}
\label{eq:spinH}
\hat{H}^\text{eff}_{g,u} = \lambda_{g,u} + {\mu}_B  \mathbf{S} \mathbf{g}_{g,u} \mathbf{B}\text{,} 
\end{equation}
where $\mu_B$ is the Bohr magneton of the electron, $\mathbf{B}$ represents the
external homogeneous magnetic field. In this simplified formula, $\lambda_{g,u}$
contains all the electron-phonon related corrections whereas $\mathbf{g}$
amalgamates all the electron-phonon related corrections and orbital moment
interaction. We note that the corrections in the Zeeman term specifically occur
along the symmetry axis of the defect~\cite{Thiering2018, Thiering2020} that
causes an observable deviation from the Bohr magneton of the free electron at
2.0023 in the Zeeman shift when the magnetic field is aligned along the symmetry axis of the defect.

\begin{figure*}[t]
\includegraphics[scale=0.308]{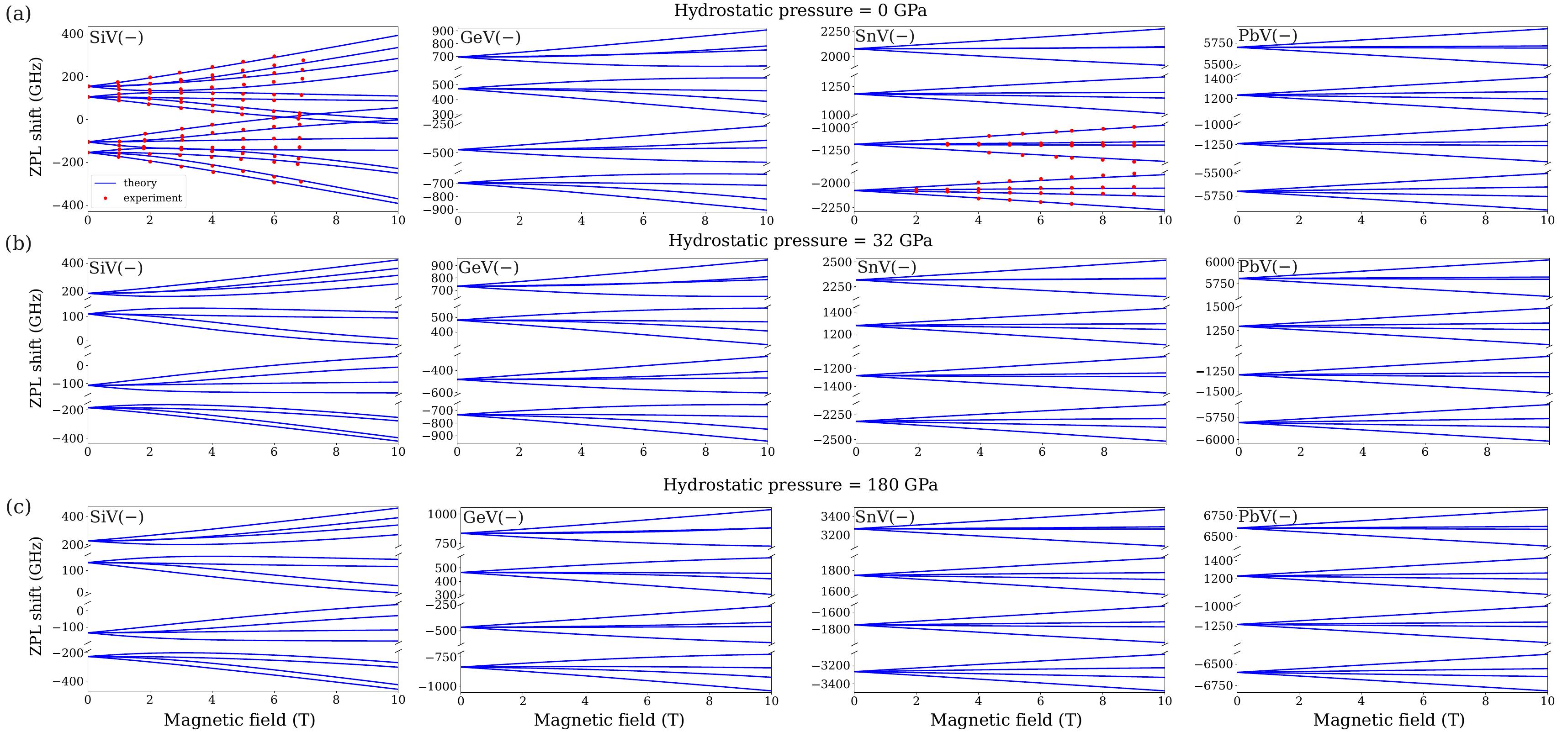}
\caption{\label{fig:Zeeman} The magnetic field dependency of the ZPL shift is
  plotted from 0 to 10~T where the magnetic field is aligned with the symmetry
  axis of the defect. We considered all four possible electronic transitions
  from excited to the ground states~\cite{Hepp2014} which resulted in four
  distinct branches of the ZPL for each color center. We performed the
  calculations for three different hydrostatic pressures of (a) 0, (b) 32, and
  (c) 180~GPa. Nevertheless, it is expected that the PbV($-$) may not exhibit
  photostable transitions at 180~GPa. The experimental data are depicted in red
  dots, while the simulated ZPL shifts are represented by blue solid lines.}
\end{figure*}

With the caveat about the approximations in the microscopic theory of spin
Hamiltonian, we calculated the splitting of the ZPL under magnetic fields
ranging from 0 to 10~T for G4V centers under hydrostatic pressures of 0, 32, and
180~GPa where the magnetic field is perfectly aligned with the symmetry axis of
the defects. We note that it is expected that the PbV($-$) may not exhibit
photostable optical transitions over 32~GPa but we still calculated it for the
sake of complete dataset. Initially, for SiV($-$) and SnV($-$) centers at zero
pressure, we achieved good agreement with experimental data~\cite{Hepp2014,
  Rugar2019} which motivated us to proceed with the calculations under
hydrostatic pressures. The results are plotted in Fig.~\ref{fig:Zeeman}. We
considered all four possible electronic transitions from excited $E_u$ to the
ground $E_g$ states~\cite{Hepp2014} which resulted in four distinct branches of
the ZPL for each color center. The effective g-factors can be read out from the
steepness of the curves. We find that the character of the Zeeman splitting
under high magnetic fields is basically not altered under high pressure. This is
very promising to apply these quantum sensors under extreme conditions.
\subsection{Estimation of the electron spin coherence times}

In this section, we present our estimation of the
electron spin coherence times, a pivotal parameter in
quantum information processing applications. We
  employ the phenomenological theory developed in Ref.~\onlinecite{Jahnke2015},
which describes the relaxation between $|E_{g\frac{3}{2}}
\rangle$ and $|E_{g\frac{1}{2}} \rangle$ orbital states mediated
  by symmetry breaking $E$ phonons. This model incorporates the electron-phonon
  coupling strength, $\chi$, as well as the phonon density of states, $\rho$, to
  describe the underlying relaxation process. The transition between two
Kramers states occur via orbital relaxation processes,
  characterized by the rates $\gamma_+$ and $\gamma_-$ [see
    Fig.~\ref{fig:JT}(c)]. In our theoretical
  framework, we incorporate the \textit{ab initio} effective spin-orbit
splitting which is
  essential for the estimation of the coherence times. It is important to note
that although symmetry-breaking $E$ phonons are mostly responsible for the
decoherence processes, their energies are much lower
than those quasilocalized $E$ phonons that cause Jahn--Teller-effect. As a
consequence, these two regimes are not intertwined; thus, it is legitimate to
apply the calculated effective $\lambda$ in this process and independently take
the electron-phonon coupling with phonons associated with the decoherence
process. Since long-wavelengths phonons responsible for decoherence cannot be
  directly calculated using our supercell formalism, we extract the parameters $\chi$ and $\rho$
from experimental data that are available at zero pressure. Furthermore we also
assume that $\chi$ and $\rho$ do not significantly change in the range of
applied hydrostatic pressure. We believe that this simple approximation suffices
to sketch the basic trends for the coherence times of the G4V($-$) centers.

The temperature as an essential parameter of spin coherence times delineates two
regions as found in Ref.~\onlinecite{Jahnke2015}: (i) the high-temperature
region where $T \gg \frac{\hbar\lambda_g}{k_\text{B}}$ and (ii) the
low-temperature region where $T \ll \frac{\hbar\lambda_g}{k_\text{B}}$. Here,
$\hbar$ and $k_\text{B}$ are the reduced Planck and Boltzmann constants,
respectively, and $\hbar \lambda_g$ is the effective spin-orbit strength in
energy unit. At high temperatures, the orbital relaxation rate scales as~\onlinecite{Jahnke2015}
\begin{equation}\label{eq:high-orbital-rate}
    \gamma_{+} \approx \gamma_{-} \approx \frac{2\pi}{\hbar} \chi \rho \lambda_g^2 k_\text{B} T \text{,}
\end{equation}
where $T$ denotes the temperature. In contrast, at low
temperatures, the orbital relaxation rate is given by
\begin{equation}\label{eq:low-orbital-rate}
    \begin{aligned}
        \gamma_{+} &\approx 2\pi \chi \rho \lambda_g^3 \exp\left(-\frac{\hbar\lambda_g}{k_\text{B} T}\right), \\
        \gamma_{-} &\approx 2\pi \chi \rho \lambda_g^3\text{.}
    \end{aligned}
\end{equation}
In this region, acoustic phonons contributing to $\gamma_+$ are exponentially
suppressed, while the nonradiative relaxation process described by $\gamma_-$
persists even at $T = 0$~K due to spontaneous decay (e.g., see
Ref.~\onlinecite{Thiering2021}).

In above equations, $\chi$ and $\rho$ always appear together as a product.
  Consequently, it is unnecessary to extract these non-calculable terms
separately; instead the combination $\chi \rho$ can be considered
as a single parameter. We applied the following strategy to extract this
parameter from experimental data. We identify the temperature region in terms of
the decoherence process at the given temperature and then we apply the
respective Eq.~\eqref{eq:high-orbital-rate} for temperature region (i) or
Eq.~\eqref{eq:low-orbital-rate} for temperature region (ii). Finally, taking the
experimental coherence time $\tau = \gamma^{-1}$ of the G4V($-$) center the
$\chi \rho$ parameter can be extracted. In this procedure, careful choice of the
experimental data is critical as the electron spin coherence time may be limited by proximate nuclear or electron spins in the system at a given temperature.
However, our theory only contains the spin-phonon related decoherence, thus the
experimental data should be selected from high quality samples to use
spin-phonon relaxation related coherence times of G4V($-$) centers.

Finally, we note that we apply the calculated effective spin-orbit parameter as
obtained by SCAN functional which generally overestimate the experimental data
at zero pressure (see Table~\ref{table:lambda}). As a consequence, the estimated
coherence times are typically shorter than
the experimentally observed values, reflecting the
  fact that larger spin-orbit splitting leads to reduced coherence times.
Nevertheless, we demonstrate below that the estimated values
are not far from the experimental ones, thereby, enabling a consistent characterization of the hydrostatic
  pressure trends using the SCAN functional dataset.

Most of the experimental data is available for SiV($-$) center at zero presure. The $\chi \rho$ value increases by raising the temperature from $2.204 \times 10^{-29}$ $T=0.1~\text{K}$~\cite{Pingault2017} to $1.164 \times 10^{-28}$ at $T=3.6~\text{K}$~\cite{Sukachev2017}. In particular, the measured spin coherence time of SiV($-$) with splitting of $\lambda_\text{exp}\approx50~\text{GHz}$ is reported by values of $\tau^\text{exp}_{3.6\text{K}}\approx115~\text{ns}$~\cite{Sukachev2017} and $\tau^\text{exp}_{0.1\text{K}}\approx13~\text{ms}$~\cite{Pingault2017}. Our SCAN functional yields $\lambda=70~\text{GHz}$~GHz which results in $\tau_{3.6\text{K}}\approx59~\text{ns}$ and $\tau_{0.1\text{K}}\approx$43~ms, respectively, that are in the same order of magnitudes of the experimental data. We further note that two measurement temperatures will fall to different temperature regions in terms of the decoherence process: $T=3.6~\text{K}$ belongs to region (i) whereas $T=0.1~\text{K}$ belongs to region (ii). As a consequence, the pressure dependence of the estimated coherence times behaves differently at the two temperatures: $\tau_{3.6\text{K}}$ values decrease with increasing the hydrostatic pressures whereas $\tau_{0.1\text{K}}$ values increase with increasing hydrostatic pressures (Table~\ref{table:spin-coh-SiV}).

Spin coherence time of 20~ms was reported~\cite{Senkalla2024} for GeV($-$) with
$\lambda_\text{exp}=181~\text{GHz}$ at $T = 300~\text{mK}$. The extracted $\chi \rho$
is $9.402 \times 10^{-31}$ and it belongs to the temperature region (ii). In
this case, the estimated spin coherence time increases from 49~ms (zero
pressure) up to 2.63~s at hydrostatic pressure of 180~GPa
(Table~\ref{table:spin-coh-GeV}).
  
Next, we study the coherence times of SnV($-$) based on the available
experimental data~\cite{Trusheim2020} at temperatures of 2.9~K and 6~K where
$\lambda_\text{exp}=850~\text{GHz}$ and the decoherence process goes in the region (ii)
at both temperatures. Subsequently, $\chi \rho$ is $1.175\times10^{-29}$ at
2.9~K and $1.986\times10^{-29}$ at 6~K. In our simulations, $\tau_{2.9\text{K}}$
values increase from 545~ns to $1.3~\mu\text{s}$ whereas $\tau_{6\text{K}}$ values
decrease from 55~ns to 35~ns (Table~\ref{table:spin-coh-SnV}).

To our knowledge, no coherence times for PbV($-$) has been reported. However, it is predicted to exhibit a millisecond spin coherence time at 9~K~\cite{Wang2021} with $\lambda_\text{exp}=3914~\text{GHz}$ which is located in temperature region (ii). As a result, $ \chi \rho=3\times10^{-34}$ may be extracted. Finally, the estimated spin coherence times raised from 1.29~ms to 1.85~ms at 9~K from zero to 32~GPa hydrostatic pressures at photostable region (Table~\ref{table:spin-coh-PbV}). 

\begin{table}[!ht]
    \caption{\label{table:spin-coh-SiV}
    Hydrostatic pressure (P) dependency of spin coherence time ($\tau$) at the given temperatures at zero to 180~GPa pressure for SiV($-$) center in diamond as obtained by SCAN functional. These parameters with $T=0.1$~K results in ($T \ll \frac{\hbar\lambda_g}{k_\text{B}}$) so Eq.~\ref{eq:low-orbital-rate} is applied whereas it yields ($T \gg \frac{\hbar\lambda_g}{k_\text{B}}$) with $T=3.6~\text{K}$ so Eq.~\ref{eq:high-orbital-rate} is applied.}
    \begin{ruledtabular}    
        \setlength\tabcolsep{0pt}
            \setlength\extrarowheight{5pt}
    \begin{tabular}{cccc}
         P (GPa) & $\lambda$ (GHz)& $\tau$ ($\tau_\text{exp}$) (ms) & $\tau$($\tau_\text{exp}$) (ns)  \\
         ~ & ~ & $T=0.1K$ & $T=3.6K$ \\\hline 
         0 & 70  & 43  (13\footnotemark[1])&  59 (115\footnotemark[2]) \\ 
        32 & 76  & 64 &  50\\ 
        72 & 82  & 98 &  43 \\ 
        120 & 88  & 147 &  38 \\ 
        180 & 93  & 220 &  34 \\  
    \end{tabular}
    \footnotetext[1]{Ref.~\onlinecite{Sukachev2017}}
    \footnotetext[2]{Ref.~\onlinecite{Pingault2017}}
    \end{ruledtabular}
\end{table}

\begin{table}[!ht]
    \caption{\label{table:spin-coh-GeV}
    Hydrostatic pressure (P) dependency of spin coherence time ($\tau$) at $T=0.3$~K at zero to 180~GPa pressure for GeV($-$) center in diamond as obtained by SCAN functional. These parameters results in ($T \ll \frac{\hbar\lambda_g}{k_\text{B}}$) so Eq.~\ref{eq:low-orbital-rate} is applied.}
    \begin{ruledtabular}    
        \setlength\tabcolsep{0pt}
            \setlength\extrarowheight{5pt}
    \begin{tabular}{ccc}
         P (GPa) & $\lambda$ (GHz)& $\tau$ ($\tau_\text{exp}$) (ms)  \\ \hline
        0 & 223 & 49 (20\footnotemark[1]) \\ 
        32 & 256 &   106 \\ 
        72 & 294 &   278  \\ 
        120 & 332 &   762 \\ 
        180 & 376 &  2626 \\ 
    \end{tabular}
\footnotetext[1]{Ref.~\onlinecite{Senkalla2024}}
    \end{ruledtabular}
\end{table}

\begin{table}[!ht]
    \caption{\label{table:spin-coh-SnV}
    Hydrostatic pressure (P) dependency of spin coherence time ($\tau$) at the given temperatures at zero to 180~GPa pressure for SnV($-$) center in diamond as obtained by SCAN functional. These parameters with $T=2.9$~K results in ($T \ll \frac{\hbar\lambda_g}{k_\text{B}}$) so Eq.~\ref{eq:low-orbital-rate} is applied whereas it yields ($T \gg \frac{\hbar\lambda_g}{k_\text{B}}$) with $T=6~\text{K}$ so Eq.~\ref{eq:high-orbital-rate} is applied.}
    \begin{ruledtabular}    
        \setlength\tabcolsep{0pt}
            \setlength\extrarowheight{5pt}
    \begin{tabular}{cccc}
         P (GPa) & $\lambda$ (GHz) &  $\tau$ ($\tau_\text{exp}$) (ns) & $\tau$ ($\tau_\text{exp}$) (ns)  \\ 
         ~&~& $T=2.9~\text{K}$ &$T=6~\text{K}$ \\\hline
         0 & 915 &  545 (540\footnotemark[1]) & 55 (59\footnotemark[1]) \\ 
        32 & 1068  & 608& 45 \\ 
        72 & 1224  & 724& 40 \\ 
        120 & 1407  & 946& 37 \\ 
        180 & 1586 & 1291 & 35 \\ 
    \end{tabular}
\footnotetext[1]{Ref.~\onlinecite{Trusheim2020}}
    \end{ruledtabular}
\end{table}

\begin{table}[!ht]
    \caption{\label{table:spin-coh-PbV}
    Hydrostatic pressure (P) dependency of spin coherence time ($\tau$) at $T=9~K$ at zero to 180~GPa pressure for PbV($-$) center in diamond as obtained by SCAN functional. These parameters results in ($T \ll \frac{\hbar\lambda_g}{k_\text{B}}$) so Eq.~\ref{eq:low-orbital-rate} is applied.}
    \begin{ruledtabular}    
        \setlength\tabcolsep{0pt}
            \setlength\extrarowheight{5pt}
    \begin{tabular}{ccc}
         P (GPa) & $\lambda$ (GHz)& $\tau$ ($\tau_\text{exp}$) (ms)  \\ \hline
         0 & 4436 & 1.29 (1.0\footnotemark[1])  \\ 
        32 & 5065 & 1.85  \\ 
        72 & 5744 & 2.87  \\ 
        120 & 6387 & 4.54  \\ 
        180 & 6593 & 5.29  \\ 
    \end{tabular}
\footnotetext[1]{Ref.~\onlinecite{Wang2021}}
    \end{ruledtabular}
\end{table}

\section{Summary}

In this work, we studied the pressure-dependent magneto-optical spectrum of G4V centers while assuming compressive hydrostatic pressures by means of plane wave supercell density functional theory calculations. We showed that the strong electron-phonon coupling significantly modifies the electronic spin-orbit, quadrupole and hyperfine interactions. In particular, we developed a theory for calculating the effective hyperfine tensors for $E\otimes e$ Jahn-Teller systems and applied it to the G4V centers. We find that all these parameters shift with hydrostatic pressures and cross-correlation measurements of these parameters may be applied to deduce the external hydrostatic pressures. We find that the maximum pressure is set at about 32~GPa for PbV($-$) center based pressure sensor because the photostability of PbV($-$) is compromised at larger pressures. The estimated pressure-dependent coherence times of G4V($-$) centers imply that they can be applied as quantum sensors under extreme conditions, including high constant magnetic fields when aligned with the symmetry axis of the defect centers.
\textcolor{black}{To put into perspective the pressure response of the G4V centers, we benchmark our computed deformation potentials against the well-characterized nitrogen-vacancy [NV($-$)] center in diamond. Several experimental works have established the NV($-$) center in diamond as a robust in situ pressure gauge under hydrostatic conditions.  In particular, the 637~nm zero‐phonon line exhibits a linear blue‐shift of approximately $5.8\ \mathrm{meV/GPa}$ up to $\sim60$~GPa~\cite{Doherty2014}, and the ground‐state zero‐field splitting $D$ increases at $\sim14.6\ \mathrm{MHz/GPa}$~\cite{Hilberer2023}.  More recent studies have extended optical detection of magnetic resonance (ODMR) to pressures exceeding 130~GPa in improved hydrostatic-loading environments~\cite{wong2021kin}.  By comparison, our calculated deformation potentials for SiV($-$), GeV($-$), SnV($-$), and PbV($-$) centers are $0.54$, $2.87$, $2.03$, and $3.94~\mathrm{meV/GPa}$, respectively [see Fig.~\ref{fig:CHGtrns}(b)], confirming that inversion‐symmetric group‐IV vacancies couple more weakly to lattice compression than the NV($-$) defect.}
\section*{Acknowledgments}
We acknowledge the discussion with Christoph Becher and the technical help from
Bal\'azs T\'oth. This work was supported by the Hungarian National Research,
Development and Innovation Office (NKFIH) for Quantum Information National
Laboratory of Hungary (grant no.\ 2022-2.1.1-NL-2022-00004), the EU QuantERA II
SensExtreme project funded by NKFIH (grant no.\
2019-2.1.7-ERA-NET-2022-00040) and Lithuanian Research Council (grant
  no. S-QUANTERA-22-1), the Horizon Europe EIC Pathfinder QuMicro project
(grant no.\ 101046911) and the Horizon Europe Quantum Flagship SPINUS project
(grant no.\ 101135699). G.\ T.\ was supported by the J\'anos Bolyai Research
Scholarship of the Hungarian Academy of Sciences and by NKFIH grant no.\
STARTING 150113.

M.M.\ performed the density functional theory calculations and contributed to the writing of the initial draft. L.R.\ and V.\v{Z}.\ contributed to the dynamic Jahn–Teller and group theory analyses. G.T.\ developed the theoretical framework for calculating the hyperfine tensors in the presence of Jahn–Teller effects. G.T.\ and A.G.\ supervised M.M.’s computational work. A.G.\ wrote the initial draft and conceptualized the study. All authors contributed to the final version of the manuscript.
\section*{Data Availability Statement}
We provide additional data in the Appendices. All the other data are available from the authors upon reasonable request~\cite{Gali2025Data}.

\appendix
\section{Charge transition level with associated variable dielectric constant under high compressive hydrostatic pressure }
\label{app:Dielectric}

In charge transition level calculation, we used the Freysoldt-Neugebauer-Van de
Walle (FNV) charge correction~\cite{Freysoldt2009, Freysoldt2011} of
$E_\text{corr}=E_\text{el}+q\Delta V$, where potential alignment correction
($q\Delta V$) corrects the shift in the reference potential due to the
introduction of the defect in the supercell and the electrostatic correction
($E_\text{el}$) corrects the energy associated with the interaction of the
defect charge with its periodic images. It considers the anisotropy of the
dielectric constant and the geometry of the supercell as
\begin{equation}\label{eq:CHG-correction}
    E_\text{el}=\frac{q^2}{2} \left(\frac{\Delta Q}{\epsilon_0}-\frac{Q_\text{model}}{\epsilon}\right)
\end{equation}
where $q$ is the charge of the defect, $\Delta Q$ is the difference between the
defect charge density and the background charge density, $\epsilon_0$ is the
vacuum permittivity, $Q_\text{model}$ is the potential due to a model charge
distribution, typically a Gaussian or point charge, representing the defect and
$\epsilon$ is the dielectric constant.

We used the sxdefectalign code~\cite{Freysoldt2009} to calculate charge
correction as a function of hydrostatic pressure. For this purpose, it is
essential to use the corresponding dielectric constant at each pressure.
Therefore, we employed the VASP package to determine the dielectric constants.
VASP calculates the dielectric constant using density functional perturbation
theory (DFPT) and the linear response approach~\cite{Gajdos2006, Baroni1986}. The
dielectric tensor ($\epsilon$) is computed by considering both electronic
($\epsilon_{\infty}$) and ionic ($\epsilon_\text{ion}$) as:
\begin{equation}\label{eq:dielectric}
    \epsilon=\epsilon_{\infty}+\epsilon_\text{ion}.
\end{equation}
We listed the charge correction to the total energy and the dielectric constants ($\epsilon$) under each hydrostatic pressure in Table~\ref{table:eps}. We find that the charge correction to the total energy for a given charge state is identical across all the considered defects. This consistency may be attributed to the ionic behavior of the dopants in the G4V defects, which appears to be independent of the dopant type.
\begin{table}[!ht]
    \caption{\label{table:eps}
    The calculated hydrostatic pressure (P) dependency of the charge correction to the total energy and dielectric constant ($\epsilon$) from zero to 180~GPa with HSE06 functional.}
    \begin{ruledtabular}    
        \setlength\tabcolsep{0pt}
            \setlength\extrarowheight{5pt}
    \begin{tabular}{cccccc}
        P (GPa)  & 0 & 32 & 72 & 120 & 180 \\ \hline
        $\epsilon$ & 5.7 & 5.5 & 5.3 & 5.1 & 4.9 \\ 
         \multicolumn{6}{c}{charge corrections (eV)}  \\  \hline
        $\text{G4V}(\pm1)$ & 0.25 & 0.27 & 0.28 & 0.28 & 0.39 \\ 
        $\text{G4V}(-2)$ & 1.00& 1.09 & 1.11 & 1.13 & 1.55 \\ 
    \end{tabular}
    \end{ruledtabular}
\end{table}

\section{Additional data for the effective spin-orbit splitting}
\label{app:JT}

\begin{table*}[!ht]
    \caption{\label{table:lambda-HS-pressure}
    The calculated ground and excited states of Jahn--Teller (JT) energy in meV ($E_\text{JT}$), the barrier energy in meV ($\delta_\text{JT}$), the effective mode in meV ($\hbar \omega$), $p$ reduction factor and $\lambda$ (GHz) with SCAN functional which is Ham reduced spin-orbit splitting for the G4V($-$) defects with hydrostatic pressure.}
        \begin{ruledtabular}
    \setlength\extrarowheight{2pt}
\setlength\tabcolsep{1pt}
    \begin{tabular}{ccclcccccclr}
         & \multicolumn{5}{c}{Ground states} & ~ & \multicolumn{5}{c}{Excited states} \\
        \cmidrule(lr){2-6} \cmidrule(lr){8-12}
        P (GPa) & $E_\text{JT}$ (meV) & $\delta_\text{JT}$ (meV) & $\hbar \omega$ (meV) & $p$ & $\lambda$ (GHz) & ~ & $E_\text{JT}$ (meV) & $\delta_\text{JT}$ (meV) & $\hbar \omega$ (meV) & $p$ & $\lambda$ (GHz) \\ \hline
        SiV($-$)  & ~ & ~ & ~ & ~ & ~ & ~ & ~ & ~ & ~ & ~ & ~ \\
        0 & 40.86 & 3.79 & 89.70 & 0.34 & 70.26 & ~ & 62.58 & 1.12 & 60.81 &  0.133 & 286.21 \\
         32 & 39.62 & 3.08 & 88.97 & 0.34 & 76.15 & ~ & 58.57 & 1.32 & 59.16 &  0.141 & 311.82 \\
         72 & 38.47 & 2.97 & 87.68 & 0.35 & 82.12 & ~ & 56.69 & 1.6 & 58.08  & 0.144 & 331.62 \\
        120 & 38.72 & 2.55 & 88.18 & 0.35 & 87.71 & ~ & 56.61 & 1.71 & 58.00 &  0.144 & 344.89 \\
         180 & 39.58 & 2.24 & 89.43 & 0.35 & 93.01 & ~ & 56.42 & 1.78 & 58.08 & 0.145 & 359.72 \\
        GeV($-$)  & ~ & ~ & ~ & ~ & ~ & ~ & ~ & ~ & ~ & ~ & ~ \\
        0 & 30.59 & 4.05 & 77.01 & 0.38 & 222.84 & ~ & 71.48 & 2.31 & 70.49 &  0.136 & 1155.54 \\
        32 & 30.00 & 3.34 & 76.71 & 0.38 & 255.53 & ~ & 68.28 & 2.48 & 68.97 & 0.141 & 1204.69 \\
        72 & 29.18 & 3.34 & 75.59 & 0.38 & 293.72 & ~ & 65.70 & 2.65 & 67.95 & 0.146 & 1237.21 \\
         120 & 29.89 & 2.99 & 76.29 & 0.38 & 331.61 & ~ & 63.42 & 2.91 & 66.82 & 0.150 & 1262.39 \\
         180 & 30.82 & 2.56 & 78.36 & 0.38 & 376.21 & ~ & 62.18 & 3.2 & 67.03 & 0.155 & 1289.74 \\
        SnV($-$) & ~ & ~ & ~ & ~ & ~ & ~ & ~ & ~ & ~ & ~ & ~ \\
        0 & 20.81 & 1.15 & 64.87 & 0.44 & 915.23 & ~ & 67.69 & 4.2 & 68.04 & 0.140 & 3214.51 \\
         32 & 20.47 & 1.10 & 64.94 & 0.44 & 1068.02 & ~ & 65.01 & 4.31 & 66.92 & 0.145 & 3490.44 \\
         72 & 20.58 & 1.09 & 64.54 & 0.44 & 1223.62 & ~ & 61.93 & 4.51 & 65.24 & 0.150 & 3938.52 \\
         120 & 21.17 & 0.36 & 66.08 & 0.44 & 1407.08 & ~ & 57.01 & 4.53 & 62.86 & 0.160 & 4456.48 \\
         180 & 22.08 & 0.29 & 67.54 & 0.43 & 1585.59 & ~ & 50.53 & 4.72 & 59.25 & 0.174 & 5275.23 \\
        PbV($-$) & ~ & ~ & ~ & ~ & ~ & ~ & ~ & ~ & ~ & ~ & ~ \\
         0 & 15.02 & 3.88 & 55.98 & 0.48 & 4436.03 & ~ & 87.32 & 6.69 & 77.85 & 0.116 & 6782.56 \\
         32 & 15.12 & 3.98 & 56.43 & 0.48 & 5064.64 & ~ & 87.28 & 6.59 & 77.82 & 0.116 & 6973.45 \\
         72 & 15.27 & 3.90 & 56.13 & 0.48 & 5743.95 & ~ & 87.20 & 6.2 & 78.16 & 0.117 & 7170.19 \\
         120 & 16.49 & 3.89 & 59.04 & 0.47 & 6387.39 & ~ & 86.21 & 6.12 & 77.67 & 0.118 & 7417.08 \\
         180 & 18.27 & 4.43 & 55.94 & 0.46 & 6592.98 & ~ & 82.25 & 5.87 & 74.88 & 0.120 & 7964.79 \\
    \end{tabular}
    \end{ruledtabular}
\end{table*}
\begin{figure*}[t]
\includegraphics[scale=1.05]{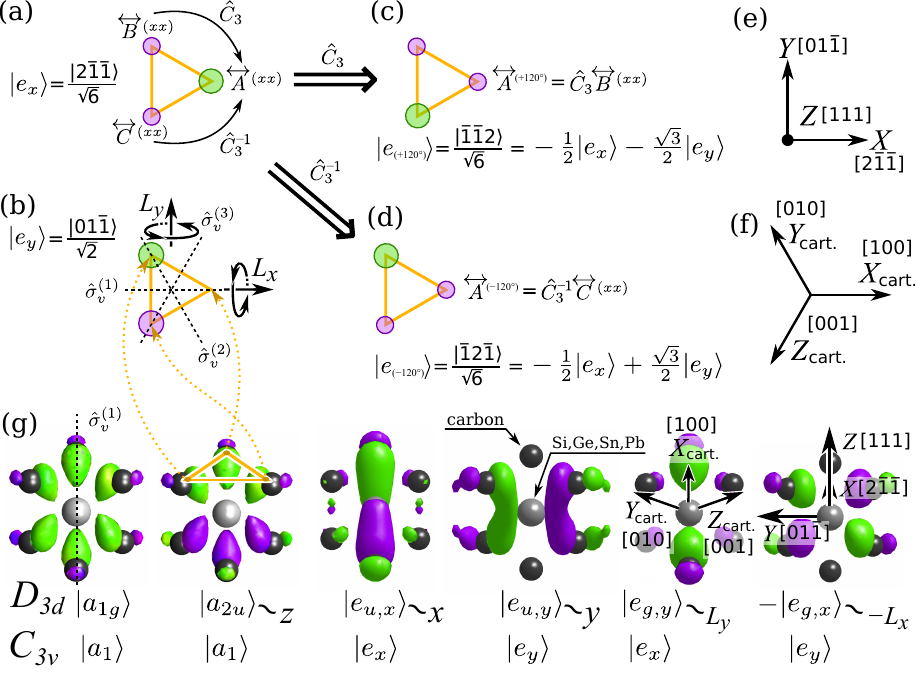}
\caption{\label{fig:rotate}
Transformation of three equivalent carbon atoms. (a) and (b) represent the electronic wavefunctions when the $|e_x\rangle$ and $|e_y\rangle$ are chosen, respectively. The red and blue circles represent the positive and negative values of $|e_x\rangle$ and  $|e_y\rangle$ wavefunctions, respectively that we also quantify inside the [...] parentheses. (c) and (d) depict the transformation laws for $\hat{C}_3$ and  $\hat{C^{-1}}_3$ rotations, respectively. (e) depicts the $z=$[111] oriented coordinates of the G4V defects. (f) depicts the Cartesian coordinates of diamond. (g) Depiction of single particle orbitals of G4V centers. Green/purple color depicts positive/negative wavefunction phase. We also show the subduced representation labels from $D_{3v}$ to $C_{3v}$.}
\end{figure*}

Here, we list the calculated Jahn--Teller parameters and electronic spin-orbit splitting under compressive hydrostatic pressure as obtained by SCAN functional for G4V($-$) color centers (see Table~\ref{table:lambda-HS-pressure}).

\textcolor{black}{Additional data supporting this Appendix, including full input files, raw computational outputs, and extended pressure-dependent results, are available from the corresponding author upon reasonable request~\cite{Mohseni2025Data}.}
\section{Hyperfine Hamiltonian for an orbital doublet of $D_{3d}$ symmetry}
\label{app:HFHam}

The general form of the hyperfine (HF) tensor associated with an orbital doublet
(transforming according to either the $e_g$ or $e_u$ irreducible representation)
under $D_{3d}$ symmetry can be systematically derived using group-theoretical
methods. For hyperfine interactions involving the central \mbox{group-IV} ion, the HF
Hamiltonian must remain invariant under all $D_{3d}$ symmetry operations,
implying that it transforms as a scalar corresponding to the $A_{1g}$
irreducible representation. Consequently, for real-valued representations of
degenerate orbitals, the perturbation introduced by the HF interaction on an
orbital doublet can be expressed through a tensor
convolution~\cite{Bersuker2012, ham1968}:
\begin{equation}
  \hat{H}_{\mathrm{HF}}
  =
  \sum_{\Gamma\gamma} \vec{S} {A}_{\Gamma\gamma} \vec{I} \hat{C}_{\Gamma\gamma},
\end{equation}
where the sum runs over the irreducible representations $\Gamma$ of the $D_{3d}$
point group. $\hat{C}_{\Gamma\gamma}$ are Clebsch-–Gordan (CG) coefficient
matrices determined in the orbital state space, and ${A}_{\Gamma\gamma}$ are
the hyperfine interaction tensors that transform as row $\gamma$ of the
irreducible representation $\Gamma$ under the operations of the $D_{3d}$ point
group. The symmetry-allowed parameters of $\hat{A}_{\Gamma\gamma}$ can be
systematically obtained using group-theoretical projection operators~\cite{elliott1979b}.

For an orbital doublet $|e_{\{g,u\}}\rangle$ 
only $\hat{C}_{\Gamma\gamma}$ with "gerade" (even) symmetry is possible because orbital operators (2$\times$2 matrices)
$|e_{\{u,g\}}\rangle\!\otimes\!\langle e_{\{u,g\}}|=A_{1g}\oplus E_{g}\oplus A_{2g}$ 
built from them must be "gerade" representations. 
Furthermore, the interaction tensors $\hat{A}_{\Gamma\gamma}$ must be Hermitian to
ensure the hyperfine energy remains real-valued and thus prevents $\hat{C}_{A_{2g}}$ to appear.
Considering these constraints,
the most general form of the hyperfine Hamiltonian in the real-valued
representation of $e$-orbitals is given by
\begin{align}
  \label{eq:HFD3d2}
  \hat{H}_{\mathrm{HF}}
  &
    =
    \left[
    A_{\perp}(\hat{S}_{x}\hat{I}_{x}+\hat{S}_{y}\hat{I}_{y})+2A_{\parallel}\hat{S}_{z}\hat{I}_{z}
    \right]
    \hat{\mathbb{I}}
  \\
  \notag
  &
    \! \! \! \! \! \! +
    A_{1}\left[
    \left(\hat{S}_{x}\hat{I}_{z}\!+\!\hat{S}_{z}\hat{I}_{x}\right)
    \Cc{E_{gy}}^{(u,g)} 
    \! - \!
    \left(\hat{S}_{y}\hat{I}_{z}\!+\!\hat{S}_{z}\hat{I}_{y}\right)
    \Cc{E_{gx}}^{(u,g)} 
    \right]
  \\
  \notag
  &
     \! \! \! \! \! \!+
    \frac{A_{2}}{2}
    \left[
    \left(\hat{S}_{y}\hat{I}_{y}\! - \!\hat{S}_{x}\hat{I}_{x}\right)
    \Cc{E_{gy}}^{(u,g)} 
    \! - \!
    \left(\hat{S}_{x}\hat{I}_{y} \!+\! \hat{S}_{y}\hat{I}_{x}\right)
    \Cc{E_{gx}}^{(u,g)} 
    \right]\text{.}
\end{align}
Here, $A_i$ are scalar parameters that characterize the interaction. The
additional factors of 2 are introduced for convenience in subsequent analysis.
The term $\hat{\mathbb{I}}$ denotes the 2$\times$2 identity matrix for $\hat{C}_{A_{1g}}$, while
$\hat{C}_{E_{gx}}$ and $\hat{C}_{E_{gy}}$ represent the CG
matrices associated with the $E_g$ irreducible representation. 

In our chosen coordinate system, these matrices correspond to the standard Pauli spin matrices for the case of $|e_u\rangle$ orbitals,
as $\hat{C}_{E_{gx}}^{(u)}= \hat{\sigma}_x = (\begin{smallmatrix} & 1\\1 \end{smallmatrix}) = |e_{u,x}\rangle\langle e_{u,y}|+|e_{u,y}\rangle\langle e_{u,x}|$ 
and $\hat{C}_{E_{gy}}^{(u)} = \hat{\sigma}_z = (\begin{smallmatrix} 1  & \\ &-1 \end{smallmatrix}) = |e_{u,x}\rangle\langle e_{u,x}|-|e_{u,y}\rangle\langle e_{u,y}|$. 
We note that this convention differs from that used in Ref.~\onlinecite{ham1968}. 
However, we add an extra negative $-1$ phase for $|e_g\rangle$ by defining 
$\hat{C}_{E_{gx}}^{(g)}$=$-\hat{\sigma}_{x}^{(g)}$=$-|e_{g,x}\rangle\langle e_{g,y}|-|e_{g,y}\rangle\langle e_{g,x}|$ 
and 
$\hat{C}_{E_{gy}}^{(g)}$=$-\hat{\sigma}_{z}^{(g)}$=$-|e_{g,y}\rangle\langle e_{g,y}|+|e_{g,y}\rangle\langle e_{g,y}|$ for the gerade representations. 
We do this because the $|e_g\rangle$ state subduces into the $|e\rangle$ representation of $C_{3v}$ differently from how the $|e_u\rangle$ state does, as we discuss in the next paragraph. 

Here, we note that the subduction of the $e$ orbitals into the $C_{3v}$ point group should not be taken lightly. Indeed, the "ungerade" (odd) case is trivial because both $|e_{ux}\rangle$ and $|e_{uy}\rangle$ are pointing to \{$x, y$\} Cartesian axes: 
$\{|e_{u,x}\rangle,|e_{u,y}\rangle\}\rightarrow\{|e_{x}\rangle,|e_{y}\rangle\}$. 
In simple terms, the orbitals maintain their symmetry but are relabeled without the $u$ subscript. However, the "gerade" case 
is vastly different because they are transforming as axial vectors:
$\{|e_{g,x}\rangle,|e_{g,y}\rangle\}\sim\{\hat{L}_{x},\hat{L}_{y}\}=i\hbar\{z\partial_{y}-\underline{y\partial_{z}},\underline{x\partial_{z}}-z\partial_{x}\}\sim\{-|e_{y}\rangle,+|e_{x}\rangle\}$,
see Ref.~\onlinecite{Thiering2024_supp} or Fig.~\ref{fig:rotate}(b) for visual interpretation. 
This has a consequence of that $\hat{\sigma}_{x}$ and $\hat{\sigma}_{z}$ 
inverts their sign when they are discussed within $C_{3v}$:
$\hat{\sigma}_{z}^{(g)}=|e_{g,x}\rangle\langle e_{g,x}|-|e_{g,y}\rangle\langle e_{g,y}|\overset{C_{3v}}{\rightarrow}|e_{y}\rangle\langle e_{y}|-|e_{x}\rangle\langle e_{x}|=-\hat{\sigma}_{z}$ and 
$\hat{\sigma}_{x}^{(g)}=|e_{g,x}\rangle\langle e_{g,y}|+|e_{g,y}\rangle\langle e_{g,x}|\overset{C_{3v}}{\rightarrow}-|e_{x}\rangle\langle e_{y}|-|e_{y}\rangle\langle e_{x}|=-\hat{\sigma}_{x}$. 
Simply put, one may select $|e_x\rangle$ as the orbital that remains invariant under mirror symmetry, i.e., $\hat{\sigma}{v}^{(1)}|e{x}\rangle = +|e_{x}\rangle$, and disregard the presence of inversion symmetry, as illustrated in Fig.\ref{fig:rotate}. However, as noted, this $|e_x\rangle$ orbital corresponds to $|e_{g,y}\rangle$ in $D_{3d}$---somewhat surprisingly---and $|e_y\rangle$ corresponds to $-|e_{g,x}\rangle$. In short, we introduced the $-1$ sign into the CG matrices of the $|e_g\rangle$ representation to ensure that Eq.~\eqref{eq:HFD3d2} reduces to the same Hamiltonian in $C_{3v}$, regardless of whether we start from the $|e_u\rangle$ or the $|e_g\rangle$ case.

From this Hamiltonian, the HF parameters can be determined by calculating the HF
tensor for the selected $|e_x\rangle$ orbital state as
\begin{align*}
\overleftarrow{S}\mathcal{A}^{x}\overrightarrow{I}=\begin{cases}
\bigl\langle e_{x}\bigr|\hat{H}_{\mathrm{HF}}\bigl|e_{x}\bigr\rangle & C_{3v}\text{ case }\\
\bigl\langle e_{u,x}\bigr|\hat{H}_{\mathrm{HF}}\bigl|e_{u,x}\bigr\rangle & \text{ungerade case}\\
\bigl\langle e_{g,y}\bigr|\hat{H}_{\mathrm{HF}}\bigl|e_{g,y}\bigr\rangle & \text{gerade case}
\end{cases}
\text{,} 
\end{align*}
where ${\mathcal{A}}^{x}$ represents the HF coupling tensor [see
Eq.~\eqref{eq:HF}], evaluated for the state $E_{x}/E_{u,x}/E_{g,y}$ in which the $|e_x\rangle/|e_{u,x}\rangle/|e_{g,y}\rangle$ orbital is
occupied by a hole. From Eq.~\eqref{eq:HFD3d2}, the HF tensor components are
given by
\begin{align}
\mathcal{A}^x &=
\begin{pmatrix}
A_{\perp} - \frac{A_{2}}{2} & 0 & A_{1} \\
0 & A_{\perp} + \frac{A_{2}}{2} & 0 \\
A_{1} & 0 & 2A_{\parallel}
\end{pmatrix}
\text{,}
\label{eq:HFtensorD3d}
\end{align}
from which one can derive Eq.~\eqref{eq:HFparameters} of the main text straightforwardly.
Specifically, one can obtain the $\mathcal{A}^{x}$ by \textit{ab initio} DFT that uniquely
defines all hyperfine parameters: $\mathcal{A}_{zz}^{x}=2A_{\parallel}$, 
$\mathcal{A}_{xx}^{x}+\mathcal{A}_{yy}^{x}=2A_{\perp}$, $\mathcal{A}_{yy}^{x}-\mathcal{A}_{xx}^{x}=A_{2}$, $\mathcal{A}_{xz}^{x}=A_{1}$.

In the complex electronic basis $|e_{u\pm}\rangle=\mp(|e_{u,x}\rangle\pm i|e_{u,y}\rangle)/\sqrt{2}$ and $|e_{g\pm}\rangle=(|e_{g,x}\rangle\pm i|e_{g,y}\rangle)/\sqrt{2}$,
Eq.~\eqref{eq:HFD3d2} can be rewritten in canonical form as
\begin{align}
  \hat{H}_{\mathrm{HF}}
  &=
  \left[
    \frac{1}{2}A_{\perp}(\hat{S}_{+}\hat{I}_{-} + \hat{S}_{-}\hat{I}_{+}) + 2A_{\parallel}\hat{S}_z \hat{I}_z
    \right]
  \hat{\mathbb{I}}
  \nonumber
  \\
  &
    \quad
    - A_{1}
    \left[
      \left(\hat{S}_z \hat{I}_{+} + \hat{S}_{+}\hat{I}_z\right)\hat{\sigma}_{+}
      +
      \left(\hat{S}_z \hat{I}_{-} + \hat{S}_{-}\hat{I}_z\right)
      \hat{\sigma}_{-}
    \right]
    \nonumber
  \\
  & \quad
    + \frac{A_{2}}{2}
    \left[
        \hat{S}_{-}\hat{I}_{-}\hat{\sigma}_{+} + \hat{S}_{+}\hat{I}_{+}\hat{\sigma}_{-}
    \right]\text{,}
    \label{eq:HFD3d3}
\end{align}
where $\hat{S}_\pm$ and $\hat{I}_\pm$ represent the spin raising or lowering
operators acting on the electronic and nuclear spin degrees of freedom,
respectively. The operators
$\hat{\sigma}_{\pm} = |e_{\{g,u\}\pm}\rangle\langle e_{\{g,u\}\mp}|$ denote the
orbital angular momentum raising and lowering operators. One may notice that 
Eq.~\eqref{eq:HFD3d3} is the same for the three $|e\rangle/|e_u\rangle/|e_g\rangle$ cases. 
Both the "gerade" and "ungerade" cases are subducing into 
$\hat{\sigma}_{\pm}^{(u)}\overset{C_{3v}}{\rightarrow}\hat{\sigma}_{\pm}\overset{C_{3v}}{\leftarrow}\hat{\sigma}_{\pm}^{(g)}$ 
the same $\hat{\sigma}_{\pm}$ ladder operator without any additional phase unlike
that was in the real valued basis. We will prove this in the next paragraph.

The "ungerade" ($u$) case will be 
$\hat{\sigma}_{\pm}^{(u)}=|e_{u\pm}\rangle\langle e_{u\mp}|=(-\hat{\sigma}_{z}^{(u)}\mp i\hat{\sigma}_{x}^{(u)})/2\overset{C_{3v}}{\rightarrow}(-\hat{\sigma}_{z}\mp i\hat{\sigma}_{x})/2=\hat{\sigma}_{\pm}$. 
The "gerade" ($g$) case will be 
$\hat{\sigma}_{\pm}^{(g)}=|e_{g\pm}\rangle\langle e_{g\mp}|=(\hat{\sigma}_{z}^{(g)}\pm i\hat{\sigma}_{x}^{(g)})/2\overset{C_{3v}}{\rightarrow}(-\hat{\sigma}_{z}\mp i\hat{\sigma}_{x})/2=\hat{\sigma}_{\pm}$.
Therefore, to obtain the $\mathcal{A}^x$ hyperfine tensor using \emph{ab initio} DFT, one must select either the $|e_{u,x}\rangle$ or $|e_{g,y}\rangle$ orbital, as both subduce to $|e_x\rangle$ in the $C_{3v}$ point group. However, since all of these orbitals yield the same Hamiltonian (Eq.\eqref{eq:HFD3d3}) in the ladder operator formalism---whether in the "gerade", "ungerade", or subduced $C_{3v}$ representation---one can directly use Eq.~\eqref{eq:HFparameters} to define the orbital-dependent $A_{1}$ and $A_{2}$ hyperfine parameters.

The solutions of Eq.~\eqref{eq:HFD3d3} can be analyzed in terms of the double
group representations of the hyperfine (HF) structure. Adding the nuclear spin
to the lower-lying SOC state $E_{3/2,\{g,u\}}$ [see~Fig.~\ref{fig:JT}] results
in two doublets of $E_{\{g,u\}}$ symmetry
\begin{align}
  \ket{E_{\{g,u\}\pm}^{1}}
  & =
    \left\{
    |E_{\{g,u\}+}^{\uparrow\Downarrow}\rangle
    ,
    -|E_{\{g,u\}-}^{\downarrow\Uparrow}\rangle
    \right\}, \\
  \ket{E_{\{g,u\}\pm}^{2}}
  &=
    \left\{
    |E_{\{g,u\}-}^{\downarrow\Downarrow}\rangle, |E_{\{g,u\}+}^{\uparrow\Uparrow}\rangle
    \right\}.
\end{align}
Here, for example, $|E_{g+}^{\uparrow\Downarrow}\rangle$ denotes a wavefunction
whose orbital part transforms as the $e_{g+}$ irreducible representation, with
the electronic spin (first arrow) in the spin-up state and the nuclear spin
(second arrow) in the spin-down state.

The double group structure of the upper SOC branch results in doublet of
$E_{\{g,u\}}$ symmetry and two singlets of $A_{\{1g,2u\}}$ and $A_{\{2g,1u\}}$
symmetries
\begin{align}
  \ket{E_{\{g,u\}\pm}^{3}}
  & =
    \left\{
    |E_{\{g,y\}+}^{\downarrow\Uparrow}\rangle, -|E_{\{g,u\}-}^{\uparrow\Downarrow}\rangle
    \right\},
  \\
  \ket{A_{\{1g,2u\}}}
  &=
    \left(|E_{\{g,u\}-}^{\uparrow\Uparrow}\rangle-|E_{\{g,u\}+}^{\downarrow\Downarrow}\rangle \right)/\sqrt{2}
  \\
  \ket{A_{\{2g,1u\}}}
  &=
    \left(|E_{\{g,u\}-}^{\uparrow\Uparrow}\rangle+|E_{\{g,u\}+}^{\downarrow\Downarrow}\rangle \right)/\sqrt{2}
\end{align}

The Hamiltonian describing SOC and HF interactions in the basis
$\{E_{\{g,u\}}^1, E_{\{g,u\}}^2, E_{\{g,u\}}^3, A_{1\{g,u\}}, A_{2\{g,u\}}\}$
are
\begin{align}
  \label{eq:3}
  \hat{H}_{\mathrm{SOC}}
  &=\frac{1}{2}
    \left(
    \begin{smallmatrix}
      -\lambda  & 0 & 0 & 0 & 0 \\
      0 & -\lambda  & 0 & 0 & 0 \\
      0 & 0 & \lambda  & 0 & 0 \\
      0 & 0 & 0 & \lambda  & 0 \\
      0 & 0 & 0 & 0 & \lambda  \\
    \end{smallmatrix}
    \right)
  \\
  \label{eq:4}
  \hat{H}_{\mathrm{HF}}
  & =
    \frac{1}{2}\left(
    \begin{smallmatrix}
 -A_{\parallel} & A_1 & A_{\perp} & 0 & 0 \\
 A_1 & A_{\parallel} &  A_1 & 0 & 0 \\
 A_{\perp} & A_1 & -A_{\parallel} & 0 & 0 \\
 0 & 0 & 0 & A_{\parallel} - A_2 & 0 \\
 0 & 0 & 0 & 0 & A_{\parallel} + A_2 \\
    \end{smallmatrix}
    \right)
\end{align}

Neglecting Zeeman interactions other than those arising from the coupling of the
magnetic field with the electron spin, and assuming the magnetic field is
aligned along the symmetry axis of the defect, the Zeeman interaction is
expressed as $\hat{H}_{Z} = \mu_{B} g_{S} \hat{S}_{z} B$. Here, $g_{S} \approx 2$,
$\mu_{B}$ represents the effective $g$-factor of the electron and the Bohr magneton, respecively,
and $B$ is the strength of the applied magnetic field. This interaction splits
the $E_{g}$ doublets by an energy of $g_{S} \mu_{B} B$ and also induces mixing
between the $A_{\{1g,2u\}}$ and $A_{\{2g,1u\}}$ states.

Using the states described above as the zeroth-order basis, the fine-structure energies of these states---up to second-order perturbation theory---are given by
\begin{align}
  \label{eq:HFEnergiesE_1}
  \notag
  E[E_{\{g,u\}\pm}^{1}]
  & = \frac{1}{2}(-\lambda - A_{\parallel} \pm \mu_{B}g_{S} B)
  \\
  & - \frac{1}{4} \left[\frac{A_1^2}{A_{\parallel} \mp \mu_{B} g_{S} B} + \frac{A_{\perp}^2}{\lambda \mp \mu_{B} g_{S} B}\right]
  \\\notag
  \label{eq:HFEnergiesE_2}
  E[E_{\{g,u\}\pm}^{2}] & = \frac{1}{2}(-\lambda + A_{\parallel} \mp \mu_{B}g_{S} B)
  \\
  & + \frac{1}{4} A_1^2 \left[\frac{1}{A_{\parallel} \mp \mu_{B} g_{S} B} - \frac{1}{\lambda - A_{\parallel}}\right]
  \\\notag
  E[E_{\{g,u\}\pm}^{3}] & = \frac{1}{2}(\lambda - A_{\parallel} \mp \mu_{B}g_{S} B)
  \\
  & + \frac{1}{4} \left[\frac{A_{\perp}^2}{\lambda \mp \mu_{B} g_{S} B} + \frac{A_1^2}{\lambda  - A_{\parallel}}\right]
  \\
  E[A_{\{1g,2u\}}]
  & = \frac{1}{2}(\lambda + A_{\parallel} - A_{2}) - \frac{(\mu_{B} g_{S} B)^{2}}{4A_{2}}
  \\
  E[A_{\{2g,1u\}}]
  & = \frac{1}{2}(\lambda + A_{\parallel} + A_{2}) + \frac{(\mu_{B} g_{S} B)^{2}}{4A_{2}} \text{,}
  \label{eq:HFEnergiesA}
\end{align}
where we omit the ${g,u}$ subscripts on $A_1$, $A_2$, $A_{\parallel}$, $A_{\perp}$, and $\lambda$ on the right-hand side of the equations for simplicity.

In experiments, one can measure the difference between the PLE signals: $A_{\text{PLE}}=\big(E[E_{u\pm}^{1}]-E[E_{g\pm}^{1}]\big)-\big(E[E_{u\pm}^{2}]-E[E_{g\pm}^{2}]\big)$. Thus, the leading terms in approximation of $A_1$,$A_\perp$,$A_\parallel \ll \mu_{B}g_{S}B \ll \lambda$ will be
\begin{align}
A_{\text{PLE}}(B) \approx \overset{A_{\text{PLE}}}{\overbrace{A_{\parallel,u}-A_{\parallel,g}}}+\frac{A_{\perp,g}^{2}}{4\lambda_{g}}-\frac{A_{\perp,u}^{2}}{4\lambda_{u}}\notag
\\
\pm\left(\frac{A_{\perp,g}^{2}}{\lambda_{g}^{2}}-\frac{A_{\perp,u}^{2}}{\lambda_{u}^{2}}\right)\frac{\mu_{B}g_{S}B}{4}  \label{eq:HF_PLE} \text{,}
\end{align}
on which we note that the magnetic field independent correction is negligible to $A_{\text{PLE}}$ thus we omit it from Eq.~\eqref{eq:hfcorrection} of the main text.

We note that one can reach avoided crossing by setting the external magnetic field as $A_{\parallel , g} \approx \mu_{B}g_{S}B$. In this case the effect of $A_{1,g} \approx 1\dots3$~MHz on the PLE spectrum will be observable, in principle, because $A_{1,u}\approx 0$~MHz, according to our calculations (see Table~\ref{table:hyperfine}). This can be readily derived where Eqs.~\eqref{eq:HFEnergiesE_1} -\eqref{eq:HFEnergiesA} become invalid and Eqs.~\eqref{eq:3} - \eqref{eq:4} should be directly diagonalized at the avoided crossing condition. However, the observation at this condition (low magnetic fields) could be very difficult, thus we omit this discussion in the main text.   


\section{Hyperfine interaction for $^{13}$C neighbors}
\label{app:hfder}
\begin{table*}[!ht]
\setlength\extrarowheight{2pt}
\caption{\label{table:hyperfineconstant}
The raw data of hyperfine tensors in both $D_{3d}$ and $C_{2h}$ symmetries for G4V($-$) defects at zero pressure as obtained in HSE06 calculations in the electronic ground state. $A_{xx}$, $A_{yy}$ and $A_{zz}$ are the symmetric hyperfine tensor elements. $A_{xy}$, $A_{xz}$ and $A_{yz}$ parameters are the asymmetric hyperfine tensor elements. The unitless reduction factor $q$ is given for each defect associated with the Jahn--Teller effect. Atoms are labeled for low $C_{2h}$ symmetry configuration in Fig.~\ref{fig:spindens}(j) and the spin density of high $D_{3d}$ symmetry configuration is illustrated in Fig.~\ref{fig:spindens}.}
\begin{ruledtabular}
    \begin{tabular}{lrrrrrrr}
        Defect & $q$  & $A_{xx}$ (MHz) & $A_{yy}$ (MHz) & $A_{zz}$ (MHz) & $A_{xy}$ (MHz) & $A_{xz}$ (MHz) & $A_{yz}$ (MHz) \\ \hline
        SiV($-$) & 0.67  & ~ & ~ & ~ & ~ & ~ & \\ 
        $C_{2h}$ & ~ & ~ & ~ & ~ & ~ & ~ & ~ \\ 
        C$_{2,3,5,6}$ & ~& 18.36 & 22.91 & 21.14 & 6.03 & $-$5.17 & $-$7.53 \\
        C$_{1,4}$ & ~& 98.46 & 100.94 & 98.45 & $-$28.79 & 28.26 & $-$28.79 \\ 
        $^{29}$Si & ~& 87.08 & 85.48 & 87.09 & $-$4.27 & 3.15 & $-$4.25 \\  
        $D_{3d}$ & ~ & ~ & ~ & ~ & ~ & ~ & ~ \\ 
        C-first neighbor & ~& 44.85 & 44.85 & 45.95 & 13.02 & $-$13.41 & $-$13.41 \\ 
        C-second neighbor& ~& $-$2.60 & $-$1.87 & $-$3.32 & $-$1.60 & 0.03 & 0.47 \\  
        $^{29}$Si  & ~ & 90.29 & 90.29 & 90.29 & $-$1.70 & $-$1.70 & $-$1.70 \\
        GeV($-$) & 0.69  & ~ & ~ & ~ & ~ & ~ & \\ 
        $C_{2h}$ & ~ & ~ & ~ & ~ & ~ & ~ & ~ \\ 
        C$_{2,3,5,6}$ & ~& 19.76 & 23.84 & 22.14 & 6.69 & $-$5.83 & $-$7.92 \\ 
        C$_{1,4}$ & ~& 113.87 & 112.83 & 81.39 & $-$31.78 & 31.75 & $-$31.79 \\ 
        $^{73}$Ge & ~& 43.24 & 42.84 & 43.24 & $-$1.63 & 0.45 & $-$1.64 \\ 
        $D_{3d}$ & ~ & ~ & ~ & ~ & ~ & ~ & ~ \\ 
        C-first neighbor & ~& 50.64 & 50.64 & 51.17 & 14.74 & $-$14.87 & $-$14.87 \\ 
        C-second neighbor& ~& $-$3.91 & $-$4.30 & $-$3.13 & 0.11 & $-$1.22 & 0.51 \\  
        $^{73}$Ge & ~& 46.61 & 46.61 & 46.61 & $-$0.93 & $-$0.93 & $-$0.93 \\ 
        SnV($-$) & 0.72 & ~ & ~ & ~ & ~ & ~ &  \\ 
        $C_{2h}$ & ~ & ~ & ~ & ~ & ~ & ~ & ~ \\ 
        C$_{2,3,5,6}$ & ~& 16.32 & 20.34 & 18.48 & 7.13 & $-$6.23 & $-$8.33 \\  
        C$_{1,4}$ & ~& 95.15 & 95.64 & 95.16 & $-$32.49 & 32.50 & $-$32.49 \\  
        $^{117}$Sn  & ~&1012.14 & 1011.20 & 1012.15 & $-$18.766 & $-$16.214 & $-$18.763\\
        $D_{3d}$ & ~ & ~ & ~ & ~ & ~ & ~ & ~ \\ 
        C-first neighbor & ~& 42.61 & 42.61 & 42.84 & 15.40 & $-$15.50 & $-$15.50 \\ 
        C-second neighbor & ~& $-$9.59 & $-$9.85 & $-$9.59 & 0.05 & 0.97 & 0.05 \\ 
        $^{117}$Sn & ~&1074.989 &1074.985&1074.99& $-$18.066& $-$18.066 & $-$18.062\\
        PbV($-$) & 0.74 & ~ & ~ & ~ & ~ & ~ &  \\ 
        $C_{2h}$ & ~ & ~ & ~ & ~ & ~ & ~ & ~ \\ 
        C$_{2,3,5,6}$ & ~& 14.83 & 18.50 & 16.65 & 7.15 & $-$6.23 & $-$8.20 \\
        C$_{1,4}$ & ~& 99.96 & 99.02 & 99.96 & $-$33.97 & 34.53 & $-$33.97 \\ 
        $^{207}$Pb  & ~& $-$1179.04 & $-$1175.31 & $-$1179.05 & 14.46 & 13.63 & 14.45 \\ 
        $D_{3d}$ & ~ & ~ & ~ & ~ & ~ & ~ & ~ \\ 
        C-first neighbor & ~& 42.83 & 42.83 & 42.51 & 16.21 & $-$16.08 & $-$16.08 \\ 
        C-second neighbor & ~& $-$12.13 & $-$10.79 & $-$11.97 & 0.30 & $-$0.06 & $-$1.26 \\ 
        $^{207}$Pb & ~& $-$1323.24 & $-$1323.24 & $-$1323.24 & 14.27 & 14.27 & 14.27 \\ 
    \end{tabular}
\end{ruledtabular}
\end{table*}
\begin{table*}[!ht]
\setlength\extrarowheight{1pt}
\setlength\tabcolsep{1pt}
    \caption{\label{table:hyperfineconstantexc}
The raw data of hyperfine tensors in both $D_{3d}$ and $C_{2h}$ symmetries for G4V($-$) defects at zero pressure as obtained in HSE06 calculations in the electronic excited state. $A_{xx}$, $A_{yy}$ and $A_{zz}$ are the symmetric hyperfine tensor elements. $A_{xy}$, $A_{xz}$ and $A_{yz}$ parameters are the asymmetric hyperfine tensor elements. The reduction factor $q$ is given for each defect associated with the Jahn--Teller effect. Atoms are labeled for low $C_{2h}$ symmetry configuration in Fig.~\ref{fig:spindens}(j) and the spin density of high $D_{3d}$ symmetry configuration is illustrated in Fig.~\ref{fig:spindens}.}
\begin{ruledtabular}
    \begin{tabular}{lccccccc}
        Defect & $q$  & $A_{xx}$ (MHz) & $A_{yy}$ (MHz) & $A_{zz}$ (MHz) & $A_{xy}$ (MHz) & $A_{xz}$ (MHz) & $A_{yz}$ (MHz) \\ \hline
        SiV($-$) & ~ & ~ & ~ & ~ & ~ & ~ & ~ \\ 
        $C_{2h}$& ~ & ~ & ~ & ~ & ~ & ~ & ~ \\ 
        C$_{1,4}$ & ~ & 40.72 & 37.07 & 36.95 & 22.24 & 22.32 & 53.44 \\ 
        C$_{2,3,5,6}$ & ~ & 21.48 & 14.03 & 14.38 & 13.83 & 14.45 & 17.54 \\ 
        $^{29}$Si  & 0.57 & 4.97 & 5.02 & 4.96 & 4.91 & 4.86 & 4.90 \\ 
        $D_{3d}$ & ~ & ~ & ~ & ~ & ~ & ~ & ~ \\ 
        C-first neighbor & ~ & $-$16.09 & $-$16.09 & $-$20.87 & $-$14.94 & $-$17.57 & $-$17.57 \\
        $^{29}$Si & ~ & 36.21 & 36.21 & 36.21 & 36.19 & 36.19 & 36.19 \\ 
        GeV($-$) & 0.57 & ~ & ~ & ~ & ~ & ~ & ~ \\ 
        $C_{2h}$& ~ & ~ & ~ & ~ & ~ & ~ & ~ \\
        C$_{1,4}$ & ~ & $-$8.57 & $-$8.57 & $-$9.73 & $-$10.30 & $-$8.50 & $-$8.50 \\ 
        C$_{2,3,5,6}$ & ~ & 5.39 & 3.91 & $-$1.21 & $-$10.82 & 13.54 & $-$6.84 \\ 
        $^{73}$Ge & ~ & 4.59 & 4.51 & 4.59 & 4.28 & 5.01 & 4.28 \\ 
        $D_{3d}$ & ~ & ~ & ~ & ~ & ~ & ~ & ~ \\  
        C-first neighbor  & ~ & 2.91 & 2.91 & 2.42 & 3.01 & 2.44 & 2.44 \\
        $^{73}$Ge & ~ & 15.437 & 15.912 & 15.439 &$-$0.507 &$-$0.165 & $-$0.508\\
        SnV($-$) & ~ & ~ & ~ & ~ & ~ & ~ & ~ \\ 
        $C_{2h}$& ~ & ~ & ~ & ~ & ~ & ~ & ~ \\
        C$_{1,4}$ & ~ & 122.80 & 123.52 & 122.78 & 89.45 & 156.41 & 89.46 \\ 
        C$_{2,3,5,6}$ & ~ & 24.32 & 25.96 & 21.86 & 16.12 & 18.35 & 30.55 \\ 
        $^{117}$Sn & 0.57 &31.523&30.845&31.5&30.596&30.347&30.559\\
        $D_{3d}$ & ~ & ~ & ~ & ~ & ~ & ~ & ~ \\ 
        C-first neighbor & ~ &13.183&12.892&14.026&9.292&$-$9.237&$-$9.067\\
        $^{117}$Sn & ~ & 183.785 & 211.081 &$-$22.512 & $-$60.844 & 146.023 & 121.711\\
        PbV($-$) & 0.56 & ~ & ~ & ~ & ~ & ~ & ~ \\ 
        $C_{2h}$& ~ & ~ & ~ & ~ & ~ & ~ & ~ \\
        C$_{1,4}$ & ~ & 53.72 & 55.04 & 53.72 & 32.67 & 75.57 & 32.67 \\ 
        C$_{2,3,5,6}$ & ~ & 13.07 & 12.16 & 11.31 & 9.06 & 14.71 & 9.97 \\
        $^{207}$Pb  & ~ & $-$18.15 & $-$18.11 & $-$18.68 & $-$18.37 & $-$17.57 & $-$17.65\\
        $D_{3d}$ & ~ & ~ & ~ & ~ & ~ & ~ & ~ \\ 
        C-first neighbor & ~ & 19.86 & 19.86 & 20.76 & 27.81 & 12.19 & 12.19 \\ 
        $^{207}$Pb  & ~ & $-$793.26 & $-$793.26 & $-$793.26 & $-$1004.34 & $-$1004.34 & $-$1004.34 \\ 
    \end{tabular}
\end{ruledtabular}
\end{table*}
\begin{table*}[!ht]
    \caption{\label{table:hyperfine-C13}
    The effective dynamic hyperfine parameters for selected $^{13}$C nuclear spins in low $C_{2h}$ symmetry for G4V($-$) defects as obtained by HSE06 calculations in the electronic ground state. The directions are defined in Fig.~\ref{fig:rotate}(e).}
    \begin{ruledtabular}
    \setlength\extrarowheight{1.2pt}
\setlength\tabcolsep{1pt}
    \begin{tabular}{lrrrrrr}
        Defect & $A_{xx}$ (MHz) & $A_{yy}$ (MHz) & $A_{zz}$ (MHz) & $A_{xy}$ (MHz) & $A_{xz}$ (MHz) & $A_{yz}$ (MHz) \\  \hline
        &&&&&&\\
         \multicolumn{7}{c}{Ground state} \\ \hline
        SiV($-$) - first neighbor & ~ & ~ & ~ & ~ & ~ & ~ \\
        $A$ & 60.4 & 43.2 & 37.5 & 0.0 & 7.2 & 0.0 \\ 
        $A_{x}$ & $-$22.0 & $-$6.8 & $-$10.6 & $-$3.0 & $-$4.6 & 2.7 \\ 
        $A_{y}$ & $-$50.7 & $-$15.6 & $-$24.4 & 2.3 & $-$10.7 & $-$2.1 \\ 
        SiV($-$) - second neighbor& ~ & ~ & ~ & ~ & ~ & ~ \\
        $A$ & $-$0.6 & $-$2.7 & $-$2.6 & $-$0.8 & 0.5 & $-$0.3 \\ 
        $A_{x}$ & 0.6 & 1.1 & 1.3 & 0.3 & 0.1 & 0.1 \\ 
        $A_{y}$ & 1.8 & 1.7 & 2.7 & 1.0 & 0.7 & $-$0.6 \\
        GeV($-$) - first neighbor& ~ & ~ & ~ & ~ & ~ & ~ \\
        $A$ & 77.6 & 37.0 & 42.0 & 0.0 & 13.9 & 0.0 \\ 
        $A_{x}$ & $-$23.1 & $-$11.5 & $-$13.0 & 1.2 & $-$4.0 & 0.4 \\ 
        $A_{y}$ & $-$52.9 & $-$26.2 & $-$29.7 & $-$1.0 & $-$9.2 & $-$0.3 \\  
        GeV($-$) - second neighbor& ~ & ~ & ~ & ~ & ~ & ~ \\
        $A$ & $-$3.2 & $-$4.3 & $-$4.2 & $-$1.1 & $-$0.2 & 0.6 \\ 
        $A_{x}$ & 1.7 & 1.8 & 2.0 & 0.4 & 0.4 & $-$0.4 \\ 
        $A_{y}$ & 1.5 & 1.9 & 2.7 & 0.3 & $-$0.1 & 0.3 \\ 
        SnV($-$) - first neighbor& ~ & ~ & ~ & ~ & ~ & ~ \\
        $A$ & 70.1 & 28.3 & 33.4 & 0.0 & 14.6 & 0.0 \\ 
        $A_{x}$ & $-$21.4 & $-$9.3 & $-$10.8 & 1.3 & $-$4.3 & 0.4 \\ 
        $A_{y}$ & $-$49.3 & $-$21.3 & $-$24.9 & $-$1.0 & $-$9.8 & $-$0.3 \\
        SnV($-$) - second neighbor& ~ & ~ & ~ & ~ & ~ & ~ \\
        $A$ & $-$4.1 & $-$5.3 & $-$4.9 & $-$0.8 & $-$0.2 & 0.5 \\ 
        $A_{x}$ & 1.9 & 2.1 & 2.1 & 0.3 & 0.4 & $-$0.3 \\ 
        $A_{y}$ & 2.1 & 2.4 & 3.2 & 0.2 & $-$0.1 & 0.2 \\
        PbV($-$) - first neighbor& ~ & ~ & ~ & ~ & ~ & ~ \\
        $A$ & 60.4 & 39.1 & 33.7 & 0.0 & 9.1 & 0.0 \\ 
        $A_{x}$ & $-$26.7 & $-$7.4 & $-$12.2 & $-$4.0 & $-$6.6 & 3.5 \\ 
        $A_{y}$ & $-$61.6 & $-$16.9 & $-$28.1 & 3.1 & $-$15.2 & $-$2.7 \\ 
        PbV($-$) - second neighbor& ~ & ~ & ~ & ~ & ~ & ~ \\
        $A$ & $-$4.9 & $-$5.9 & $-$5.2 & $-$0.6 & $-$0.1 & 0.4 \\ 
        $A_{x}$ & 2.0 & 2.2 & 2.1 & 0.2 & 0.3 & $-$0.3 \\ 
        $A_{y}$ & 2.9 & 3.1 & 3.9 & 0.1 & $-$0.0 & 0.1 \\ 
        &&&&&&\\
         \multicolumn{7}{c}{Excited state} \\ \hline
        SiV($-$) & ~ & ~ & ~ & ~ & ~ & ~ \\ 
        $A$ & 15.91 & 21.81 & 10.57 & -5.13 & -2.58 & 4.46 \\ 
        $A_{x}$ & 1.53 & -4.93 & 0.10 & 2.09 & 0.32 & -1.73 \\ 
        $A_{y}$ & -1.10 & -6.58 & 0.30 & 7.53 & 2.31 & -3.08 \\ 
        GeV($-$) & ~ & ~ & ~ & ~ & ~ & ~ \\ 
        $A$ & 4.80 & 14.61 & 1.43 & 0.94 & -3.18 & 0.44 \\ 
        $A_{x}$ & 4.19 & 3.52 & 2.19 & 6.20 & -1.10 & -3.53 \\ 
        $A_{y}$ & 4.93 & 12.90 & 5.07 & -3.26 & -1.33 & 2.83 \\ 
        SnV($-$) & ~ & ~ & ~ & ~ & ~ & ~ \\ 
        $A$ & 54.96 & 33.86 & 28.43 & 0.03 & 8.83 & -0.02 \\ 
        $A_{x}$ & -18.26 & -3.71 & -7.19 & -2.77 & -4.75 & 2.38 \\ 
        a$A_{y}$ & -42.01 & -8.31 & -16.44 & 2.15 & -10.99 & -1.85 \\ 
        PbV($-$) & ~ & ~ & ~ & ~ & ~ & ~ \\ 
        $A$ & 27.98 & 11.04 & 10.78 & 0.09 & 6.77 & 1.29 \\ 
        $A_{x}$ & -7.83 & 0.23 & -0.87 & 1.13 & -2.18 & 0.12 \\ 
        $A_{y}$ & -19.11 & 1.60 & -2.62 & -0.81 & -6.72 & 0.75 \\ 
    \end{tabular}
\end{ruledtabular}
\end{table*}
In the previous Appendix~\ref{app:HFHam} we discussed thoroughly the case of the
central atom. However, we need a more generic form for $^{13}$C because they are not
sitting on the central high symmetry point that of $D_{3d}$. As before, hyperfine 
tensor elements can be evaluated by spin-polarized DFT whenever the $|e_{x}\rangle$ or
$|e_{y}\rangle$ orbital is being occupied for defects exhibiting $D_{3d}$ symmetry.
In simple words, hyperfine tensor elements can be evaluated whenever the 
$|e_x\rangle$ or $|e_y\rangle$ state is selected. Therefore, the hyperfine 
interaction depicted can be extended for orbital degrees of freedom as
\begin{equation}
    \label{eq:Hyper1}
    \begin{aligned}
    \hat{H}_\text{HF}^{\mathrm{^{13}C}}\! = & \overleftarrow{S}\overleftrightarrow{A}^{(xx)}\overrightarrow{I}|e_{x}\rangle\langle e_{x}|+\overleftarrow{S}\overleftrightarrow{A}^{(xy)}\overrightarrow{I}|e_{x}\rangle\langle e_{y}| 
        \\
    + & \overleftarrow{S}\overleftrightarrow{A}^{(yx)}\overrightarrow{I}|e_{y}\rangle\langle e_{x}|+\overleftarrow{S}\overleftrightarrow{A}^{(yy)}\overrightarrow{I}|e_{y}\rangle\langle e_{y}|\text{,}
    \end{aligned}
\end{equation}
where $\overleftrightarrow{A}^{(xx)}\bigr|_{ij}=\mathcal{A}_{ij}^{x}$ and $\overleftrightarrow{A}^{(yy)}\bigr|_{ij}=\mathcal{A}_{ij}^{y}$ tensors
can be determined by applying Eq.~\eqref{eq:HF} upon spin densities of $|e_x\rangle$
and $|e_y\rangle$, respectively. However, $\overleftrightarrow{A}^{(xy)}$,
$\overleftrightarrow{A}^{(yx)}$ off-diagonal terms cannot be computed directly. 
We also cannot act similarly to what we did in Eq.~\eqref{eq:HFD3d2} because
$^{13}$C ions are not sitting in the central position.

First, we introduce Pauli matrices ($\hat{\sigma}_{x}=(\begin{smallmatrix} & 1\\1 \end{smallmatrix})=|e_{x}\rangle\langle e_{y}|+|e_{y}\rangle\langle e_{x}|$, $\hat{\sigma}_{y}=(\begin{smallmatrix}&\!\!-i\\i&\end{smallmatrix})$, $\hat{\sigma}_{z}=(\begin{smallmatrix}1\\ & \!\!-1 \end{smallmatrix})=|e_{x}\rangle\langle e_{x}|-|e_{y}\rangle\langle e_{y}|$) for orbital degrees of freedom and thus Eq.~\eqref{eq:Hyper1} can be transformed into a compact form as
\begin{equation}
\label{eq:Hyper3}
\begin{aligned}
 \\
\hat{H}_\text{HF}^{\mathrm{^{13}C}}\! = \overleftarrow{S}\overleftrightarrow{A}\overrightarrow{I}
\!+\!
\overleftarrow{S}\overleftrightarrow{A}_{x}\overrightarrow{I}\hat{\sigma}_{z}
\!+\!
\overleftarrow{S}\overleftrightarrow{A}_{y}\overrightarrow{I}\hat{\sigma}_{x} 
\!+\!
\underset{=0}{\underbrace{\overleftarrow{S}\overleftrightarrow{A}_{z}\overrightarrow{I}i\hat{\sigma}_{y}}}
\end{aligned}
\end{equation}
by introducing the following hyperfine tensors as
\begin{subequations}
\label{eq:Hyper4}
\begin{align}
\overleftrightarrow{A}&=\frac{1}{2}\Bigl(\overleftrightarrow{A}^{(xx)}+\overleftrightarrow{A}^{(yy)}\Bigr)  
\\
\overleftrightarrow{A}_{x}&=\frac{1}{2}\Bigl(\overleftrightarrow{A}^{(xx)}-\overleftrightarrow{A}^{(yy)}\Bigr)
\\
\overleftrightarrow{A}_{y}&=\frac{1}{2}\Bigl(\overleftrightarrow{A}^{(xy)}+\overleftrightarrow{A}^{(yx)}\Bigr) 
\\
\overleftrightarrow{A}_{z}&=\frac{1}{2}\Bigl(\overleftrightarrow{A}^{(xy)}-\overleftrightarrow{A}^{(yx)}\Bigr)=0
\text{,}
\end{align}
\end{subequations}
where $\overleftrightarrow{A}_{z}$ is zero because hyperfine interaction is real-valued thus symmetric.

Next, we discuss how to calculate these hyperfine interactions in practice within spin-polarized DFT calculations. In the usual Born-Oppenheimer \textit{ab initio} DFT calculations, the double degenerate $|e_{\{g,u\}}\rangle$ orbital is occupied by a single electron in the ground state for G4V($-$) defect. We analyze the system by assuming that $|e_{{\{g,u\}}(x)}\rangle$ is occupied by an electron in the symmetry configuration $D_{3d}$. Among the six equivalent immediate carbon neighbor atoms together with inversion symmetry, two pairs of three carbon atoms can be rotated to each other about the $C_3$ symmetry axis (see Fig.~\ref{fig:rotate}), e.g.,  $\mathrm{C_1}$, $\mathrm{C_2}$, $\mathrm{C_3}$ carbon atoms exhibit $\overleftrightarrow{A}^{(xx)}$, $\overleftrightarrow{B}^{(xx)}$, $\overleftrightarrow{C}^{(xx)}$ hyperfine tensors, respectively. In this example, we define the following hyperfine matrices for C$_1$,
\begin{equation}
    \label{eq:Hyper5}
\overleftrightarrow{A}^{(xx)}=\overleftrightarrow{A}^{(0^{\circ})}=\left\langle +1;0\right|\hat{H}_\text{HF}^{\mathrm{^{13}C}}\left|+1;0\right\rangle =\overleftrightarrow{A}+\overleftrightarrow{A}_{x} \text{,}
\end{equation}
where $\left\langle +1;0\right|$ and $\left|+1;0\right\rangle$ indicate being fully in $|e_{x}\rangle$ orbital than $|e_{y}\rangle$, with using the notation of $\left|+1,0\right\rangle=|e_{x}\rangle$ and $\left|0,+1\right\rangle=|e_{y}\rangle$.

The hyperfine tensors of $\mathrm{C_2}$ and $\mathrm{C_3}$ can be rotated to that of  $\mathrm{C_1}$ by $\pm120^\circ$ rotations as
\begin{subequations}
\label{eq:Hyper6}
\begin{align}
\begin{split}
\hat{C}_{3}\overleftrightarrow{B}^{(xx)}=\overleftrightarrow{A}^{(+120^{\circ})}= & 
\biggl\langle-\frac{1}{2};-\frac{\sqrt{3}}{2}\biggr|\hat{H}_\text{HF}^{\mathrm{^{13}C}}\biggl|-\frac{1}{2};-\frac{\sqrt{3}}{2}\biggr\rangle
\\
 & =\overleftrightarrow{A}-\frac{1}{2}\overleftrightarrow{A}_{x}+\frac{\sqrt{3}}{2}\overleftrightarrow{A}_{y} \text{,}
\end{split}
\label{eq:Hyper6a}
\\
\label{eq:Hyper6b}
\begin{split}
\hat{C}_{3}^{-1}\overleftrightarrow{C}^{(xx)}=\overleftrightarrow{A}^{(-120^{\circ})}= & 
\biggl\langle-\frac{1}{2};+\frac{\sqrt{3}}{2}\biggr|\hat{H}_\text{HF}^{\mathrm{^{13}C}}\biggl|-\frac{1}{2};+\frac{\sqrt{3}}{2}\biggr\rangle
\\
 & =\overleftrightarrow{A}-\frac{1}{2}\overleftrightarrow{A}_{x}-\frac{\sqrt{3}}{2}\overleftrightarrow{A}_{y} \text{.}
\end{split}
\end{align}
\end{subequations}
At first glance, $\hat{C}_{3}\overleftrightarrow{B}^{(xx)}$ and $\hat{C}_{3}^{-1}\overleftrightarrow{C}^{(xx)}$ matrices should agree with ${A}^{(xx)}$ of C$_1$.
However, the rotation also acts on the orbitals thus the hyperfine tensor should be evaluated on $|e_{(\pm120^{\circ})}\rangle=\frac{1}{2}|e_{x}\rangle\pm\frac{\sqrt{3}}{2}|e_{y}\rangle$ orbitals that we label as $\overleftrightarrow{A}^{(\pm 120^{\circ})}$. Eqs. \eqref{eq:Hyper5}-\eqref{eq:Hyper6a}-\eqref{eq:Hyper6b} allow us to determine the orbital flipping matrices directly as
\begin{subequations}
\label{eq:Hyper7}
\begin{align}
\overleftrightarrow{A}=\frac{1}{3}\Bigl(\overleftrightarrow{A}^{(xx)}+\hat{C}_{3}\overleftrightarrow{B}^{(xx)}+\hat{C}_{3}^{-1}\overleftrightarrow{C}^{(xx)}\Bigr)
\text{,} \label{eq:Hyper7a}
\\
\overleftrightarrow{A}_{x}= \frac{q}{4}\Bigl(2\overleftrightarrow{A}^{(xx)}-\hat{C}_{3}\overleftrightarrow{B}^{(xx)}-\hat{C}_{3}^{-1}\overleftrightarrow{C}^{(xx)}\Bigr)
\text{,} \label{eq:Hyper7b}
\\
\overleftrightarrow{A}_{y}= \frac{q}{\sqrt{3}}\Bigl(\hat{C}_{3}\overleftrightarrow{B}^{(xx)}-\hat{C}_{3}^{-1}\overleftrightarrow{C}^{(xx)}\Bigr)
\text{,} \label{eq:Hyper7c}
\end{align}
\end{subequations}
where we also included the effect of $q=(1+p)/2$ vibronic reduction factor~\cite{Bersuker2006, Bersuker2012} that partially quenches the strength of $\{\hat{\sigma}_{z},\hat{\sigma}_{x}\}$ orbital operators transforming as $E$ representation of $D_{3d}$. The $p$ reduction factor reduces the orbital operators transforming as $A_2$ such as $\hat{L}_z=\hat{\sigma}_{y}$ of Eq.~\eqref{eq:SOC-hammiltonian}. We note that the $\overleftrightarrow{A}$ tensor can be evaluated by putting two half electrons on $|e_x\rangle$ and $|e_y\rangle$ orbitals which method was already used in Ref.~\onlinecite{Harris2023}. 

Finally, we list of the electronic hyperfine tensor data in Tables~\ref{table:hyperfineconstant} and~\ref{table:hyperfineconstantexc} as obtained in HSE06 calculations that we list for the sake of reproduction of our results and the derived hyperfine tensor data in Tables~\ref{table:hyperfine-C13} and \ref{table:hyperfine}. We note that the hyperfine constants of the second neighbor atoms are almost negligible in the electronic excited state.

\bibliography{G4V}

\begin{thebibliography}{88}%
\makeatletter
\providecommand \@ifxundefined [1]{%
 \@ifx{#1\undefined}
}%
\providecommand \@ifnum [1]{%
 \ifnum #1\expandafter \@firstoftwo
 \else \expandafter \@secondoftwo
 \fi
}%
\providecommand \@ifx [1]{%
 \ifx #1\expandafter \@firstoftwo
 \else \expandafter \@secondoftwo
 \fi
}%
\providecommand \natexlab [1]{#1}%
\providecommand \enquote  [1]{``#1''}%
\providecommand \bibnamefont  [1]{#1}%
\providecommand \bibfnamefont [1]{#1}%
\providecommand \citenamefont [1]{#1}%
\providecommand \href@noop [0]{\@secondoftwo}%
\providecommand \href [0]{\begingroup \@sanitize@url \@href}%
\providecommand \@href[1]{\@@startlink{#1}\@@href}%
\providecommand \@@href[1]{\endgroup#1\@@endlink}%
\providecommand \@sanitize@url [0]{\catcode `\\12\catcode `\$12\catcode `\&12\catcode `\#12\catcode `\^12\catcode `\_12\catcode `\%12\relax}%
\providecommand \@@startlink[1]{}%
\providecommand \@@endlink[0]{}%
\providecommand \url  [0]{\begingroup\@sanitize@url \@url }%
\providecommand \@url [1]{\endgroup\@href {#1}{\urlprefix }}%
\providecommand \urlprefix  [0]{URL }%
\providecommand \Eprint [0]{\href }%
\providecommand \doibase [0]{https://doi.org/}%
\providecommand \selectlanguage [0]{\@gobble}%
\providecommand \bibinfo  [0]{\@secondoftwo}%
\providecommand \bibfield  [0]{\@secondoftwo}%
\providecommand \translation [1]{[#1]}%
\providecommand \BibitemOpen [0]{}%
\providecommand \bibitemStop [0]{}%
\providecommand \bibitemNoStop [0]{.\EOS\space}%
\providecommand \EOS [0]{\spacefactor3000\relax}%
\providecommand \BibitemShut  [1]{\csname bibitem#1\endcsname}%
\let\auto@bib@innerbib\@empty
\bibitem [{\citenamefont {Vavilov}\ and\ \citenamefont {Gippius}(1980)}]{Vavilov1980}%
  \BibitemOpen
  \bibfield  {author} {\bibinfo {author} {\bibfnamefont {V.~S.}\ \bibnamefont {Vavilov}}\ and\ \bibinfo {author} {\bibfnamefont {A.}~\bibnamefont {Gippius}},\ }\bibfield  {title} {\bibinfo {title} {Investigation of the cathodoluminescence of epitaxial diamond films},\ }\href@noop {} {\bibfield  {journal} {\bibinfo  {journal} {Sov. Phys. Semicond.}\ }\textbf {\bibinfo {volume} {14}},\ \bibinfo {pages} {1078} (\bibinfo {year} {1980})}\BibitemShut {NoStop}%
\bibitem [{\citenamefont {Hepp}\ \emph {et~al.}(2014)\citenamefont {Hepp}, \citenamefont {M\"uller}, \citenamefont {Waselowski}, \citenamefont {Becker}, \citenamefont {Pingault}, \citenamefont {Sternschulte}, \citenamefont {Steinm\"uller-Nethl}, \citenamefont {Gali}, \citenamefont {Maze}, \citenamefont {Atat\"ure},\ and\ \citenamefont {Becher}}]{Hepp2014}%
  \BibitemOpen
  \bibfield  {author} {\bibinfo {author} {\bibfnamefont {C.}~\bibnamefont {Hepp}}, \bibinfo {author} {\bibfnamefont {T.}~\bibnamefont {M\"uller}}, \bibinfo {author} {\bibfnamefont {V.}~\bibnamefont {Waselowski}}, \bibinfo {author} {\bibfnamefont {J.~N.}\ \bibnamefont {Becker}}, \bibinfo {author} {\bibfnamefont {B.}~\bibnamefont {Pingault}}, \bibinfo {author} {\bibfnamefont {H.}~\bibnamefont {Sternschulte}}, \bibinfo {author} {\bibfnamefont {D.}~\bibnamefont {Steinm\"uller-Nethl}}, \bibinfo {author} {\bibfnamefont {A.}~\bibnamefont {Gali}}, \bibinfo {author} {\bibfnamefont {J.~R.}\ \bibnamefont {Maze}}, \bibinfo {author} {\bibfnamefont {M.}~\bibnamefont {Atat\"ure}},\ and\ \bibinfo {author} {\bibfnamefont {C.}~\bibnamefont {Becher}},\ }\bibfield  {title} {\bibinfo {title} {Electronic structure of the silicon vacancy color center in diamond},\ }\href {https://doi.org/10.1103/PhysRevLett.112.036405} {\bibfield  {journal} {\bibinfo  {journal} {Phys. Rev. Lett.}\ }\textbf {\bibinfo {volume} {112}},\ \bibinfo
  {pages} {036405} (\bibinfo {year} {2014})}\BibitemShut {NoStop}%
\bibitem [{\citenamefont {M{\"u}ller}\ \emph {et~al.}(2014)\citenamefont {M{\"u}ller}, \citenamefont {Hepp}, \citenamefont {Pingault}, \citenamefont {Neu}, \citenamefont {Gsell}, \citenamefont {Schreck}, \citenamefont {Sternschulte}, \citenamefont {Steinm{\"u}ller-Nethl}, \citenamefont {Becher},\ and\ \citenamefont {Atat{\"u}re}}]{Muller2014}%
  \BibitemOpen
  \bibfield  {author} {\bibinfo {author} {\bibfnamefont {T.}~\bibnamefont {M{\"u}ller}}, \bibinfo {author} {\bibfnamefont {C.}~\bibnamefont {Hepp}}, \bibinfo {author} {\bibfnamefont {B.}~\bibnamefont {Pingault}}, \bibinfo {author} {\bibfnamefont {E.}~\bibnamefont {Neu}}, \bibinfo {author} {\bibfnamefont {S.}~\bibnamefont {Gsell}}, \bibinfo {author} {\bibfnamefont {M.}~\bibnamefont {Schreck}}, \bibinfo {author} {\bibfnamefont {H.}~\bibnamefont {Sternschulte}}, \bibinfo {author} {\bibfnamefont {D.}~\bibnamefont {Steinm{\"u}ller-Nethl}}, \bibinfo {author} {\bibfnamefont {C.}~\bibnamefont {Becher}},\ and\ \bibinfo {author} {\bibfnamefont {M.}~\bibnamefont {Atat{\"u}re}},\ }\bibfield  {title} {\bibinfo {title} {Optical signatures of silicon-vacancy spins in diamond},\ }\href {https://doi.org/10.1038/ncomms4328} {\bibfield  {journal} {\bibinfo  {journal} {Nature Communications}\ }\textbf {\bibinfo {volume} {5}},\ \bibinfo {pages} {3328} (\bibinfo {year} {2014})}\BibitemShut {NoStop}%
\bibitem [{\citenamefont {Rogers}\ \emph {et~al.}(2014{\natexlab{a}})\citenamefont {Rogers}, \citenamefont {Jahnke}, \citenamefont {Doherty}, \citenamefont {Dietrich}, \citenamefont {McGuinness}, \citenamefont {M\"uller}, \citenamefont {Teraji}, \citenamefont {Sumiya}, \citenamefont {Isoya}, \citenamefont {Manson},\ and\ \citenamefont {Jelezko}}]{Rogers2014}%
  \BibitemOpen
  \bibfield  {author} {\bibinfo {author} {\bibfnamefont {L.~J.}\ \bibnamefont {Rogers}}, \bibinfo {author} {\bibfnamefont {K.~D.}\ \bibnamefont {Jahnke}}, \bibinfo {author} {\bibfnamefont {M.~W.}\ \bibnamefont {Doherty}}, \bibinfo {author} {\bibfnamefont {A.}~\bibnamefont {Dietrich}}, \bibinfo {author} {\bibfnamefont {L.~P.}\ \bibnamefont {McGuinness}}, \bibinfo {author} {\bibfnamefont {C.}~\bibnamefont {M\"uller}}, \bibinfo {author} {\bibfnamefont {T.}~\bibnamefont {Teraji}}, \bibinfo {author} {\bibfnamefont {H.}~\bibnamefont {Sumiya}}, \bibinfo {author} {\bibfnamefont {J.}~\bibnamefont {Isoya}}, \bibinfo {author} {\bibfnamefont {N.~B.}\ \bibnamefont {Manson}},\ and\ \bibinfo {author} {\bibfnamefont {F.}~\bibnamefont {Jelezko}},\ }\bibfield  {title} {\bibinfo {title} {Electronic structure of the negatively charged silicon-vacancy center in diamond},\ }\href {https://doi.org/10.1103/PhysRevB.89.235101} {\bibfield  {journal} {\bibinfo  {journal} {Phys. Rev. B}\ }\textbf {\bibinfo {volume} {89}},\ \bibinfo
  {pages} {235101} (\bibinfo {year} {2014}{\natexlab{a}})}\BibitemShut {NoStop}%
\bibitem [{\citenamefont {Ekimov}\ \emph {et~al.}(2015)\citenamefont {Ekimov}, \citenamefont {Lyapin}, \citenamefont {Boldyrev}, \citenamefont {Kondrin}, \citenamefont {Khmelnitskiy}, \citenamefont {Gavva}, \citenamefont {Kotereva},\ and\ \citenamefont {Popova}}]{Ekimov2015}%
  \BibitemOpen
  \bibfield  {author} {\bibinfo {author} {\bibfnamefont {E.~A.}\ \bibnamefont {Ekimov}}, \bibinfo {author} {\bibfnamefont {S.~G.}\ \bibnamefont {Lyapin}}, \bibinfo {author} {\bibfnamefont {K.~N.}\ \bibnamefont {Boldyrev}}, \bibinfo {author} {\bibfnamefont {M.~V.}\ \bibnamefont {Kondrin}}, \bibinfo {author} {\bibfnamefont {R.}~\bibnamefont {Khmelnitskiy}}, \bibinfo {author} {\bibfnamefont {V.~A.}\ \bibnamefont {Gavva}}, \bibinfo {author} {\bibfnamefont {T.~V.}\ \bibnamefont {Kotereva}},\ and\ \bibinfo {author} {\bibfnamefont {M.~N.}\ \bibnamefont {Popova}},\ }\bibfield  {title} {\bibinfo {title} {Germanium--vacancy color center in isotopically enriched diamonds synthesized at high pressures},\ }\href {https://doi.org/10.1134/S0021364015230034} {\bibfield  {journal} {\bibinfo  {journal} {JETP Letters}\ }\textbf {\bibinfo {volume} {102}},\ \bibinfo {pages} {701} (\bibinfo {year} {2015})}\BibitemShut {NoStop}%
\bibitem [{\citenamefont {Iwasaki}\ \emph {et~al.}(2015)\citenamefont {Iwasaki}, \citenamefont {Ishibashi}, \citenamefont {Miyamoto}, \citenamefont {Doi}, \citenamefont {Kobayashi}, \citenamefont {Miyazaki}, \citenamefont {Tahara}, \citenamefont {Jahnke}, \citenamefont {Rogers}, \citenamefont {Naydenov}, \citenamefont {Jelezko}, \citenamefont {Yamasaki}, \citenamefont {Nagamachi}, \citenamefont {Inubushi}, \citenamefont {Mizuochi},\ and\ \citenamefont {Hatano}}]{Iwasaki2015}%
  \BibitemOpen
  \bibfield  {author} {\bibinfo {author} {\bibfnamefont {T.}~\bibnamefont {Iwasaki}}, \bibinfo {author} {\bibfnamefont {F.}~\bibnamefont {Ishibashi}}, \bibinfo {author} {\bibfnamefont {Y.}~\bibnamefont {Miyamoto}}, \bibinfo {author} {\bibfnamefont {Y.}~\bibnamefont {Doi}}, \bibinfo {author} {\bibfnamefont {S.}~\bibnamefont {Kobayashi}}, \bibinfo {author} {\bibfnamefont {T.}~\bibnamefont {Miyazaki}}, \bibinfo {author} {\bibfnamefont {K.}~\bibnamefont {Tahara}}, \bibinfo {author} {\bibfnamefont {K.~D.}\ \bibnamefont {Jahnke}}, \bibinfo {author} {\bibfnamefont {L.~J.}\ \bibnamefont {Rogers}}, \bibinfo {author} {\bibfnamefont {B.}~\bibnamefont {Naydenov}}, \bibinfo {author} {\bibfnamefont {F.}~\bibnamefont {Jelezko}}, \bibinfo {author} {\bibfnamefont {S.}~\bibnamefont {Yamasaki}}, \bibinfo {author} {\bibfnamefont {S.}~\bibnamefont {Nagamachi}}, \bibinfo {author} {\bibfnamefont {T.}~\bibnamefont {Inubushi}}, \bibinfo {author} {\bibfnamefont {N.}~\bibnamefont {Mizuochi}},\ and\ \bibinfo {author} {\bibfnamefont
  {M.}~\bibnamefont {Hatano}},\ }\bibfield  {title} {\bibinfo {title} {Germanium-vacancy single color centers in diamond},\ }\href {https://doi.org/10.1038/srep12882} {\bibfield  {journal} {\bibinfo  {journal} {Scientific Reports}\ }\textbf {\bibinfo {volume} {5}},\ \bibinfo {pages} {12882} (\bibinfo {year} {2015})}\BibitemShut {NoStop}%
\bibitem [{\citenamefont {Iwasaki}\ \emph {et~al.}(2017)\citenamefont {Iwasaki}, \citenamefont {Miyamoto}, \citenamefont {Taniguchi}, \citenamefont {Siyushev}, \citenamefont {Metsch}, \citenamefont {Jelezko},\ and\ \citenamefont {Hatano}}]{Iwasaki2017}%
  \BibitemOpen
  \bibfield  {author} {\bibinfo {author} {\bibfnamefont {T.}~\bibnamefont {Iwasaki}}, \bibinfo {author} {\bibfnamefont {Y.}~\bibnamefont {Miyamoto}}, \bibinfo {author} {\bibfnamefont {T.}~\bibnamefont {Taniguchi}}, \bibinfo {author} {\bibfnamefont {P.}~\bibnamefont {Siyushev}}, \bibinfo {author} {\bibfnamefont {M.~H.}\ \bibnamefont {Metsch}}, \bibinfo {author} {\bibfnamefont {F.}~\bibnamefont {Jelezko}},\ and\ \bibinfo {author} {\bibfnamefont {M.}~\bibnamefont {Hatano}},\ }\bibfield  {title} {\bibinfo {title} {Tin-vacancy quantum emitters in diamond},\ }\href {https://doi.org/10.1103/PhysRevLett.119.253601} {\bibfield  {journal} {\bibinfo  {journal} {Phys. Rev. Lett.}\ }\textbf {\bibinfo {volume} {119}},\ \bibinfo {pages} {253601} (\bibinfo {year} {2017})}\BibitemShut {NoStop}%
\bibitem [{\citenamefont {Meesala}\ \emph {et~al.}(2018)\citenamefont {Meesala}, \citenamefont {Sohn}, \citenamefont {Pingault}, \citenamefont {Shao}, \citenamefont {Atikian}, \citenamefont {Holzgrafe}, \citenamefont {G\"undo\ifmmode~\breve{g}\else \u{g}\fi{}an}, \citenamefont {Stavrakas}, \citenamefont {Sipahigil}, \citenamefont {Chia}, \citenamefont {Evans}, \citenamefont {Burek}, \citenamefont {Zhang}, \citenamefont {Wu}, \citenamefont {Pacheco}, \citenamefont {Abraham}, \citenamefont {Bielejec}, \citenamefont {Lukin}, \citenamefont {Atat\"ure},\ and\ \citenamefont {Lon\ifmmode~\check{c}\else \v{c}\fi{}ar}}]{Meesala2018}%
  \BibitemOpen
  \bibfield  {author} {\bibinfo {author} {\bibfnamefont {S.}~\bibnamefont {Meesala}}, \bibinfo {author} {\bibfnamefont {Y.-I.}\ \bibnamefont {Sohn}}, \bibinfo {author} {\bibfnamefont {B.}~\bibnamefont {Pingault}}, \bibinfo {author} {\bibfnamefont {L.}~\bibnamefont {Shao}}, \bibinfo {author} {\bibfnamefont {H.~A.}\ \bibnamefont {Atikian}}, \bibinfo {author} {\bibfnamefont {J.}~\bibnamefont {Holzgrafe}}, \bibinfo {author} {\bibfnamefont {M.}~\bibnamefont {G\"undo\ifmmode~\breve{g}\else \u{g}\fi{}an}}, \bibinfo {author} {\bibfnamefont {C.}~\bibnamefont {Stavrakas}}, \bibinfo {author} {\bibfnamefont {A.}~\bibnamefont {Sipahigil}}, \bibinfo {author} {\bibfnamefont {C.}~\bibnamefont {Chia}}, \bibinfo {author} {\bibfnamefont {R.}~\bibnamefont {Evans}}, \bibinfo {author} {\bibfnamefont {M.~J.}\ \bibnamefont {Burek}}, \bibinfo {author} {\bibfnamefont {M.}~\bibnamefont {Zhang}}, \bibinfo {author} {\bibfnamefont {L.}~\bibnamefont {Wu}}, \bibinfo {author} {\bibfnamefont {J.~L.}\ \bibnamefont {Pacheco}}, \bibinfo {author}
  {\bibfnamefont {J.}~\bibnamefont {Abraham}}, \bibinfo {author} {\bibfnamefont {E.}~\bibnamefont {Bielejec}}, \bibinfo {author} {\bibfnamefont {M.~D.}\ \bibnamefont {Lukin}}, \bibinfo {author} {\bibfnamefont {M.}~\bibnamefont {Atat\"ure}},\ and\ \bibinfo {author} {\bibfnamefont {M.}~\bibnamefont {Lon\ifmmode~\check{c}\else \v{c}\fi{}ar}},\ }\bibfield  {title} {\bibinfo {title} {Strain engineering of the silicon-vacancy center in diamond},\ }\href {https://doi.org/10.1103/PhysRevB.97.205444} {\bibfield  {journal} {\bibinfo  {journal} {Phys. Rev. B}\ }\textbf {\bibinfo {volume} {97}},\ \bibinfo {pages} {205444} (\bibinfo {year} {2018})}\BibitemShut {NoStop}%
\bibitem [{\citenamefont {Ditalia~Tchernij}\ \emph {et~al.}(2018)\citenamefont {Ditalia~Tchernij}, \citenamefont {L{\"u}hmann}, \citenamefont {Herzig}, \citenamefont {K{\"u}pper}, \citenamefont {Damin}, \citenamefont {Santonocito}, \citenamefont {Signorile}, \citenamefont {Traina}, \citenamefont {Moreva}, \citenamefont {Celegato}, \citenamefont {Pezzagna}, \citenamefont {Degiovanni}, \citenamefont {Olivero}, \citenamefont {Jak{\v{s}}i{\'{c}}}, \citenamefont {Meijer}, \citenamefont {Genovese},\ and\ \citenamefont {Forneris}}]{Tchernij2018}%
  \BibitemOpen
  \bibfield  {author} {\bibinfo {author} {\bibfnamefont {S.}~\bibnamefont {Ditalia~Tchernij}}, \bibinfo {author} {\bibfnamefont {T.}~\bibnamefont {L{\"u}hmann}}, \bibinfo {author} {\bibfnamefont {T.}~\bibnamefont {Herzig}}, \bibinfo {author} {\bibfnamefont {J.}~\bibnamefont {K{\"u}pper}}, \bibinfo {author} {\bibfnamefont {A.}~\bibnamefont {Damin}}, \bibinfo {author} {\bibfnamefont {S.}~\bibnamefont {Santonocito}}, \bibinfo {author} {\bibfnamefont {M.}~\bibnamefont {Signorile}}, \bibinfo {author} {\bibfnamefont {P.}~\bibnamefont {Traina}}, \bibinfo {author} {\bibfnamefont {E.}~\bibnamefont {Moreva}}, \bibinfo {author} {\bibfnamefont {F.}~\bibnamefont {Celegato}}, \bibinfo {author} {\bibfnamefont {S.}~\bibnamefont {Pezzagna}}, \bibinfo {author} {\bibfnamefont {I.~P.}\ \bibnamefont {Degiovanni}}, \bibinfo {author} {\bibfnamefont {P.}~\bibnamefont {Olivero}}, \bibinfo {author} {\bibfnamefont {M.}~\bibnamefont {Jak{\v{s}}i{\'{c}}}}, \bibinfo {author} {\bibfnamefont {J.}~\bibnamefont {Meijer}}, \bibinfo {author}
  {\bibfnamefont {P.~M.}\ \bibnamefont {Genovese}},\ and\ \bibinfo {author} {\bibfnamefont {J.}~\bibnamefont {Forneris}},\ }\bibfield  {title} {\bibinfo {title} {Single-photon emitters in lead-implanted single-crystal diamond},\ }\href {https://doi.org/10.1021/acsphotonics.8b01013} {\bibfield  {journal} {\bibinfo  {journal} {ACS Photonics}\ }\textbf {\bibinfo {volume} {5}},\ \bibinfo {pages} {4864} (\bibinfo {year} {2018})}\BibitemShut {NoStop}%
\bibitem [{\citenamefont {Trusheim}\ \emph {et~al.}(2019)\citenamefont {Trusheim}, \citenamefont {Wan}, \citenamefont {Chen}, \citenamefont {Ciccarino}, \citenamefont {Flick}, \citenamefont {Sundararaman}, \citenamefont {Malladi}, \citenamefont {Bersin}, \citenamefont {Walsh}, \citenamefont {Lienhard}, \citenamefont {Bakhru}, \citenamefont {Narang},\ and\ \citenamefont {Englund}}]{Trusheim2019}%
  \BibitemOpen
  \bibfield  {author} {\bibinfo {author} {\bibfnamefont {M.~E.}\ \bibnamefont {Trusheim}}, \bibinfo {author} {\bibfnamefont {N.~H.}\ \bibnamefont {Wan}}, \bibinfo {author} {\bibfnamefont {K.~C.}\ \bibnamefont {Chen}}, \bibinfo {author} {\bibfnamefont {C.~J.}\ \bibnamefont {Ciccarino}}, \bibinfo {author} {\bibfnamefont {J.}~\bibnamefont {Flick}}, \bibinfo {author} {\bibfnamefont {R.}~\bibnamefont {Sundararaman}}, \bibinfo {author} {\bibfnamefont {G.}~\bibnamefont {Malladi}}, \bibinfo {author} {\bibfnamefont {E.}~\bibnamefont {Bersin}}, \bibinfo {author} {\bibfnamefont {M.}~\bibnamefont {Walsh}}, \bibinfo {author} {\bibfnamefont {B.}~\bibnamefont {Lienhard}}, \bibinfo {author} {\bibfnamefont {H.}~\bibnamefont {Bakhru}}, \bibinfo {author} {\bibfnamefont {P.}~\bibnamefont {Narang}},\ and\ \bibinfo {author} {\bibfnamefont {D.}~\bibnamefont {Englund}},\ }\bibfield  {title} {\bibinfo {title} {Lead-related quantum emitters in diamond},\ }\href {https://doi.org/10.1103/PhysRevB.99.075430} {\bibfield  {journal}
  {\bibinfo  {journal} {Phys. Rev. B}\ }\textbf {\bibinfo {volume} {99}},\ \bibinfo {pages} {075430} (\bibinfo {year} {2019})}\BibitemShut {NoStop}%
\bibitem [{\citenamefont {Chen}\ \emph {et~al.}(2019)\citenamefont {Chen}, \citenamefont {Zheludev},\ and\ \citenamefont {Gao}}]{Chen2019}%
  \BibitemOpen
  \bibfield  {author} {\bibinfo {author} {\bibfnamefont {D.}~\bibnamefont {Chen}}, \bibinfo {author} {\bibfnamefont {N.~I.}\ \bibnamefont {Zheludev}},\ and\ \bibinfo {author} {\bibfnamefont {W.}~\bibnamefont {Gao}},\ }\bibfield  {title} {\bibinfo {title} {Building blocks for quantum network based on group‐iv split‐vacancy centers in diamond},\ }\href {https://api.semanticscholar.org/CorpusID:209985241} {\bibfield  {journal} {\bibinfo  {journal} {Advanced Quantum Technologies}\ }\textbf {\bibinfo {volume} {3}} (\bibinfo {year} {2019})}\BibitemShut {NoStop}%
\bibitem [{\citenamefont {Trusheim}\ \emph {et~al.}(2020)\citenamefont {Trusheim}, \citenamefont {Pingault}, \citenamefont {Wan}, \citenamefont {G\"undo\ifmmode~\breve{g}\else \u{g}\fi{}an}, \citenamefont {De~Santis}, \citenamefont {Debroux}, \citenamefont {Gangloff}, \citenamefont {Purser}, \citenamefont {Chen}, \citenamefont {Walsh}, \citenamefont {Rose}, \citenamefont {Becker}, \citenamefont {Lienhard}, \citenamefont {Bersin}, \citenamefont {Paradeisanos}, \citenamefont {Wang}, \citenamefont {Lyzwa}, \citenamefont {Montblanch}, \citenamefont {Malladi}, \citenamefont {Bakhru}, \citenamefont {Ferrari}, \citenamefont {Walmsley}, \citenamefont {Atat\"ure},\ and\ \citenamefont {Englund}}]{Trusheim2020}%
  \BibitemOpen
  \bibfield  {author} {\bibinfo {author} {\bibfnamefont {M.~E.}\ \bibnamefont {Trusheim}}, \bibinfo {author} {\bibfnamefont {B.}~\bibnamefont {Pingault}}, \bibinfo {author} {\bibfnamefont {N.~H.}\ \bibnamefont {Wan}}, \bibinfo {author} {\bibfnamefont {M.}~\bibnamefont {G\"undo\ifmmode~\breve{g}\else \u{g}\fi{}an}}, \bibinfo {author} {\bibfnamefont {L.}~\bibnamefont {De~Santis}}, \bibinfo {author} {\bibfnamefont {R.}~\bibnamefont {Debroux}}, \bibinfo {author} {\bibfnamefont {D.}~\bibnamefont {Gangloff}}, \bibinfo {author} {\bibfnamefont {C.}~\bibnamefont {Purser}}, \bibinfo {author} {\bibfnamefont {K.~C.}\ \bibnamefont {Chen}}, \bibinfo {author} {\bibfnamefont {M.}~\bibnamefont {Walsh}}, \bibinfo {author} {\bibfnamefont {J.~J.}\ \bibnamefont {Rose}}, \bibinfo {author} {\bibfnamefont {J.~N.}\ \bibnamefont {Becker}}, \bibinfo {author} {\bibfnamefont {B.}~\bibnamefont {Lienhard}}, \bibinfo {author} {\bibfnamefont {E.}~\bibnamefont {Bersin}}, \bibinfo {author} {\bibfnamefont {I.}~\bibnamefont {Paradeisanos}},
  \bibinfo {author} {\bibfnamefont {G.}~\bibnamefont {Wang}}, \bibinfo {author} {\bibfnamefont {D.}~\bibnamefont {Lyzwa}}, \bibinfo {author} {\bibfnamefont {A.~R.-P.}\ \bibnamefont {Montblanch}}, \bibinfo {author} {\bibfnamefont {G.}~\bibnamefont {Malladi}}, \bibinfo {author} {\bibfnamefont {H.}~\bibnamefont {Bakhru}}, \bibinfo {author} {\bibfnamefont {A.~C.}\ \bibnamefont {Ferrari}}, \bibinfo {author} {\bibfnamefont {I.~A.}\ \bibnamefont {Walmsley}}, \bibinfo {author} {\bibfnamefont {M.}~\bibnamefont {Atat\"ure}},\ and\ \bibinfo {author} {\bibfnamefont {D.}~\bibnamefont {Englund}},\ }\bibfield  {title} {\bibinfo {title} {Transform-limited photons from a coherent tin-vacancy spin in diamond},\ }\href {https://doi.org/10.1103/PhysRevLett.124.023602} {\bibfield  {journal} {\bibinfo  {journal} {Phys. Rev. Lett.}\ }\textbf {\bibinfo {volume} {124}},\ \bibinfo {pages} {023602} (\bibinfo {year} {2020})}\BibitemShut {NoStop}%
\bibitem [{\citenamefont {G\"orlitz}\ \emph {et~al.}(2020)\citenamefont {G\"orlitz}, \citenamefont {Herrmann}, \citenamefont {Thiering}, \citenamefont {Fuchs}, \citenamefont {Gandil}, \citenamefont {Iwasaki}, \citenamefont {Taniguchi}, \citenamefont {Kieschnick}, \citenamefont {Meijer}, \citenamefont {Hatano}, \citenamefont {Gali},\ and\ \citenamefont {Becher}}]{Gorlitz2020}%
  \BibitemOpen
  \bibfield  {author} {\bibinfo {author} {\bibfnamefont {J.}~\bibnamefont {G\"orlitz}}, \bibinfo {author} {\bibfnamefont {D.}~\bibnamefont {Herrmann}}, \bibinfo {author} {\bibfnamefont {G.}~\bibnamefont {Thiering}}, \bibinfo {author} {\bibfnamefont {P.}~\bibnamefont {Fuchs}}, \bibinfo {author} {\bibfnamefont {M.}~\bibnamefont {Gandil}}, \bibinfo {author} {\bibfnamefont {T.}~\bibnamefont {Iwasaki}}, \bibinfo {author} {\bibfnamefont {T.}~\bibnamefont {Taniguchi}}, \bibinfo {author} {\bibfnamefont {M.}~\bibnamefont {Kieschnick}}, \bibinfo {author} {\bibfnamefont {J.}~\bibnamefont {Meijer}}, \bibinfo {author} {\bibfnamefont {M.}~\bibnamefont {Hatano}}, \bibinfo {author} {\bibfnamefont {A.}~\bibnamefont {Gali}},\ and\ \bibinfo {author} {\bibfnamefont {C.}~\bibnamefont {Becher}},\ }\bibfield  {title} {\bibinfo {title} {Spectroscopic investigations of negatively charged tin-vacancy centres in diamond},\ }\href {https://doi.org/10.1088/1367-2630/ab6631} {\bibfield  {journal} {\bibinfo  {journal} {New Journal of
  Physics}\ }\textbf {\bibinfo {volume} {22}},\ \bibinfo {pages} {013048} (\bibinfo {year} {2020})}\BibitemShut {NoStop}%
\bibitem [{\citenamefont {Krivobok}\ \emph {et~al.}(2020)\citenamefont {Krivobok}, \citenamefont {Ekimov}, \citenamefont {Lyapin}, \citenamefont {Nikolaev}, \citenamefont {Skakov}, \citenamefont {Razgulov},\ and\ \citenamefont {Kondrin}}]{Krivobok2020}%
  \BibitemOpen
  \bibfield  {author} {\bibinfo {author} {\bibfnamefont {V.~S.}\ \bibnamefont {Krivobok}}, \bibinfo {author} {\bibfnamefont {E.~A.}\ \bibnamefont {Ekimov}}, \bibinfo {author} {\bibfnamefont {S.~G.}\ \bibnamefont {Lyapin}}, \bibinfo {author} {\bibfnamefont {S.~N.}\ \bibnamefont {Nikolaev}}, \bibinfo {author} {\bibfnamefont {Y.~A.}\ \bibnamefont {Skakov}}, \bibinfo {author} {\bibfnamefont {A.~A.}\ \bibnamefont {Razgulov}},\ and\ \bibinfo {author} {\bibfnamefont {M.~V.}\ \bibnamefont {Kondrin}},\ }\bibfield  {title} {\bibinfo {title} {Observation of a 1.979-ev spectral line of a germanium-related color center in microdiamonds and nanodiamonds},\ }\href {https://doi.org/10.1103/PhysRevB.101.144103} {\bibfield  {journal} {\bibinfo  {journal} {Phys. Rev. B}\ }\textbf {\bibinfo {volume} {101}},\ \bibinfo {pages} {144103} (\bibinfo {year} {2020})}\BibitemShut {NoStop}%
\bibitem [{\citenamefont {Rugar}\ \emph {et~al.}(2021)\citenamefont {Rugar}, \citenamefont {Aghaeimeibodi}, \citenamefont {Riedel}, \citenamefont {Dory}, \citenamefont {Lu}, \citenamefont {McQuade}, \citenamefont {Shen}, \citenamefont {Melosh},\ and\ \citenamefont {Vu\ifmmode \check{c}\else \v{c}\fi{}kovi\ifmmode~\acute{c}\else \'{c}\fi{}}}]{Rugar2021}%
  \BibitemOpen
  \bibfield  {author} {\bibinfo {author} {\bibfnamefont {A.~E.}\ \bibnamefont {Rugar}}, \bibinfo {author} {\bibfnamefont {S.}~\bibnamefont {Aghaeimeibodi}}, \bibinfo {author} {\bibfnamefont {D.}~\bibnamefont {Riedel}}, \bibinfo {author} {\bibfnamefont {C.}~\bibnamefont {Dory}}, \bibinfo {author} {\bibfnamefont {H.}~\bibnamefont {Lu}}, \bibinfo {author} {\bibfnamefont {P.~J.}\ \bibnamefont {McQuade}}, \bibinfo {author} {\bibfnamefont {Z.-X.}\ \bibnamefont {Shen}}, \bibinfo {author} {\bibfnamefont {N.~A.}\ \bibnamefont {Melosh}},\ and\ \bibinfo {author} {\bibfnamefont {J.}~\bibnamefont {Vu\ifmmode \check{c}\else \v{c}\fi{}kovi\ifmmode~\acute{c}\else \'{c}\fi{}}},\ }\bibfield  {title} {\bibinfo {title} {Quantum photonic interface for tin-vacancy centers in diamond},\ }\href {https://doi.org/10.1103/PhysRevX.11.031021} {\bibfield  {journal} {\bibinfo  {journal} {Phys. Rev. X}\ }\textbf {\bibinfo {volume} {11}},\ \bibinfo {pages} {031021} (\bibinfo {year} {2021})}\BibitemShut {NoStop}%
\bibitem [{\citenamefont {Aghaeimeibodi}\ \emph {et~al.}(2021)\citenamefont {Aghaeimeibodi}, \citenamefont {Riedel}, \citenamefont {Rugar}, \citenamefont {Dory},\ and\ \citenamefont {Vu\ifmmode \check{c}\else \v{c}\fi{}kovi\ifmmode~\acute{c}\else \'{c}\fi{}}}]{Aghaeimeibodi2021}%
  \BibitemOpen
  \bibfield  {author} {\bibinfo {author} {\bibfnamefont {S.}~\bibnamefont {Aghaeimeibodi}}, \bibinfo {author} {\bibfnamefont {D.}~\bibnamefont {Riedel}}, \bibinfo {author} {\bibfnamefont {A.~E.}\ \bibnamefont {Rugar}}, \bibinfo {author} {\bibfnamefont {C.}~\bibnamefont {Dory}},\ and\ \bibinfo {author} {\bibfnamefont {J.}~\bibnamefont {Vu\ifmmode \check{c}\else \v{c}\fi{}kovi\ifmmode~\acute{c}\else \'{c}\fi{}}},\ }\bibfield  {title} {\bibinfo {title} {Electrical tuning of tin-vacancy centers in diamond},\ }\href {https://doi.org/10.1103/PhysRevApplied.15.064010} {\bibfield  {journal} {\bibinfo  {journal} {Phys. Rev. Appl.}\ }\textbf {\bibinfo {volume} {15}},\ \bibinfo {pages} {064010} (\bibinfo {year} {2021})}\BibitemShut {NoStop}%
\bibitem [{\citenamefont {Wang}\ \emph {et~al.}(2021)\citenamefont {Wang}, \citenamefont {Taniguchi}, \citenamefont {Miyamoto}, \citenamefont {Hatano},\ and\ \citenamefont {Iwasaki}}]{Wang2021}%
  \BibitemOpen
  \bibfield  {author} {\bibinfo {author} {\bibfnamefont {P.}~\bibnamefont {Wang}}, \bibinfo {author} {\bibfnamefont {T.}~\bibnamefont {Taniguchi}}, \bibinfo {author} {\bibfnamefont {Y.}~\bibnamefont {Miyamoto}}, \bibinfo {author} {\bibfnamefont {M.}~\bibnamefont {Hatano}},\ and\ \bibinfo {author} {\bibfnamefont {T.}~\bibnamefont {Iwasaki}},\ }\bibfield  {title} {\bibinfo {title} {Low-temperature spectroscopic investigation of lead-vacancy centers in diamond fabricated by high-pressure and high-temperature treatment},\ }\href {https://doi.org/10.1021/acsphotonics.1c00840} {\bibfield  {journal} {\bibinfo  {journal} {ACS Photonics}\ }\textbf {\bibinfo {volume} {8}},\ \bibinfo {pages} {2947} (\bibinfo {year} {2021})}\BibitemShut {NoStop}%
\bibitem [{\citenamefont {Vindolet}\ \emph {et~al.}(2022)\citenamefont {Vindolet}, \citenamefont {Adam}, \citenamefont {Toraille}, \citenamefont {Chipaux}, \citenamefont {Hilberer}, \citenamefont {Dupuy}, \citenamefont {Razinkovas}, \citenamefont {Alkauskas}, \citenamefont {Thiering}, \citenamefont {Gali}, \citenamefont {De~Feudis}, \citenamefont {Ngandeu~Ngambou}, \citenamefont {Achard}, \citenamefont {Tallaire}, \citenamefont {Schmidt}, \citenamefont {Becher},\ and\ \citenamefont {Roch}}]{Vindolet2022}%
  \BibitemOpen
  \bibfield  {author} {\bibinfo {author} {\bibfnamefont {B.}~\bibnamefont {Vindolet}}, \bibinfo {author} {\bibfnamefont {M.-P.}\ \bibnamefont {Adam}}, \bibinfo {author} {\bibfnamefont {L.}~\bibnamefont {Toraille}}, \bibinfo {author} {\bibfnamefont {M.}~\bibnamefont {Chipaux}}, \bibinfo {author} {\bibfnamefont {A.}~\bibnamefont {Hilberer}}, \bibinfo {author} {\bibfnamefont {G.}~\bibnamefont {Dupuy}}, \bibinfo {author} {\bibfnamefont {L.}~\bibnamefont {Razinkovas}}, \bibinfo {author} {\bibfnamefont {A.}~\bibnamefont {Alkauskas}}, \bibinfo {author} {\bibfnamefont {G.}~\bibnamefont {Thiering}}, \bibinfo {author} {\bibfnamefont {A.}~\bibnamefont {Gali}}, \bibinfo {author} {\bibfnamefont {M.}~\bibnamefont {De~Feudis}}, \bibinfo {author} {\bibfnamefont {M.~W.}\ \bibnamefont {Ngandeu~Ngambou}}, \bibinfo {author} {\bibfnamefont {J.}~\bibnamefont {Achard}}, \bibinfo {author} {\bibfnamefont {A.}~\bibnamefont {Tallaire}}, \bibinfo {author} {\bibfnamefont {M.}~\bibnamefont {Schmidt}}, \bibinfo {author} {\bibfnamefont
  {C.}~\bibnamefont {Becher}},\ and\ \bibinfo {author} {\bibfnamefont {J.-F.}\ \bibnamefont {Roch}},\ }\bibfield  {title} {\bibinfo {title} {Optical properties of siv and gev color centers in nanodiamonds under hydrostatic pressures up to 180 gpa},\ }\href {https://doi.org/10.1103/PhysRevB.106.214109} {\bibfield  {journal} {\bibinfo  {journal} {Phys. Rev. B}\ }\textbf {\bibinfo {volume} {106}},\ \bibinfo {pages} {214109} (\bibinfo {year} {2022})}\BibitemShut {NoStop}%
\bibitem [{\citenamefont {Wang}\ \emph {et~al.}(2024)\citenamefont {Wang}, \citenamefont {Kazak}, \citenamefont {Senkalla}, \citenamefont {Siyushev}, \citenamefont {Abe}, \citenamefont {Taniguchi}, \citenamefont {Onoda}, \citenamefont {Kato}, \citenamefont {Makino}, \citenamefont {Hatano}, \citenamefont {Jelezko},\ and\ \citenamefont {Iwasaki}}]{Wang2024}%
  \BibitemOpen
  \bibfield  {author} {\bibinfo {author} {\bibfnamefont {P.}~\bibnamefont {Wang}}, \bibinfo {author} {\bibfnamefont {L.}~\bibnamefont {Kazak}}, \bibinfo {author} {\bibfnamefont {K.}~\bibnamefont {Senkalla}}, \bibinfo {author} {\bibfnamefont {P.}~\bibnamefont {Siyushev}}, \bibinfo {author} {\bibfnamefont {R.}~\bibnamefont {Abe}}, \bibinfo {author} {\bibfnamefont {T.}~\bibnamefont {Taniguchi}}, \bibinfo {author} {\bibfnamefont {S.}~\bibnamefont {Onoda}}, \bibinfo {author} {\bibfnamefont {H.}~\bibnamefont {Kato}}, \bibinfo {author} {\bibfnamefont {T.}~\bibnamefont {Makino}}, \bibinfo {author} {\bibfnamefont {M.}~\bibnamefont {Hatano}}, \bibinfo {author} {\bibfnamefont {F.}~\bibnamefont {Jelezko}},\ and\ \bibinfo {author} {\bibfnamefont {T.}~\bibnamefont {Iwasaki}},\ }\bibfield  {title} {\bibinfo {title} {Transform-limited photon emission from a lead-vacancy center in diamond above 10 k},\ }\href {https://doi.org/10.1103/PhysRevLett.132.073601} {\bibfield  {journal} {\bibinfo  {journal} {Phys. Rev. Lett.}\
  }\textbf {\bibinfo {volume} {132}},\ \bibinfo {pages} {073601} (\bibinfo {year} {2024})}\BibitemShut {NoStop}%
\bibitem [{\citenamefont {Bhaskar}\ \emph {et~al.}(2020)\citenamefont {Bhaskar}, \citenamefont {Riedinger}, \citenamefont {Machielse}, \citenamefont {Levonian}, \citenamefont {Nguyen}, \citenamefont {Knall}, \citenamefont {Park}, \citenamefont {Englund}, \citenamefont {Lon{\v{c}}ar}, \citenamefont {Sukachev},\ and\ \citenamefont {Lukin}}]{Bhaskar2020}%
  \BibitemOpen
  \bibfield  {author} {\bibinfo {author} {\bibfnamefont {M.~K.}\ \bibnamefont {Bhaskar}}, \bibinfo {author} {\bibfnamefont {R.}~\bibnamefont {Riedinger}}, \bibinfo {author} {\bibfnamefont {B.}~\bibnamefont {Machielse}}, \bibinfo {author} {\bibfnamefont {D.~S.}\ \bibnamefont {Levonian}}, \bibinfo {author} {\bibfnamefont {C.~T.}\ \bibnamefont {Nguyen}}, \bibinfo {author} {\bibfnamefont {E.~N.}\ \bibnamefont {Knall}}, \bibinfo {author} {\bibfnamefont {H.}~\bibnamefont {Park}}, \bibinfo {author} {\bibfnamefont {D.}~\bibnamefont {Englund}}, \bibinfo {author} {\bibfnamefont {M.}~\bibnamefont {Lon{\v{c}}ar}}, \bibinfo {author} {\bibfnamefont {D.~D.}\ \bibnamefont {Sukachev}},\ and\ \bibinfo {author} {\bibfnamefont {M.~D.}\ \bibnamefont {Lukin}},\ }\bibfield  {title} {\bibinfo {title} {Experimental demonstration of memory-enhanced quantum communication},\ }\href {https://doi.org/10.1038/s41586-020-2103-5} {\bibfield  {journal} {\bibinfo  {journal} {Nature}\ }\textbf {\bibinfo {volume} {580}},\ \bibinfo {pages} {60}
  (\bibinfo {year} {2020})}\BibitemShut {NoStop}%
\bibitem [{\citenamefont {Bersin}\ \emph {et~al.}(2024)\citenamefont {Bersin}, \citenamefont {Sutula}, \citenamefont {Huan}, \citenamefont {Suleymanzade}, \citenamefont {Assumpcao}, \citenamefont {Wei}, \citenamefont {Stas}, \citenamefont {Knaut}, \citenamefont {Knall}, \citenamefont {Langrock}, \citenamefont {Sinclair}, \citenamefont {Murphy}, \citenamefont {Riedinger}, \citenamefont {Yeh}, \citenamefont {Xin}, \citenamefont {Bandyopadhyay}, \citenamefont {Sukachev}, \citenamefont {Machielse}, \citenamefont {Levonian}, \citenamefont {Bhaskar}, \citenamefont {Hamilton}, \citenamefont {Park}, \citenamefont {Lon\ifmmode~\check{c}\else \v{c}\fi{}ar}, \citenamefont {Fejer}, \citenamefont {Dixon}, \citenamefont {Englund},\ and\ \citenamefont {Lukin}}]{Bersin2024}%
  \BibitemOpen
  \bibfield  {author} {\bibinfo {author} {\bibfnamefont {E.}~\bibnamefont {Bersin}}, \bibinfo {author} {\bibfnamefont {M.}~\bibnamefont {Sutula}}, \bibinfo {author} {\bibfnamefont {Y.~Q.}\ \bibnamefont {Huan}}, \bibinfo {author} {\bibfnamefont {A.}~\bibnamefont {Suleymanzade}}, \bibinfo {author} {\bibfnamefont {D.~R.}\ \bibnamefont {Assumpcao}}, \bibinfo {author} {\bibfnamefont {Y.-C.}\ \bibnamefont {Wei}}, \bibinfo {author} {\bibfnamefont {P.-J.}\ \bibnamefont {Stas}}, \bibinfo {author} {\bibfnamefont {C.~M.}\ \bibnamefont {Knaut}}, \bibinfo {author} {\bibfnamefont {E.~N.}\ \bibnamefont {Knall}}, \bibinfo {author} {\bibfnamefont {C.}~\bibnamefont {Langrock}}, \bibinfo {author} {\bibfnamefont {N.}~\bibnamefont {Sinclair}}, \bibinfo {author} {\bibfnamefont {R.}~\bibnamefont {Murphy}}, \bibinfo {author} {\bibfnamefont {R.}~\bibnamefont {Riedinger}}, \bibinfo {author} {\bibfnamefont {M.}~\bibnamefont {Yeh}}, \bibinfo {author} {\bibfnamefont {C.}~\bibnamefont {Xin}}, \bibinfo {author} {\bibfnamefont
  {S.}~\bibnamefont {Bandyopadhyay}}, \bibinfo {author} {\bibfnamefont {D.~D.}\ \bibnamefont {Sukachev}}, \bibinfo {author} {\bibfnamefont {B.}~\bibnamefont {Machielse}}, \bibinfo {author} {\bibfnamefont {D.~S.}\ \bibnamefont {Levonian}}, \bibinfo {author} {\bibfnamefont {M.~K.}\ \bibnamefont {Bhaskar}}, \bibinfo {author} {\bibfnamefont {S.}~\bibnamefont {Hamilton}}, \bibinfo {author} {\bibfnamefont {H.}~\bibnamefont {Park}}, \bibinfo {author} {\bibfnamefont {M.}~\bibnamefont {Lon\ifmmode~\check{c}\else \v{c}\fi{}ar}}, \bibinfo {author} {\bibfnamefont {M.~M.}\ \bibnamefont {Fejer}}, \bibinfo {author} {\bibfnamefont {P.~B.}\ \bibnamefont {Dixon}}, \bibinfo {author} {\bibfnamefont {D.~R.}\ \bibnamefont {Englund}},\ and\ \bibinfo {author} {\bibfnamefont {M.~D.}\ \bibnamefont {Lukin}},\ }\bibfield  {title} {\bibinfo {title} {Telecom networking with a diamond quantum memory},\ }\href {https://doi.org/10.1103/PRXQuantum.5.010303} {\bibfield  {journal} {\bibinfo  {journal} {PRX Quantum}\ }\textbf {\bibinfo {volume}
  {5}},\ \bibinfo {pages} {010303} (\bibinfo {year} {2024})}\BibitemShut {NoStop}%
\bibitem [{\citenamefont {Knaut}\ \emph {et~al.}(2024)\citenamefont {Knaut}, \citenamefont {Suleymanzade}, \citenamefont {Wei}, \citenamefont {Assumpcao}, \citenamefont {Stas}, \citenamefont {Huan}, \citenamefont {Machielse}, \citenamefont {Knall}, \citenamefont {Sutula}, \citenamefont {Baranes}, \citenamefont {Sinclair}, \citenamefont {De-Eknamkul}, \citenamefont {Levonian}, \citenamefont {Bhaskar}, \citenamefont {Park}, \citenamefont {Lon{\v c}ar},\ and\ \citenamefont {Lukin}}]{Knaut2024}%
  \BibitemOpen
  \bibfield  {author} {\bibinfo {author} {\bibfnamefont {C.~M.}\ \bibnamefont {Knaut}}, \bibinfo {author} {\bibfnamefont {A.}~\bibnamefont {Suleymanzade}}, \bibinfo {author} {\bibfnamefont {Y.-C.}\ \bibnamefont {Wei}}, \bibinfo {author} {\bibfnamefont {D.~R.}\ \bibnamefont {Assumpcao}}, \bibinfo {author} {\bibfnamefont {P.-J.}\ \bibnamefont {Stas}}, \bibinfo {author} {\bibfnamefont {Y.~Q.}\ \bibnamefont {Huan}}, \bibinfo {author} {\bibfnamefont {B.}~\bibnamefont {Machielse}}, \bibinfo {author} {\bibfnamefont {E.~N.}\ \bibnamefont {Knall}}, \bibinfo {author} {\bibfnamefont {M.}~\bibnamefont {Sutula}}, \bibinfo {author} {\bibfnamefont {G.}~\bibnamefont {Baranes}}, \bibinfo {author} {\bibfnamefont {N.}~\bibnamefont {Sinclair}}, \bibinfo {author} {\bibfnamefont {C.}~\bibnamefont {De-Eknamkul}}, \bibinfo {author} {\bibfnamefont {D.~S.}\ \bibnamefont {Levonian}}, \bibinfo {author} {\bibfnamefont {M.~K.}\ \bibnamefont {Bhaskar}}, \bibinfo {author} {\bibfnamefont {H.}~\bibnamefont {Park}}, \bibinfo {author}
  {\bibfnamefont {M.}~\bibnamefont {Lon{\v c}ar}},\ and\ \bibinfo {author} {\bibfnamefont {M.~D.}\ \bibnamefont {Lukin}},\ }\bibfield  {title} {\bibinfo {title} {Entanglement of nanophotonic quantum memory nodes in a telecom network},\ }\href@noop {} {\bibfield  {journal} {\bibinfo  {journal} {Nature}\ }\textbf {\bibinfo {volume} {629}},\ \bibinfo {pages} {573} (\bibinfo {year} {2024})}\BibitemShut {NoStop}%
\bibitem [{\citenamefont {Goss}\ \emph {et~al.}(1996)\citenamefont {Goss}, \citenamefont {Jones}, \citenamefont {Breuer}, \citenamefont {Briddon},\ and\ \citenamefont {\"Oberg}}]{Goss1996}%
  \BibitemOpen
  \bibfield  {author} {\bibinfo {author} {\bibfnamefont {J.~P.}\ \bibnamefont {Goss}}, \bibinfo {author} {\bibfnamefont {R.}~\bibnamefont {Jones}}, \bibinfo {author} {\bibfnamefont {S.~J.}\ \bibnamefont {Breuer}}, \bibinfo {author} {\bibfnamefont {P.~R.}\ \bibnamefont {Briddon}},\ and\ \bibinfo {author} {\bibfnamefont {S.}~\bibnamefont {\"Oberg}},\ }\bibfield  {title} {\bibinfo {title} {The twelve-line 1.682 ev luminescence center in diamond and the vacancy-silicon complex},\ }\href {https://doi.org/10.1103/PhysRevLett.77.3041} {\bibfield  {journal} {\bibinfo  {journal} {Phys. Rev. Lett.}\ }\textbf {\bibinfo {volume} {77}},\ \bibinfo {pages} {3041} (\bibinfo {year} {1996})}\BibitemShut {NoStop}%
\bibitem [{\citenamefont {Gali}\ and\ \citenamefont {Maze}(2013)}]{Gali2013}%
  \BibitemOpen
  \bibfield  {author} {\bibinfo {author} {\bibfnamefont {A.}~\bibnamefont {Gali}}\ and\ \bibinfo {author} {\bibfnamefont {J.~R.}\ \bibnamefont {Maze}},\ }\bibfield  {title} {\bibinfo {title} {Ab initio study of the split silicon-vacancy defect in diamond: Electronic structure and related properties},\ }\href {https://doi.org/10.1103/PhysRevB.88.235205} {\bibfield  {journal} {\bibinfo  {journal} {Phys. Rev. B}\ }\textbf {\bibinfo {volume} {88}},\ \bibinfo {pages} {235205} (\bibinfo {year} {2013})}\BibitemShut {NoStop}%
\bibitem [{\citenamefont {Thiering}\ and\ \citenamefont {Gali}(2018)}]{Thiering2018}%
  \BibitemOpen
  \bibfield  {author} {\bibinfo {author} {\bibfnamefont {G.}~\bibnamefont {Thiering}}\ and\ \bibinfo {author} {\bibfnamefont {A.}~\bibnamefont {Gali}},\ }\bibfield  {title} {\bibinfo {title} {Ab initio magneto-optical spectrum of group-iv vacancy color centers in diamond},\ }\href {https://doi.org/10.1103/PhysRevX.8.021063} {\bibfield  {journal} {\bibinfo  {journal} {Phys. Rev. X}\ }\textbf {\bibinfo {volume} {8}},\ \bibinfo {pages} {021063} (\bibinfo {year} {2018})}\BibitemShut {NoStop}%
\bibitem [{\citenamefont {Doherty}\ \emph {et~al.}(2013)\citenamefont {Doherty}, \citenamefont {Manson}, \citenamefont {Delaney}, \citenamefont {Jelezko}, \citenamefont {Wrachtrup},\ and\ \citenamefont {Hollenberg}}]{Doherty2013}%
  \BibitemOpen
  \bibfield  {author} {\bibinfo {author} {\bibfnamefont {M.~W.}\ \bibnamefont {Doherty}}, \bibinfo {author} {\bibfnamefont {N.~B.}\ \bibnamefont {Manson}}, \bibinfo {author} {\bibfnamefont {P.}~\bibnamefont {Delaney}}, \bibinfo {author} {\bibfnamefont {F.}~\bibnamefont {Jelezko}}, \bibinfo {author} {\bibfnamefont {J.}~\bibnamefont {Wrachtrup}},\ and\ \bibinfo {author} {\bibfnamefont {L.~C.}\ \bibnamefont {Hollenberg}},\ }\bibfield  {title} {\bibinfo {title} {The nitrogen-vacancy colour centre in diamond},\ }\href {https://doi.org/https://doi.org/10.1016/j.physrep.2013.02.001} {\bibfield  {journal} {\bibinfo  {journal} {Physics Reports}\ }\textbf {\bibinfo {volume} {528}},\ \bibinfo {pages} {1} (\bibinfo {year} {2013})}\BibitemShut {NoStop}%
\bibitem [{\citenamefont {Gali}(2019)}]{Gali2019}%
  \BibitemOpen
  \bibfield  {author} {\bibinfo {author} {\bibfnamefont {A.}~\bibnamefont {Gali}},\ }\bibfield  {title} {\bibinfo {title} {Ab initio theory of the nitrogen-vacancy center in diamond},\ }\href {https://doi.org/doi:10.1515/nanoph-2019-0154} {\bibfield  {journal} {\bibinfo  {journal} {Nanophotonics}\ }\textbf {\bibinfo {volume} {8}},\ \bibinfo {pages} {1907} (\bibinfo {year} {2019})}\BibitemShut {NoStop}%
\bibitem [{\citenamefont {Pingault}\ \emph {et~al.}(2014)\citenamefont {Pingault}, \citenamefont {Becker}, \citenamefont {Schulte}, \citenamefont {Arend}, \citenamefont {Hepp}, \citenamefont {Godde}, \citenamefont {Tartakovskii}, \citenamefont {Markham}, \citenamefont {Becher},\ and\ \citenamefont {Atat\"ure}}]{Pingault2014}%
  \BibitemOpen
  \bibfield  {author} {\bibinfo {author} {\bibfnamefont {B.}~\bibnamefont {Pingault}}, \bibinfo {author} {\bibfnamefont {J.~N.}\ \bibnamefont {Becker}}, \bibinfo {author} {\bibfnamefont {C.~H.~H.}\ \bibnamefont {Schulte}}, \bibinfo {author} {\bibfnamefont {C.}~\bibnamefont {Arend}}, \bibinfo {author} {\bibfnamefont {C.}~\bibnamefont {Hepp}}, \bibinfo {author} {\bibfnamefont {T.}~\bibnamefont {Godde}}, \bibinfo {author} {\bibfnamefont {A.~I.}\ \bibnamefont {Tartakovskii}}, \bibinfo {author} {\bibfnamefont {M.}~\bibnamefont {Markham}}, \bibinfo {author} {\bibfnamefont {C.}~\bibnamefont {Becher}},\ and\ \bibinfo {author} {\bibfnamefont {M.}~\bibnamefont {Atat\"ure}},\ }\bibfield  {title} {\bibinfo {title} {All-optical formation of coherent dark states of silicon-vacancy spins in diamond},\ }\href {https://doi.org/10.1103/PhysRevLett.113.263601} {\bibfield  {journal} {\bibinfo  {journal} {Phys. Rev. Lett.}\ }\textbf {\bibinfo {volume} {113}},\ \bibinfo {pages} {263601} (\bibinfo {year} {2014})}\BibitemShut
  {NoStop}%
\bibitem [{\citenamefont {Becker}\ \emph {et~al.}(2016)\citenamefont {Becker}, \citenamefont {G{\"o}rlitz}, \citenamefont {Arend}, \citenamefont {Markham},\ and\ \citenamefont {Becher}}]{Becker2016}%
  \BibitemOpen
  \bibfield  {author} {\bibinfo {author} {\bibfnamefont {J.~N.}\ \bibnamefont {Becker}}, \bibinfo {author} {\bibfnamefont {J.}~\bibnamefont {G{\"o}rlitz}}, \bibinfo {author} {\bibfnamefont {C.}~\bibnamefont {Arend}}, \bibinfo {author} {\bibfnamefont {M.}~\bibnamefont {Markham}},\ and\ \bibinfo {author} {\bibfnamefont {C.}~\bibnamefont {Becher}},\ }\bibfield  {title} {\bibinfo {title} {Ultrafast all-optical coherent control of single silicon vacancy colour centres in diamond},\ }\href {https://doi.org/10.1038/ncomms13512} {\bibfield  {journal} {\bibinfo  {journal} {Nature Communications}\ }\textbf {\bibinfo {volume} {7}},\ \bibinfo {pages} {13512} (\bibinfo {year} {2016})}\BibitemShut {NoStop}%
\bibitem [{\citenamefont {Siyushev}\ \emph {et~al.}(2017)\citenamefont {Siyushev}, \citenamefont {Metsch}, \citenamefont {Ijaz}, \citenamefont {Binder}, \citenamefont {Bhaskar}, \citenamefont {Sukachev}, \citenamefont {Sipahigil}, \citenamefont {Evans}, \citenamefont {Nguyen}, \citenamefont {Lukin}, \citenamefont {Hemmer}, \citenamefont {Palyanov}, \citenamefont {Kupriyanov}, \citenamefont {Borzdov}, \citenamefont {Rogers},\ and\ \citenamefont {Jelezko}}]{Siyushev2017}%
  \BibitemOpen
  \bibfield  {author} {\bibinfo {author} {\bibfnamefont {P.}~\bibnamefont {Siyushev}}, \bibinfo {author} {\bibfnamefont {M.~H.}\ \bibnamefont {Metsch}}, \bibinfo {author} {\bibfnamefont {A.}~\bibnamefont {Ijaz}}, \bibinfo {author} {\bibfnamefont {J.~M.}\ \bibnamefont {Binder}}, \bibinfo {author} {\bibfnamefont {M.~K.}\ \bibnamefont {Bhaskar}}, \bibinfo {author} {\bibfnamefont {D.~D.}\ \bibnamefont {Sukachev}}, \bibinfo {author} {\bibfnamefont {A.}~\bibnamefont {Sipahigil}}, \bibinfo {author} {\bibfnamefont {R.~E.}\ \bibnamefont {Evans}}, \bibinfo {author} {\bibfnamefont {C.~T.}\ \bibnamefont {Nguyen}}, \bibinfo {author} {\bibfnamefont {M.~D.}\ \bibnamefont {Lukin}}, \bibinfo {author} {\bibfnamefont {P.~R.}\ \bibnamefont {Hemmer}}, \bibinfo {author} {\bibfnamefont {Y.~N.}\ \bibnamefont {Palyanov}}, \bibinfo {author} {\bibfnamefont {I.~N.}\ \bibnamefont {Kupriyanov}}, \bibinfo {author} {\bibfnamefont {Y.~M.}\ \bibnamefont {Borzdov}}, \bibinfo {author} {\bibfnamefont {L.~J.}\ \bibnamefont {Rogers}},\ and\
  \bibinfo {author} {\bibfnamefont {F.}~\bibnamefont {Jelezko}},\ }\bibfield  {title} {\bibinfo {title} {Optical and microwave control of germanium-vacancy center spins in diamond},\ }\href {https://doi.org/10.1103/PhysRevB.96.081201} {\bibfield  {journal} {\bibinfo  {journal} {Phys. Rev. B}\ }\textbf {\bibinfo {volume} {96}},\ \bibinfo {pages} {081201} (\bibinfo {year} {2017})}\BibitemShut {NoStop}%
\bibitem [{\citenamefont {Becker}\ \emph {et~al.}(2018)\citenamefont {Becker}, \citenamefont {Pingault}, \citenamefont {Gro\ss{}}, \citenamefont {G\"undo\ifmmode~\breve{g}\else \u{g}\fi{}an}, \citenamefont {Kukharchyk}, \citenamefont {Markham}, \citenamefont {Edmonds}, \citenamefont {Atat\"ure}, \citenamefont {Bushev},\ and\ \citenamefont {Becher}}]{Becker2018}%
  \BibitemOpen
  \bibfield  {author} {\bibinfo {author} {\bibfnamefont {J.~N.}\ \bibnamefont {Becker}}, \bibinfo {author} {\bibfnamefont {B.}~\bibnamefont {Pingault}}, \bibinfo {author} {\bibfnamefont {D.}~\bibnamefont {Gro\ss{}}}, \bibinfo {author} {\bibfnamefont {M.}~\bibnamefont {G\"undo\ifmmode~\breve{g}\else \u{g}\fi{}an}}, \bibinfo {author} {\bibfnamefont {N.}~\bibnamefont {Kukharchyk}}, \bibinfo {author} {\bibfnamefont {M.}~\bibnamefont {Markham}}, \bibinfo {author} {\bibfnamefont {A.}~\bibnamefont {Edmonds}}, \bibinfo {author} {\bibfnamefont {M.}~\bibnamefont {Atat\"ure}}, \bibinfo {author} {\bibfnamefont {P.}~\bibnamefont {Bushev}},\ and\ \bibinfo {author} {\bibfnamefont {C.}~\bibnamefont {Becher}},\ }\bibfield  {title} {\bibinfo {title} {All-optical control of the silicon-vacancy spin in diamond at millikelvin temperatures},\ }\href {https://doi.org/10.1103/PhysRevLett.120.053603} {\bibfield  {journal} {\bibinfo  {journal} {Phys. Rev. Lett.}\ }\textbf {\bibinfo {volume} {120}},\ \bibinfo {pages} {053603} (\bibinfo
  {year} {2018})}\BibitemShut {NoStop}%
\bibitem [{\citenamefont {Weinzetl}\ \emph {et~al.}(2019)\citenamefont {Weinzetl}, \citenamefont {G\"orlitz}, \citenamefont {Becker}, \citenamefont {Walmsley}, \citenamefont {Poem}, \citenamefont {Nunn},\ and\ \citenamefont {Becher}}]{Weinzetl2019}%
  \BibitemOpen
  \bibfield  {author} {\bibinfo {author} {\bibfnamefont {C.}~\bibnamefont {Weinzetl}}, \bibinfo {author} {\bibfnamefont {J.}~\bibnamefont {G\"orlitz}}, \bibinfo {author} {\bibfnamefont {J.~N.}\ \bibnamefont {Becker}}, \bibinfo {author} {\bibfnamefont {I.~A.}\ \bibnamefont {Walmsley}}, \bibinfo {author} {\bibfnamefont {E.}~\bibnamefont {Poem}}, \bibinfo {author} {\bibfnamefont {J.}~\bibnamefont {Nunn}},\ and\ \bibinfo {author} {\bibfnamefont {C.}~\bibnamefont {Becher}},\ }\bibfield  {title} {\bibinfo {title} {Coherent control and wave mixing in an ensemble of silicon-vacancy centers in diamond},\ }\href {https://doi.org/10.1103/PhysRevLett.122.063601} {\bibfield  {journal} {\bibinfo  {journal} {Phys. Rev. Lett.}\ }\textbf {\bibinfo {volume} {122}},\ \bibinfo {pages} {063601} (\bibinfo {year} {2019})}\BibitemShut {NoStop}%
\bibitem [{\citenamefont {Debroux}\ \emph {et~al.}(2021)\citenamefont {Debroux}, \citenamefont {Michaels}, \citenamefont {Purser}, \citenamefont {Wan}, \citenamefont {Trusheim}, \citenamefont {Arjona~Mart\'{\i}nez}, \citenamefont {Parker}, \citenamefont {Stramma}, \citenamefont {Chen}, \citenamefont {de~Santis}, \citenamefont {Alexeev}, \citenamefont {Ferrari}, \citenamefont {Englund}, \citenamefont {Gangloff},\ and\ \citenamefont {Atat\"ure}}]{Debroux2021}%
  \BibitemOpen
  \bibfield  {author} {\bibinfo {author} {\bibfnamefont {R.}~\bibnamefont {Debroux}}, \bibinfo {author} {\bibfnamefont {C.~P.}\ \bibnamefont {Michaels}}, \bibinfo {author} {\bibfnamefont {C.~M.}\ \bibnamefont {Purser}}, \bibinfo {author} {\bibfnamefont {N.}~\bibnamefont {Wan}}, \bibinfo {author} {\bibfnamefont {M.~E.}\ \bibnamefont {Trusheim}}, \bibinfo {author} {\bibfnamefont {J.}~\bibnamefont {Arjona~Mart\'{\i}nez}}, \bibinfo {author} {\bibfnamefont {R.~A.}\ \bibnamefont {Parker}}, \bibinfo {author} {\bibfnamefont {A.~M.}\ \bibnamefont {Stramma}}, \bibinfo {author} {\bibfnamefont {K.~C.}\ \bibnamefont {Chen}}, \bibinfo {author} {\bibfnamefont {L.}~\bibnamefont {de~Santis}}, \bibinfo {author} {\bibfnamefont {E.~M.}\ \bibnamefont {Alexeev}}, \bibinfo {author} {\bibfnamefont {A.~C.}\ \bibnamefont {Ferrari}}, \bibinfo {author} {\bibfnamefont {D.}~\bibnamefont {Englund}}, \bibinfo {author} {\bibfnamefont {D.~A.}\ \bibnamefont {Gangloff}},\ and\ \bibinfo {author} {\bibfnamefont {M.}~\bibnamefont {Atat\"ure}},\
  }\bibfield  {title} {\bibinfo {title} {Quantum control of the tin-vacancy spin qubit in diamond},\ }\href {https://doi.org/10.1103/PhysRevX.11.041041} {\bibfield  {journal} {\bibinfo  {journal} {Phys. Rev. X}\ }\textbf {\bibinfo {volume} {11}},\ \bibinfo {pages} {041041} (\bibinfo {year} {2021})}\BibitemShut {NoStop}%
\bibitem [{\citenamefont {Harris}\ \emph {et~al.}(2023)\citenamefont {Harris}, \citenamefont {Michaels}, \citenamefont {Chen}, \citenamefont {Parker}, \citenamefont {Titze}, \citenamefont {Arjona~Mart\'{\i}nez}, \citenamefont {Sutula}, \citenamefont {Christen}, \citenamefont {Stramma}, \citenamefont {Roth}, \citenamefont {Purser}, \citenamefont {Appel}, \citenamefont {Li}, \citenamefont {Trusheim}, \citenamefont {Palmer}, \citenamefont {Markham}, \citenamefont {Bielejec}, \citenamefont {Atat\"ure},\ and\ \citenamefont {Englund}}]{Harris2023}%
  \BibitemOpen
  \bibfield  {author} {\bibinfo {author} {\bibfnamefont {I.~B.}\ \bibnamefont {Harris}}, \bibinfo {author} {\bibfnamefont {C.~P.}\ \bibnamefont {Michaels}}, \bibinfo {author} {\bibfnamefont {K.~C.}\ \bibnamefont {Chen}}, \bibinfo {author} {\bibfnamefont {R.~A.}\ \bibnamefont {Parker}}, \bibinfo {author} {\bibfnamefont {M.}~\bibnamefont {Titze}}, \bibinfo {author} {\bibfnamefont {J.}~\bibnamefont {Arjona~Mart\'{\i}nez}}, \bibinfo {author} {\bibfnamefont {M.}~\bibnamefont {Sutula}}, \bibinfo {author} {\bibfnamefont {I.~R.}\ \bibnamefont {Christen}}, \bibinfo {author} {\bibfnamefont {A.~M.}\ \bibnamefont {Stramma}}, \bibinfo {author} {\bibfnamefont {W.}~\bibnamefont {Roth}}, \bibinfo {author} {\bibfnamefont {C.~M.}\ \bibnamefont {Purser}}, \bibinfo {author} {\bibfnamefont {M.~H.}\ \bibnamefont {Appel}}, \bibinfo {author} {\bibfnamefont {C.}~\bibnamefont {Li}}, \bibinfo {author} {\bibfnamefont {M.~E.}\ \bibnamefont {Trusheim}}, \bibinfo {author} {\bibfnamefont {N.~L.}\ \bibnamefont {Palmer}}, \bibinfo {author}
  {\bibfnamefont {M.~L.}\ \bibnamefont {Markham}}, \bibinfo {author} {\bibfnamefont {E.~S.}\ \bibnamefont {Bielejec}}, \bibinfo {author} {\bibfnamefont {M.}~\bibnamefont {Atat\"ure}},\ and\ \bibinfo {author} {\bibfnamefont {D.}~\bibnamefont {Englund}},\ }\bibfield  {title} {\bibinfo {title} {Hyperfine spectroscopy of isotopically engineered group-iv color centers in diamond},\ }\href {https://doi.org/10.1103/PRXQuantum.4.040301} {\bibfield  {journal} {\bibinfo  {journal} {PRX Quantum}\ }\textbf {\bibinfo {volume} {4}},\ \bibinfo {pages} {040301} (\bibinfo {year} {2023})}\BibitemShut {NoStop}%
\bibitem [{\citenamefont {Altmann}\ and\ \citenamefont {Herzig}(1994)}]{altmann1994}%
  \BibitemOpen
  \bibfield  {author} {\bibinfo {author} {\bibfnamefont {S.~L.}\ \bibnamefont {Altmann}}\ and\ \bibinfo {author} {\bibfnamefont {P.}~\bibnamefont {Herzig}},\ }\href@noop {} {\emph {\bibinfo {title} {Point-Group Theory Tables}}}\ (\bibinfo  {publisher} {Clarendon Press ; Oxford University Press},\ \bibinfo {address} {Oxford : New York},\ \bibinfo {year} {1994})\BibitemShut {NoStop}%
\bibitem [{\citenamefont {Kohn}\ and\ \citenamefont {Sham}(1965)}]{PhysRev.140.A1133}%
  \BibitemOpen
  \bibfield  {author} {\bibinfo {author} {\bibfnamefont {W.}~\bibnamefont {Kohn}}\ and\ \bibinfo {author} {\bibfnamefont {L.~J.}\ \bibnamefont {Sham}},\ }\bibfield  {title} {\bibinfo {title} {Self-consistent equations including exchange and correlation effects},\ }\href {https://doi.org/10.1103/PhysRev.140.A1133} {\bibfield  {journal} {\bibinfo  {journal} {Phys. Rev.}\ }\textbf {\bibinfo {volume} {140}},\ \bibinfo {pages} {A1133} (\bibinfo {year} {1965})}\BibitemShut {NoStop}%
\bibitem [{\citenamefont {Hohenberg}\ and\ \citenamefont {Kohn}(1964)}]{PhysRev.136.B864}%
  \BibitemOpen
  \bibfield  {author} {\bibinfo {author} {\bibfnamefont {P.}~\bibnamefont {Hohenberg}}\ and\ \bibinfo {author} {\bibfnamefont {W.}~\bibnamefont {Kohn}},\ }\bibfield  {title} {\bibinfo {title} {Inhomogeneous electron gas},\ }\href {https://doi.org/10.1103/PhysRev.136.B864} {\bibfield  {journal} {\bibinfo  {journal} {Phys. Rev.}\ }\textbf {\bibinfo {volume} {136}},\ \bibinfo {pages} {B864} (\bibinfo {year} {1964})}\BibitemShut {NoStop}%
\bibitem [{\citenamefont {Bl\"ochl}(1994)}]{PhysRevB.50.17953}%
  \BibitemOpen
  \bibfield  {author} {\bibinfo {author} {\bibfnamefont {P.~E.}\ \bibnamefont {Bl\"ochl}},\ }\bibfield  {title} {\bibinfo {title} {Projector augmented-wave method},\ }\href {https://doi.org/10.1103/PhysRevB.50.17953} {\bibfield  {journal} {\bibinfo  {journal} {Phys. Rev. B}\ }\textbf {\bibinfo {volume} {50}},\ \bibinfo {pages} {17953} (\bibinfo {year} {1994})}\BibitemShut {NoStop}%
\bibitem [{\citenamefont {Kresse}\ and\ \citenamefont {Joubert}(1999)}]{PhysRevB.59.1758}%
  \BibitemOpen
  \bibfield  {author} {\bibinfo {author} {\bibfnamefont {G.}~\bibnamefont {Kresse}}\ and\ \bibinfo {author} {\bibfnamefont {D.}~\bibnamefont {Joubert}},\ }\bibfield  {title} {\bibinfo {title} {From ultrasoft pseudopotentials to the projector augmented-wave method},\ }\href {https://doi.org/10.1103/PhysRevB.59.1758} {\bibfield  {journal} {\bibinfo  {journal} {Phys. Rev. B}\ }\textbf {\bibinfo {volume} {59}},\ \bibinfo {pages} {1758} (\bibinfo {year} {1999})}\BibitemShut {NoStop}%
\bibitem [{\citenamefont {Kresse}\ and\ \citenamefont {Furthm\"uller}(1996)}]{PhysRevB.54.11169}%
  \BibitemOpen
  \bibfield  {author} {\bibinfo {author} {\bibfnamefont {G.}~\bibnamefont {Kresse}}\ and\ \bibinfo {author} {\bibfnamefont {J.}~\bibnamefont {Furthm\"uller}},\ }\bibfield  {title} {\bibinfo {title} {Efficient iterative schemes for ab initio total-energy calculations using a plane-wave basis set},\ }\href {https://doi.org/10.1103/PhysRevB.54.11169} {\bibfield  {journal} {\bibinfo  {journal} {Phys. Rev. B}\ }\textbf {\bibinfo {volume} {54}},\ \bibinfo {pages} {11169} (\bibinfo {year} {1996})}\BibitemShut {NoStop}%
\bibitem [{\citenamefont {Kresse}\ and\ \citenamefont {Furthm{\"u}ller}(1996)}]{Kresse1996}%
  \BibitemOpen
  \bibfield  {author} {\bibinfo {author} {\bibfnamefont {G.}~\bibnamefont {Kresse}}\ and\ \bibinfo {author} {\bibfnamefont {J.}~\bibnamefont {Furthm{\"u}ller}},\ }\bibfield  {title} {\bibinfo {title} {Efficiency of ab-initio total energy calculations for metals and semiconductors using a plane-wave basis set},\ }\href {http://www.sciencedirect.com/science/article/pii/0927025696000080} {\bibfield  {journal} {\bibinfo  {journal} {Computational Materials Science}\ }\textbf {\bibinfo {volume} {6}},\ \bibinfo {pages} {15} (\bibinfo {year} {1996})}\BibitemShut {NoStop}%
\bibitem [{\citenamefont {Sun}\ \emph {et~al.}(2015)\citenamefont {Sun}, \citenamefont {Ruzsinszky},\ and\ \citenamefont {Perdew}}]{Sun2015}%
  \BibitemOpen
  \bibfield  {author} {\bibinfo {author} {\bibfnamefont {J.}~\bibnamefont {Sun}}, \bibinfo {author} {\bibfnamefont {A.}~\bibnamefont {Ruzsinszky}},\ and\ \bibinfo {author} {\bibfnamefont {J.~P.}\ \bibnamefont {Perdew}},\ }\bibfield  {title} {\bibinfo {title} {Strongly constrained and appropriately normed semilocal density functional},\ }\href {https://doi.org/10.1103/PhysRevLett.115.036402} {\bibfield  {journal} {\bibinfo  {journal} {Phys. Rev. Lett.}\ }\textbf {\bibinfo {volume} {115}},\ \bibinfo {pages} {036402} (\bibinfo {year} {2015})}\BibitemShut {NoStop}%
\bibitem [{\citenamefont {Isaacs}\ and\ \citenamefont {Wolverton}(2018)}]{Isaacs2018SCAN}%
  \BibitemOpen
  \bibfield  {author} {\bibinfo {author} {\bibfnamefont {E.~B.}\ \bibnamefont {Isaacs}}\ and\ \bibinfo {author} {\bibfnamefont {C.}~\bibnamefont {Wolverton}},\ }\bibfield  {title} {\bibinfo {title} {Performance of the strongly constrained and appropriately normed density functional for solid-state materials},\ }\href {https://doi.org/10.1103/PhysRevMaterials.2.063801} {\bibfield  {journal} {\bibinfo  {journal} {Phys. Rev. Mater.}\ }\textbf {\bibinfo {volume} {2}},\ \bibinfo {pages} {063801} (\bibinfo {year} {2018})}\BibitemShut {NoStop}%
\bibitem [{\citenamefont {Mejia-Rodriguez}\ and\ \citenamefont {Trickey}(2018)}]{Rodriguez2018SCAN}%
  \BibitemOpen
  \bibfield  {author} {\bibinfo {author} {\bibfnamefont {D.}~\bibnamefont {Mejia-Rodriguez}}\ and\ \bibinfo {author} {\bibfnamefont {S.~B.}\ \bibnamefont {Trickey}},\ }\bibfield  {title} {\bibinfo {title} {Deorbitalized meta-gga exchange-correlation functionals in solids},\ }\href {https://doi.org/10.1103/PhysRevB.98.115161} {\bibfield  {journal} {\bibinfo  {journal} {Phys. Rev. B}\ }\textbf {\bibinfo {volume} {98}},\ \bibinfo {pages} {115161} (\bibinfo {year} {2018})}\BibitemShut {NoStop}%
\bibitem [{\citenamefont {Maciaszek}\ \emph {et~al.}(2023)\citenamefont {Maciaszek}, \citenamefont {\v{Z}alandauskas}, \citenamefont {Silkinis}, \citenamefont {Alkauskas},\ and\ \citenamefont {Razinkovas}}]{scandiamond}%
  \BibitemOpen
  \bibfield  {author} {\bibinfo {author} {\bibfnamefont {M.}~\bibnamefont {Maciaszek}}, \bibinfo {author} {\bibfnamefont {V.}~\bibnamefont {\v{Z}alandauskas}}, \bibinfo {author} {\bibfnamefont {R.}~\bibnamefont {Silkinis}}, \bibinfo {author} {\bibfnamefont {A.}~\bibnamefont {Alkauskas}},\ and\ \bibinfo {author} {\bibfnamefont {L.}~\bibnamefont {Razinkovas}},\ }\bibfield  {title} {\bibinfo {title} {The application of the {SCAN} density functional to color centers in diamond},\ }\href {https://doi.org/10.1063/5.0154319} {\bibfield  {journal} {\bibinfo  {journal} {J. Chem. Phys.}\ }\textbf {\bibinfo {volume} {159}},\ \bibinfo {pages} {084708} (\bibinfo {year} {2023})}\BibitemShut {NoStop}%
\bibitem [{\citenamefont {Heyd}\ \emph {et~al.}(2003)\citenamefont {Heyd}, \citenamefont {Scuseria},\ and\ \citenamefont {Ernzerhof}}]{Heyd2023}%
  \BibitemOpen
  \bibfield  {author} {\bibinfo {author} {\bibfnamefont {J.}~\bibnamefont {Heyd}}, \bibinfo {author} {\bibfnamefont {G.~E.}\ \bibnamefont {Scuseria}},\ and\ \bibinfo {author} {\bibfnamefont {M.}~\bibnamefont {Ernzerhof}},\ }\bibfield  {title} {\bibinfo {title} {Hybrid functionals based on a screened coulomb potential},\ }\href {https://doi.org/10.1063/1.1564060} {\bibfield  {journal} {\bibinfo  {journal} {The Journal of Chemical Physics}\ }\textbf {\bibinfo {volume} {118}},\ \bibinfo {pages} {8207} (\bibinfo {year} {2003})}\BibitemShut {NoStop}%
\bibitem [{\citenamefont {Gali}(2009)}]{Gali2009}%
  \BibitemOpen
  \bibfield  {author} {\bibinfo {author} {\bibfnamefont {A.}~\bibnamefont {Gali}},\ }\bibfield  {title} {\bibinfo {title} {Theory of the neutral nitrogen-vacancy center in diamond and its application to the realization of a qubit},\ }\href {https://doi.org/10.1103/PhysRevB.79.235210} {\bibfield  {journal} {\bibinfo  {journal} {Phys. Rev. B}\ }\textbf {\bibinfo {volume} {79}},\ \bibinfo {pages} {235210} (\bibinfo {year} {2009})}\BibitemShut {NoStop}%
\bibitem [{\citenamefont {Condon}\ and\ \citenamefont {Shortley}(1935)}]{CondonShortley1935}%
  \BibitemOpen
  \bibfield  {author} {\bibinfo {author} {\bibfnamefont {E.~U.}\ \bibnamefont {Condon}}\ and\ \bibinfo {author} {\bibfnamefont {G.~H.}\ \bibnamefont {Shortley}},\ }\href@noop {} {\emph {\bibinfo {title} {The theory of atomic spectra}}}\ (\bibinfo  {publisher} {Cambridge University Press},\ \bibinfo {address} {Cambridge, UK ;},\ \bibinfo {year} {1935})\BibitemShut {NoStop}%
\bibitem [{Hep()}]{Hepp2014_supp}%
  \BibitemOpen
  \href@noop {} {}\bibinfo {note} {See Supplemental Material of Ref.~\cite{Hepp2014}}\BibitemShut {NoStop}%
\bibitem [{\citenamefont {Häußler}\ \emph {et~al.}(2017)\citenamefont {Häußler}, \citenamefont {Thiering}, \citenamefont {Dietrich}, \citenamefont {Waasem}, \citenamefont {Teraji}, \citenamefont {Isoya}, \citenamefont {Iwasaki}, \citenamefont {Hatano}, \citenamefont {Jelezko}, \citenamefont {Gali},\ and\ \citenamefont {Kubanek}}]{Haubler_2017}%
  \BibitemOpen
  \bibfield  {author} {\bibinfo {author} {\bibfnamefont {S.}~\bibnamefont {Häußler}}, \bibinfo {author} {\bibfnamefont {G.}~\bibnamefont {Thiering}}, \bibinfo {author} {\bibfnamefont {A.}~\bibnamefont {Dietrich}}, \bibinfo {author} {\bibfnamefont {N.}~\bibnamefont {Waasem}}, \bibinfo {author} {\bibfnamefont {T.}~\bibnamefont {Teraji}}, \bibinfo {author} {\bibfnamefont {J.}~\bibnamefont {Isoya}}, \bibinfo {author} {\bibfnamefont {T.}~\bibnamefont {Iwasaki}}, \bibinfo {author} {\bibfnamefont {M.}~\bibnamefont {Hatano}}, \bibinfo {author} {\bibfnamefont {F.}~\bibnamefont {Jelezko}}, \bibinfo {author} {\bibfnamefont {A.}~\bibnamefont {Gali}},\ and\ \bibinfo {author} {\bibfnamefont {A.}~\bibnamefont {Kubanek}},\ }\bibfield  {title} {\bibinfo {title} {Photoluminescence excitation spectroscopy of siv−and gev−color center in diamond},\ }\href {https://doi.org/10.1088/1367-2630/aa73e5} {\bibfield  {journal} {\bibinfo  {journal} {New Journal of Physics}\ }\textbf {\bibinfo {volume} {19}},\ \bibinfo {pages}
  {063036} (\bibinfo {year} {2017})}\BibitemShut {NoStop}%
\bibitem [{\citenamefont {Ham}(1965)}]{Ham1965}%
  \BibitemOpen
  \bibfield  {author} {\bibinfo {author} {\bibfnamefont {F.~S.}\ \bibnamefont {Ham}},\ }\bibfield  {title} {\bibinfo {title} {Dynamical jahn-teller effect in paramagnetic resonance spectra: Orbital reduction factors and partial quenching of spin-orbit interaction},\ }\href {https://doi.org/10.1103/PhysRev.138.A1727} {\bibfield  {journal} {\bibinfo  {journal} {Phys. Rev.}\ }\textbf {\bibinfo {volume} {138}},\ \bibinfo {pages} {A1727} (\bibinfo {year} {1965})}\BibitemShut {NoStop}%
\bibitem [{\citenamefont {Bl\"ochl}(2000)}]{Blochl2000}%
  \BibitemOpen
  \bibfield  {author} {\bibinfo {author} {\bibfnamefont {P.~E.}\ \bibnamefont {Bl\"ochl}},\ }\bibfield  {title} {\bibinfo {title} {First-principles calculations of defects in oxygen-deficient silica exposed to hydrogen},\ }\href {https://doi.org/10.1103/PhysRevB.62.6158} {\bibfield  {journal} {\bibinfo  {journal} {Phys. Rev. B}\ }\textbf {\bibinfo {volume} {62}},\ \bibinfo {pages} {6158} (\bibinfo {year} {2000})}\BibitemShut {NoStop}%
\bibitem [{\citenamefont {Sz\'asz}\ \emph {et~al.}(2013)\citenamefont {Sz\'asz}, \citenamefont {Hornos}, \citenamefont {Marsman},\ and\ \citenamefont {Gali}}]{Szasz2013}%
  \BibitemOpen
  \bibfield  {author} {\bibinfo {author} {\bibfnamefont {K.}~\bibnamefont {Sz\'asz}}, \bibinfo {author} {\bibfnamefont {T.}~\bibnamefont {Hornos}}, \bibinfo {author} {\bibfnamefont {M.}~\bibnamefont {Marsman}},\ and\ \bibinfo {author} {\bibfnamefont {A.}~\bibnamefont {Gali}},\ }\bibfield  {title} {\bibinfo {title} {Hyperfine coupling of point defects in semiconductors by hybrid density functional calculations: The role of core spin polarization},\ }\href {https://doi.org/10.1103/PhysRevB.88.075202} {\bibfield  {journal} {\bibinfo  {journal} {Phys. Rev. B}\ }\textbf {\bibinfo {volume} {88}},\ \bibinfo {pages} {075202} (\bibinfo {year} {2013})}\BibitemShut {NoStop}%
\bibitem [{Thi()}]{Thiering2024_supp}%
  \BibitemOpen
  \href@noop {} {}\bibinfo {note} {See Supplemental Material of Ref.~\cite{Thiering2024}}\BibitemShut {NoStop}%
\bibitem [{\citenamefont {Ham}(1968)}]{ham1968}%
  \BibitemOpen
  \bibfield  {author} {\bibinfo {author} {\bibfnamefont {F.~S.}\ \bibnamefont {Ham}},\ }\bibfield  {title} {\bibinfo {title} {Effect of {{Linear Jahn-Teller Coupling}} on {{Paramagnetic Resonance}} in a {{E}} 2 {{State}}},\ }\href {https://doi.org/10.1103/PhysRev.166.307} {\bibfield  {journal} {\bibinfo  {journal} {Physical Review}\ }\textbf {\bibinfo {volume} {166}},\ \bibinfo {pages} {307} (\bibinfo {year} {1968})}\BibitemShut {NoStop}%
\bibitem [{\citenamefont {Bersuker}(2006)}]{Bersuker2006}%
  \BibitemOpen
  \bibfield  {author} {\bibinfo {author} {\bibfnamefont {I.~B.}\ \bibnamefont {Bersuker}},\ }\href@noop {} {\emph {\bibinfo {title} {The Jahn-Teller effect}}}\ (\bibinfo  {publisher} {Cambridge University Press},\ \bibinfo {address} {Cambridge, UK ;},\ \bibinfo {year} {2006})\BibitemShut {NoStop}%
\bibitem [{\citenamefont {Bersuker}\ and\ \citenamefont {Polinger}(2012)}]{Bersuker2012}%
  \BibitemOpen
  \bibfield  {author} {\bibinfo {author} {\bibfnamefont {I.}~\bibnamefont {Bersuker}}\ and\ \bibinfo {author} {\bibfnamefont {V.}~\bibnamefont {Polinger}},\ }\href {https://books.google.hu/books?id=F6_oCAAAQBAJ} {\emph {\bibinfo {title} {Vibronic Interactions in Molecules and Crystals}}},\ Springer Series in Chemical Physics\ (\bibinfo  {publisher} {Springer Berlin Heidelberg},\ \bibinfo {year} {2012})\BibitemShut {NoStop}%
\bibitem [{\citenamefont {Steiner}\ \emph {et~al.}(2016)\citenamefont {Steiner}, \citenamefont {Khmelevskyi}, \citenamefont {Marsmann},\ and\ \citenamefont {Kresse}}]{Steiner2016}%
  \BibitemOpen
  \bibfield  {author} {\bibinfo {author} {\bibfnamefont {S.}~\bibnamefont {Steiner}}, \bibinfo {author} {\bibfnamefont {S.}~\bibnamefont {Khmelevskyi}}, \bibinfo {author} {\bibfnamefont {M.}~\bibnamefont {Marsmann}},\ and\ \bibinfo {author} {\bibfnamefont {G.}~\bibnamefont {Kresse}},\ }\bibfield  {title} {\bibinfo {title} {Calculation of the magnetic anisotropy with projected-augmented-wave methodology and the case study of disordered ${\mathrm{fe}}_{1\ensuremath{-}x}{\mathrm{co}}_{x}$ alloys},\ }\href {https://doi.org/10.1103/PhysRevB.93.224425} {\bibfield  {journal} {\bibinfo  {journal} {Phys. Rev. B}\ }\textbf {\bibinfo {volume} {93}},\ \bibinfo {pages} {224425} (\bibinfo {year} {2016})}\BibitemShut {NoStop}%
\bibitem [{\citenamefont {Bruyndonckx}\ \emph {et~al.}(1997)\citenamefont {Bruyndonckx}, \citenamefont {Daul}, \citenamefont {Manoharan},\ and\ \citenamefont {Deiss}}]{Bruyndonckx1997}%
  \BibitemOpen
  \bibfield  {author} {\bibinfo {author} {\bibfnamefont {R.}~\bibnamefont {Bruyndonckx}}, \bibinfo {author} {\bibfnamefont {C.}~\bibnamefont {Daul}}, \bibinfo {author} {\bibfnamefont {P.~T.}\ \bibnamefont {Manoharan}},\ and\ \bibinfo {author} {\bibfnamefont {E.}~\bibnamefont {Deiss}},\ }\bibfield  {title} {\bibinfo {title} {A nonempirical approach to ground-state jahn−teller distortion: Case study of vcl4},\ }\href {https://doi.org/10.1021/ic961220+} {\bibfield  {journal} {\bibinfo  {journal} {Inorganic Chemistry}\ }\textbf {\bibinfo {volume} {36}},\ \bibinfo {pages} {4251} (\bibinfo {year} {1997})}\BibitemShut {NoStop}%
\bibitem [{\citenamefont {De\'ak}\ \emph {et~al.}(2010)\citenamefont {De\'ak}, \citenamefont {Aradi}, \citenamefont {Frauenheim}, \citenamefont {Janz\'en},\ and\ \citenamefont {Gali}}]{Deak2010}%
  \BibitemOpen
  \bibfield  {author} {\bibinfo {author} {\bibfnamefont {P.}~\bibnamefont {De\'ak}}, \bibinfo {author} {\bibfnamefont {B.}~\bibnamefont {Aradi}}, \bibinfo {author} {\bibfnamefont {T.}~\bibnamefont {Frauenheim}}, \bibinfo {author} {\bibfnamefont {E.}~\bibnamefont {Janz\'en}},\ and\ \bibinfo {author} {\bibfnamefont {A.}~\bibnamefont {Gali}},\ }\bibfield  {title} {\bibinfo {title} {Accurate defect levels obtained from the hse06 range-separated hybrid functional},\ }\href {https://doi.org/10.1103/PhysRevB.81.153203} {\bibfield  {journal} {\bibinfo  {journal} {Phys. Rev. B}\ }\textbf {\bibinfo {volume} {81}},\ \bibinfo {pages} {153203} (\bibinfo {year} {2010})}\BibitemShut {NoStop}%
\bibitem [{\citenamefont {Freysoldt}\ \emph {et~al.}(2014)\citenamefont {Freysoldt}, \citenamefont {Grabowski}, \citenamefont {Hickel}, \citenamefont {Neugebauer}, \citenamefont {Kresse}, \citenamefont {Janotti},\ and\ \citenamefont {Van~de Walle}}]{Freysoldt2014}%
  \BibitemOpen
  \bibfield  {author} {\bibinfo {author} {\bibfnamefont {C.}~\bibnamefont {Freysoldt}}, \bibinfo {author} {\bibfnamefont {B.}~\bibnamefont {Grabowski}}, \bibinfo {author} {\bibfnamefont {T.}~\bibnamefont {Hickel}}, \bibinfo {author} {\bibfnamefont {J.}~\bibnamefont {Neugebauer}}, \bibinfo {author} {\bibfnamefont {G.}~\bibnamefont {Kresse}}, \bibinfo {author} {\bibfnamefont {A.}~\bibnamefont {Janotti}},\ and\ \bibinfo {author} {\bibfnamefont {C.~G.}\ \bibnamefont {Van~de Walle}},\ }\bibfield  {title} {\bibinfo {title} {First-principles calculations for point defects in solids},\ }\href {https://doi.org/10.1103/RevModPhys.86.253} {\bibfield  {journal} {\bibinfo  {journal} {Rev. Mod. Phys.}\ }\textbf {\bibinfo {volume} {86}},\ \bibinfo {pages} {253} (\bibinfo {year} {2014})}\BibitemShut {NoStop}%
\bibitem [{\citenamefont {Cs\'or\'e}\ \emph {et~al.}(2022)\citenamefont {Cs\'or\'e}, \citenamefont {Ivanov}, \citenamefont {Son},\ and\ \citenamefont {Gali}}]{csore2022}%
  \BibitemOpen
  \bibfield  {author} {\bibinfo {author} {\bibfnamefont {A.}~\bibnamefont {Cs\'or\'e}}, \bibinfo {author} {\bibfnamefont {I.~G.}\ \bibnamefont {Ivanov}}, \bibinfo {author} {\bibfnamefont {N.~T.}\ \bibnamefont {Son}},\ and\ \bibinfo {author} {\bibfnamefont {A.}~\bibnamefont {Gali}},\ }\bibfield  {title} {\bibinfo {title} {Fluorescence spectrum and charge state control of divacancy qubits via illumination at elevated temperatures in $4h$ silicon carbide},\ }\href {https://doi.org/10.1103/PhysRevB.105.165108} {\bibfield  {journal} {\bibinfo  {journal} {Phys. Rev. B}\ }\textbf {\bibinfo {volume} {105}},\ \bibinfo {pages} {165108} (\bibinfo {year} {2022})}\BibitemShut {NoStop}%
\bibitem [{\citenamefont {Thiering}\ and\ \citenamefont {Gali}(2017)}]{Thiering2017}%
  \BibitemOpen
  \bibfield  {author} {\bibinfo {author} {\bibfnamefont {G.}~\bibnamefont {Thiering}}\ and\ \bibinfo {author} {\bibfnamefont {A.}~\bibnamefont {Gali}},\ }\bibfield  {title} {\bibinfo {title} {Ab initio calculation of spin-orbit coupling for an nv center in diamond exhibiting dynamic jahn-teller effect},\ }\href {https://doi.org/10.1103/PhysRevB.96.081115} {\bibfield  {journal} {\bibinfo  {journal} {Phys. Rev. B}\ }\textbf {\bibinfo {volume} {96}},\ \bibinfo {pages} {081115} (\bibinfo {year} {2017})}\BibitemShut {NoStop}%
\bibitem [{\citenamefont {Green}\ \emph {et~al.}(2017)\citenamefont {Green}, \citenamefont {Mottishaw}, \citenamefont {Breeze}, \citenamefont {Edmonds}, \citenamefont {D'Haenens-Johansson}, \citenamefont {Doherty}, \citenamefont {Williams}, \citenamefont {Twitchen},\ and\ \citenamefont {Newton}}]{Green2017}%
  \BibitemOpen
  \bibfield  {author} {\bibinfo {author} {\bibfnamefont {B.~L.}\ \bibnamefont {Green}}, \bibinfo {author} {\bibfnamefont {S.}~\bibnamefont {Mottishaw}}, \bibinfo {author} {\bibfnamefont {B.~G.}\ \bibnamefont {Breeze}}, \bibinfo {author} {\bibfnamefont {A.~M.}\ \bibnamefont {Edmonds}}, \bibinfo {author} {\bibfnamefont {U.~F.~S.}\ \bibnamefont {D'Haenens-Johansson}}, \bibinfo {author} {\bibfnamefont {M.~W.}\ \bibnamefont {Doherty}}, \bibinfo {author} {\bibfnamefont {S.~D.}\ \bibnamefont {Williams}}, \bibinfo {author} {\bibfnamefont {D.~J.}\ \bibnamefont {Twitchen}},\ and\ \bibinfo {author} {\bibfnamefont {M.~E.}\ \bibnamefont {Newton}},\ }\bibfield  {title} {\bibinfo {title} {Neutral silicon-vacancy center in diamond: Spin polarization and lifetimes},\ }\href {https://doi.org/10.1103/PhysRevLett.119.096402} {\bibfield  {journal} {\bibinfo  {journal} {Phys. Rev. Lett.}\ }\textbf {\bibinfo {volume} {119}},\ \bibinfo {pages} {096402} (\bibinfo {year} {2017})}\BibitemShut {NoStop}%
\bibitem [{\citenamefont {Metsch}\ \emph {et~al.}(2019)\citenamefont {Metsch}, \citenamefont {Senkalla}, \citenamefont {Tratzmiller}, \citenamefont {Scheuer}, \citenamefont {Kern}, \citenamefont {Achard}, \citenamefont {Tallaire}, \citenamefont {Plenio}, \citenamefont {Siyushev},\ and\ \citenamefont {Jelezko}}]{Metsch2019}%
  \BibitemOpen
  \bibfield  {author} {\bibinfo {author} {\bibfnamefont {M.~H.}\ \bibnamefont {Metsch}}, \bibinfo {author} {\bibfnamefont {K.}~\bibnamefont {Senkalla}}, \bibinfo {author} {\bibfnamefont {B.}~\bibnamefont {Tratzmiller}}, \bibinfo {author} {\bibfnamefont {J.}~\bibnamefont {Scheuer}}, \bibinfo {author} {\bibfnamefont {M.}~\bibnamefont {Kern}}, \bibinfo {author} {\bibfnamefont {J.}~\bibnamefont {Achard}}, \bibinfo {author} {\bibfnamefont {A.}~\bibnamefont {Tallaire}}, \bibinfo {author} {\bibfnamefont {M.~B.}\ \bibnamefont {Plenio}}, \bibinfo {author} {\bibfnamefont {P.}~\bibnamefont {Siyushev}},\ and\ \bibinfo {author} {\bibfnamefont {F.}~\bibnamefont {Jelezko}},\ }\bibfield  {title} {\bibinfo {title} {Initialization and readout of nuclear spins via a negatively charged silicon-vacancy center in diamond},\ }\href {https://doi.org/10.1103/PhysRevLett.122.190503} {\bibfield  {journal} {\bibinfo  {journal} {Phys. Rev. Lett.}\ }\textbf {\bibinfo {volume} {122}},\ \bibinfo {pages} {190503} (\bibinfo {year}
  {2019})}\BibitemShut {NoStop}%
\bibitem [{\citenamefont {Monge}\ \emph {et~al.}(2023)\citenamefont {Monge}, \citenamefont {Delord}, \citenamefont {Thiering}, \citenamefont {Gali},\ and\ \citenamefont {Meriles}}]{Monge2023}%
  \BibitemOpen
  \bibfield  {author} {\bibinfo {author} {\bibfnamefont {R.}~\bibnamefont {Monge}}, \bibinfo {author} {\bibfnamefont {T.}~\bibnamefont {Delord}}, \bibinfo {author} {\bibfnamefont {G.}~\bibnamefont {Thiering}}, \bibinfo {author} {\bibfnamefont {A.}~\bibnamefont {Gali}},\ and\ \bibinfo {author} {\bibfnamefont {C.~A.}\ \bibnamefont {Meriles}},\ }\bibfield  {title} {\bibinfo {title} {Resonant versus nonresonant spin readout of a nitrogen-vacancy center in diamond under cryogenic conditions},\ }\href {https://doi.org/10.1103/PhysRevLett.131.236901} {\bibfield  {journal} {\bibinfo  {journal} {Phys. Rev. Lett.}\ }\textbf {\bibinfo {volume} {131}},\ \bibinfo {pages} {236901} (\bibinfo {year} {2023})}\BibitemShut {NoStop}%
\bibitem [{\citenamefont {Thiering}\ and\ \citenamefont {Gali}(2025)}]{Thiering2024}%
  \BibitemOpen
  \bibfield  {author} {\bibinfo {author} {\bibfnamefont {G.}~\bibnamefont {Thiering}}\ and\ \bibinfo {author} {\bibfnamefont {A.}~\bibnamefont {Gali}},\ }\bibfield  {title} {\bibinfo {title} {Nuclear spin relaxation in solid-state-defect quantum bits via electron-phonon coupling in the optical excited state},\ }\href {https://doi.org/10.1103/1q3f-7zvl} {\bibfield  {journal} {\bibinfo  {journal} {Phys. Rev. Appl.}\ ,\ } (\bibinfo {year} {2025})}\BibitemShut {NoStop}%
\bibitem [{\citenamefont {Kepesidis}\ \emph {et~al.}(2016)\citenamefont {Kepesidis}, \citenamefont {Lemonde}, \citenamefont {Norambuena}, \citenamefont {Maze},\ and\ \citenamefont {Rabl}}]{Norambuena2016}%
  \BibitemOpen
  \bibfield  {author} {\bibinfo {author} {\bibfnamefont {K.~V.}\ \bibnamefont {Kepesidis}}, \bibinfo {author} {\bibfnamefont {M.-A.}\ \bibnamefont {Lemonde}}, \bibinfo {author} {\bibfnamefont {A.}~\bibnamefont {Norambuena}}, \bibinfo {author} {\bibfnamefont {J.~R.}\ \bibnamefont {Maze}},\ and\ \bibinfo {author} {\bibfnamefont {P.}~\bibnamefont {Rabl}},\ }\bibfield  {title} {\bibinfo {title} {Cooling phonons with phonons: Acoustic reservoir engineering with silicon-vacancy centers in diamond},\ }\href {https://doi.org/10.1103/PhysRevB.94.214115} {\bibfield  {journal} {\bibinfo  {journal} {Phys. Rev. B}\ }\textbf {\bibinfo {volume} {94}},\ \bibinfo {pages} {214115} (\bibinfo {year} {2016})}\BibitemShut {NoStop}%
\bibitem [{\citenamefont {Rogers}\ \emph {et~al.}(2014{\natexlab{b}})\citenamefont {Rogers}, \citenamefont {Jahnke}, \citenamefont {Metsch}, \citenamefont {Sipahigil}, \citenamefont {Binder}, \citenamefont {Teraji}, \citenamefont {Sumiya}, \citenamefont {Isoya}, \citenamefont {Lukin}, \citenamefont {Hemmer},\ and\ \citenamefont {Jelezko}}]{Rogers2014b}%
  \BibitemOpen
  \bibfield  {author} {\bibinfo {author} {\bibfnamefont {L.~J.}\ \bibnamefont {Rogers}}, \bibinfo {author} {\bibfnamefont {K.~D.}\ \bibnamefont {Jahnke}}, \bibinfo {author} {\bibfnamefont {M.~H.}\ \bibnamefont {Metsch}}, \bibinfo {author} {\bibfnamefont {A.}~\bibnamefont {Sipahigil}}, \bibinfo {author} {\bibfnamefont {J.~M.}\ \bibnamefont {Binder}}, \bibinfo {author} {\bibfnamefont {T.}~\bibnamefont {Teraji}}, \bibinfo {author} {\bibfnamefont {H.}~\bibnamefont {Sumiya}}, \bibinfo {author} {\bibfnamefont {J.}~\bibnamefont {Isoya}}, \bibinfo {author} {\bibfnamefont {M.~D.}\ \bibnamefont {Lukin}}, \bibinfo {author} {\bibfnamefont {P.}~\bibnamefont {Hemmer}},\ and\ \bibinfo {author} {\bibfnamefont {F.}~\bibnamefont {Jelezko}},\ }\bibfield  {title} {\bibinfo {title} {All-optical initialization, readout, and coherent preparation of single silicon-vacancy spins in diamond},\ }\href {https://doi.org/10.1103/PhysRevLett.113.263602} {\bibfield  {journal} {\bibinfo  {journal} {Phys. Rev. Lett.}\ }\textbf {\bibinfo
  {volume} {113}},\ \bibinfo {pages} {263602} (\bibinfo {year} {2014}{\natexlab{b}})}\BibitemShut {NoStop}%
\bibitem [{\citenamefont {Haouas}\ \emph {et~al.}(2016)\citenamefont {Haouas}, \citenamefont {Taulelle},\ and\ \citenamefont {Martineau}}]{HAOUAS2016}%
  \BibitemOpen
  \bibfield  {author} {\bibinfo {author} {\bibfnamefont {M.}~\bibnamefont {Haouas}}, \bibinfo {author} {\bibfnamefont {F.}~\bibnamefont {Taulelle}},\ and\ \bibinfo {author} {\bibfnamefont {C.}~\bibnamefont {Martineau}},\ }\bibfield  {title} {\bibinfo {title} {Recent advances in application of 27al nmr spectroscopy to materials science},\ }\href {https://doi.org/https://doi.org/10.1016/j.pnmrs.2016.01.003} {\bibfield  {journal} {\bibinfo  {journal} {Progress in Nuclear Magnetic Resonance Spectroscopy}\ }\textbf {\bibinfo {volume} {94-95}},\ \bibinfo {pages} {11} (\bibinfo {year} {2016})}\BibitemShut {NoStop}%
\bibitem [{\citenamefont {Pyykk\"o}(2008)}]{Pekka2008}%
  \BibitemOpen
  \bibfield  {author} {\bibinfo {author} {\bibfnamefont {P.}~\bibnamefont {Pyykk\"o}},\ }\bibfield  {title} {\bibinfo {title} {Year-2008 nuclear quadrupole moments},\ }\href {https://doi.org/10.1080/00268970802018367} {\bibfield  {journal} {\bibinfo  {journal} {Molecular Physics}\ }\textbf {\bibinfo {volume} {106}},\ \bibinfo {pages} {1965} (\bibinfo {year} {2008})},\ \Eprint {https://arxiv.org/abs/https://doi.org/10.1080/00268970802018367} {https://doi.org/10.1080/00268970802018367} \BibitemShut {NoStop}%
\bibitem [{\citenamefont {Thiering}\ and\ \citenamefont {Gali}(2020)}]{Thiering2020}%
  \BibitemOpen
  \bibfield  {author} {\bibinfo {author} {\bibfnamefont {G.~m.~H.}\ \bibnamefont {Thiering}}\ and\ \bibinfo {author} {\bibfnamefont {A.}~\bibnamefont {Gali}},\ }\bibfield  {title} {\bibinfo {title} {Erratum: Ab initio magneto-optical spectrum of group-iv vacancy color centers in diamond [phys. rev. x 8, 021063 (2018)]},\ }\href {https://doi.org/10.1103/PhysRevX.10.039901} {\bibfield  {journal} {\bibinfo  {journal} {Phys. Rev. X}\ }\textbf {\bibinfo {volume} {10}},\ \bibinfo {pages} {039901} (\bibinfo {year} {2020})}\BibitemShut {NoStop}%
\bibitem [{\citenamefont {Stevens}(1953)}]{Stevens542}%
  \BibitemOpen
  \bibfield  {author} {\bibinfo {author} {\bibfnamefont {K.~W.~H.}\ \bibnamefont {Stevens}},\ }\bibfield  {title} {\bibinfo {title} {On the magnetic properties of covalent xy 6 complexes},\ }\href {https://doi.org/10.1098/rspa.1953.0166} {\bibfield  {journal} {\bibinfo  {journal} {Proceedings of the Royal Society of London A: Mathematical, Physical and Engineering Sciences}\ }\textbf {\bibinfo {volume} {219}},\ \bibinfo {pages} {542} (\bibinfo {year} {1953})}\BibitemShut {NoStop}%
\bibitem [{\citenamefont {Rugar}\ \emph {et~al.}(2019)\citenamefont {Rugar}, \citenamefont {Dory}, \citenamefont {Sun},\ and\ \citenamefont {Vu\ifmmode \check{c}\else \v{c}\fi{}kovi\ifmmode~\acute{c}\else \'{c}\fi{}}}]{Rugar2019}%
  \BibitemOpen
  \bibfield  {author} {\bibinfo {author} {\bibfnamefont {A.~E.}\ \bibnamefont {Rugar}}, \bibinfo {author} {\bibfnamefont {C.}~\bibnamefont {Dory}}, \bibinfo {author} {\bibfnamefont {S.}~\bibnamefont {Sun}},\ and\ \bibinfo {author} {\bibfnamefont {J.}~\bibnamefont {Vu\ifmmode \check{c}\else \v{c}\fi{}kovi\ifmmode~\acute{c}\else \'{c}\fi{}}},\ }\bibfield  {title} {\bibinfo {title} {Characterization of optical and spin properties of single tin-vacancy centers in diamond nanopillars},\ }\href {https://doi.org/10.1103/PhysRevB.99.205417} {\bibfield  {journal} {\bibinfo  {journal} {Phys. Rev. B}\ }\textbf {\bibinfo {volume} {99}},\ \bibinfo {pages} {205417} (\bibinfo {year} {2019})}\BibitemShut {NoStop}%
\bibitem [{\citenamefont {Jahnke}\ \emph {et~al.}(2015)\citenamefont {Jahnke}, \citenamefont {Sipahigil}, \citenamefont {Binder}, \citenamefont {Doherty}, \citenamefont {Metsch}, \citenamefont {Rogers}, \citenamefont {Manson}, \citenamefont {Lukin},\ and\ \citenamefont {Jelezko}}]{Jahnke2015}%
  \BibitemOpen
  \bibfield  {author} {\bibinfo {author} {\bibfnamefont {K.~D.}\ \bibnamefont {Jahnke}}, \bibinfo {author} {\bibfnamefont {A.}~\bibnamefont {Sipahigil}}, \bibinfo {author} {\bibfnamefont {J.~M.}\ \bibnamefont {Binder}}, \bibinfo {author} {\bibfnamefont {M.~W.}\ \bibnamefont {Doherty}}, \bibinfo {author} {\bibfnamefont {M.}~\bibnamefont {Metsch}}, \bibinfo {author} {\bibfnamefont {L.~J.}\ \bibnamefont {Rogers}}, \bibinfo {author} {\bibfnamefont {N.~B.}\ \bibnamefont {Manson}}, \bibinfo {author} {\bibfnamefont {M.~D.}\ \bibnamefont {Lukin}},\ and\ \bibinfo {author} {\bibfnamefont {F.}~\bibnamefont {Jelezko}},\ }\bibfield  {title} {\bibinfo {title} {Electron–phonon processes of the silicon-vacancy centre in diamond},\ }\href {https://doi.org/10.1088/1367-2630/17/4/043011} {\bibfield  {journal} {\bibinfo  {journal} {New Journal of Physics}\ }\textbf {\bibinfo {volume} {17}},\ \bibinfo {pages} {043011} (\bibinfo {year} {2015})}\BibitemShut {NoStop}%
\bibitem [{\citenamefont {Thiering}\ and\ \citenamefont {Gali}(2021)}]{Thiering2021}%
  \BibitemOpen
  \bibfield  {author} {\bibinfo {author} {\bibfnamefont {G.}~\bibnamefont {Thiering}}\ and\ \bibinfo {author} {\bibfnamefont {A.}~\bibnamefont {Gali}},\ }\bibfield  {title} {\bibinfo {title} {Magneto-optical spectra of the split nickel-vacancy defect in diamond},\ }\href {https://doi.org/10.1103/PhysRevResearch.3.043052} {\bibfield  {journal} {\bibinfo  {journal} {Phys. Rev. Res.}\ }\textbf {\bibinfo {volume} {3}},\ \bibinfo {pages} {043052} (\bibinfo {year} {2021})}\BibitemShut {NoStop}%
\bibitem [{\citenamefont {Pingault}\ \emph {et~al.}(2017)\citenamefont {Pingault}, \citenamefont {Jarausch}, \citenamefont {Hepp}, \citenamefont {Klintberg}, \citenamefont {Becker}, \citenamefont {Markham}, \citenamefont {Becher},\ and\ \citenamefont {Atat{\"u}re}}]{Pingault2017}%
  \BibitemOpen
  \bibfield  {author} {\bibinfo {author} {\bibfnamefont {B.}~\bibnamefont {Pingault}}, \bibinfo {author} {\bibfnamefont {D.-D.}\ \bibnamefont {Jarausch}}, \bibinfo {author} {\bibfnamefont {C.}~\bibnamefont {Hepp}}, \bibinfo {author} {\bibfnamefont {L.}~\bibnamefont {Klintberg}}, \bibinfo {author} {\bibfnamefont {J.~N.}\ \bibnamefont {Becker}}, \bibinfo {author} {\bibfnamefont {M.}~\bibnamefont {Markham}}, \bibinfo {author} {\bibfnamefont {C.}~\bibnamefont {Becher}},\ and\ \bibinfo {author} {\bibfnamefont {M.}~\bibnamefont {Atat{\"u}re}},\ }\bibfield  {title} {\bibinfo {title} {Coherent control of the silicon-vacancy spin in diamond},\ }\href {https://doi.org/10.1038/ncomms15579} {\bibfield  {journal} {\bibinfo  {journal} {Nature Communications}\ }\textbf {\bibinfo {volume} {8}},\ \bibinfo {pages} {15579} (\bibinfo {year} {2017})}\BibitemShut {NoStop}%
\bibitem [{\citenamefont {Sukachev}\ \emph {et~al.}(2017)\citenamefont {Sukachev}, \citenamefont {Sipahigil}, \citenamefont {Nguyen}, \citenamefont {Bhaskar}, \citenamefont {Evans}, \citenamefont {Jelezko},\ and\ \citenamefont {Lukin}}]{Sukachev2017}%
  \BibitemOpen
  \bibfield  {author} {\bibinfo {author} {\bibfnamefont {D.~D.}\ \bibnamefont {Sukachev}}, \bibinfo {author} {\bibfnamefont {A.}~\bibnamefont {Sipahigil}}, \bibinfo {author} {\bibfnamefont {C.~T.}\ \bibnamefont {Nguyen}}, \bibinfo {author} {\bibfnamefont {M.~K.}\ \bibnamefont {Bhaskar}}, \bibinfo {author} {\bibfnamefont {R.~E.}\ \bibnamefont {Evans}}, \bibinfo {author} {\bibfnamefont {F.}~\bibnamefont {Jelezko}},\ and\ \bibinfo {author} {\bibfnamefont {M.~D.}\ \bibnamefont {Lukin}},\ }\bibfield  {title} {\bibinfo {title} {Silicon-vacancy spin qubit in diamond: A quantum memory exceeding 10 ms with single-shot state readout},\ }\href {https://doi.org/10.1103/PhysRevLett.119.223602} {\bibfield  {journal} {\bibinfo  {journal} {Phys. Rev. Lett.}\ }\textbf {\bibinfo {volume} {119}},\ \bibinfo {pages} {223602} (\bibinfo {year} {2017})}\BibitemShut {NoStop}%
\bibitem [{\citenamefont {Senkalla}\ \emph {et~al.}(2024)\citenamefont {Senkalla}, \citenamefont {Genov}, \citenamefont {Metsch}, \citenamefont {Siyushev},\ and\ \citenamefont {Jelezko}}]{Senkalla2024}%
  \BibitemOpen
  \bibfield  {author} {\bibinfo {author} {\bibfnamefont {K.}~\bibnamefont {Senkalla}}, \bibinfo {author} {\bibfnamefont {G.}~\bibnamefont {Genov}}, \bibinfo {author} {\bibfnamefont {M.~H.}\ \bibnamefont {Metsch}}, \bibinfo {author} {\bibfnamefont {P.}~\bibnamefont {Siyushev}},\ and\ \bibinfo {author} {\bibfnamefont {F.}~\bibnamefont {Jelezko}},\ }\bibfield  {title} {\bibinfo {title} {Germanium vacancy in diamond quantum memory exceeding 20 ms},\ }\href {https://doi.org/10.1103/PhysRevLett.132.026901} {\bibfield  {journal} {\bibinfo  {journal} {Phys. Rev. Lett.}\ }\textbf {\bibinfo {volume} {132}},\ \bibinfo {pages} {026901} (\bibinfo {year} {2024})}\BibitemShut {NoStop}%
\bibitem [{\citenamefont {Doherty}\ \emph {et~al.}(2014)\citenamefont {Doherty}, \citenamefont {Struzhkin}, \citenamefont {Simpson}, \citenamefont {McGuinness}, \citenamefont {Meng}, \citenamefont {Stacey}, \citenamefont {Karle}, \citenamefont {Hemley}, \citenamefont {Manson}, \citenamefont {Hollenberg},\ and\ \citenamefont {Prawer}}]{Doherty2014}%
  \BibitemOpen
  \bibfield  {author} {\bibinfo {author} {\bibfnamefont {M.~W.}\ \bibnamefont {Doherty}}, \bibinfo {author} {\bibfnamefont {V.~V.}\ \bibnamefont {Struzhkin}}, \bibinfo {author} {\bibfnamefont {D.~A.}\ \bibnamefont {Simpson}}, \bibinfo {author} {\bibfnamefont {L.~P.}\ \bibnamefont {McGuinness}}, \bibinfo {author} {\bibfnamefont {Y.}~\bibnamefont {Meng}}, \bibinfo {author} {\bibfnamefont {A.}~\bibnamefont {Stacey}}, \bibinfo {author} {\bibfnamefont {T.~J.}\ \bibnamefont {Karle}}, \bibinfo {author} {\bibfnamefont {R.~J.}\ \bibnamefont {Hemley}}, \bibinfo {author} {\bibfnamefont {N.~B.}\ \bibnamefont {Manson}}, \bibinfo {author} {\bibfnamefont {L.~C.~L.}\ \bibnamefont {Hollenberg}},\ and\ \bibinfo {author} {\bibfnamefont {S.}~\bibnamefont {Prawer}},\ }\bibfield  {title} {\bibinfo {title} {Electronic properties and metrology applications of the diamond ${\mathrm{nv}}^{\ensuremath{-}}$ center under pressure},\ }\href {https://doi.org/10.1103/PhysRevLett.112.047601} {\bibfield  {journal} {\bibinfo  {journal} {Phys.
  Rev. Lett.}\ }\textbf {\bibinfo {volume} {112}},\ \bibinfo {pages} {047601} (\bibinfo {year} {2014})}\BibitemShut {NoStop}%
\bibitem [{\citenamefont {Hilberer}\ \emph {et~al.}(2023)\citenamefont {Hilberer}, \citenamefont {Toraille}, \citenamefont {Dailledouze}, \citenamefont {Adam}, \citenamefont {Hanlon}, \citenamefont {Weck}, \citenamefont {Schmidt}, \citenamefont {Loubeyre},\ and\ \citenamefont {Roch}}]{Hilberer2023}%
  \BibitemOpen
  \bibfield  {author} {\bibinfo {author} {\bibfnamefont {A.}~\bibnamefont {Hilberer}}, \bibinfo {author} {\bibfnamefont {L.}~\bibnamefont {Toraille}}, \bibinfo {author} {\bibfnamefont {C.}~\bibnamefont {Dailledouze}}, \bibinfo {author} {\bibfnamefont {M.-P.}\ \bibnamefont {Adam}}, \bibinfo {author} {\bibfnamefont {L.}~\bibnamefont {Hanlon}}, \bibinfo {author} {\bibfnamefont {G.}~\bibnamefont {Weck}}, \bibinfo {author} {\bibfnamefont {M.}~\bibnamefont {Schmidt}}, \bibinfo {author} {\bibfnamefont {P.}~\bibnamefont {Loubeyre}},\ and\ \bibinfo {author} {\bibfnamefont {J.-F. m.~c.}\ \bibnamefont {Roch}},\ }\bibfield  {title} {\bibinfo {title} {Enabling quantum sensing under extreme pressure: Nitrogen-vacancy magnetometry up to 130 gpa},\ }\href {https://doi.org/10.1103/PhysRevB.107.L220102} {\bibfield  {journal} {\bibinfo  {journal} {Phys. Rev. B}\ }\textbf {\bibinfo {volume} {107}},\ \bibinfo {pages} {L220102} (\bibinfo {year} {2023})}\BibitemShut {NoStop}%
\bibitem [{\citenamefont {Ho}\ \emph {et~al.}(2021)\citenamefont {Ho}, \citenamefont {Wong}, \citenamefont {Leung}, \citenamefont {Pang}, \citenamefont {Leung}, \citenamefont {Yip}, \citenamefont {Zhang}, \citenamefont {Xie}, \citenamefont {Goh},\ and\ \citenamefont {Yang}}]{wong2021kin}%
  \BibitemOpen
  \bibfield  {author} {\bibinfo {author} {\bibfnamefont {K.~O.}\ \bibnamefont {Ho}}, \bibinfo {author} {\bibfnamefont {K.~C.}\ \bibnamefont {Wong}}, \bibinfo {author} {\bibfnamefont {M.~Y.}\ \bibnamefont {Leung}}, \bibinfo {author} {\bibfnamefont {Y.~Y.}\ \bibnamefont {Pang}}, \bibinfo {author} {\bibfnamefont {W.~K.}\ \bibnamefont {Leung}}, \bibinfo {author} {\bibfnamefont {K.~Y.}\ \bibnamefont {Yip}}, \bibinfo {author} {\bibfnamefont {W.}~\bibnamefont {Zhang}}, \bibinfo {author} {\bibfnamefont {J.}~\bibnamefont {Xie}}, \bibinfo {author} {\bibfnamefont {S.~K.}\ \bibnamefont {Goh}},\ and\ \bibinfo {author} {\bibfnamefont {S.}~\bibnamefont {Yang}},\ }\bibfield  {title} {\bibinfo {title} {Recent developments of quantum sensing under pressurized environment using the nitrogen vacancy (nv) center in diamond},\ }\href {https://doi.org/10.1063/5.0052233} {\bibfield  {journal} {\bibinfo  {journal} {Journal of Applied Physics}\ }\textbf {\bibinfo {volume} {129}},\ \bibinfo {pages} {241101} (\bibinfo {year} {2021})},\
  \Eprint {https://arxiv.org/abs/https://pubs.aip.org/aip/jap/article-pdf/doi/10.1063/5.0052233/19891745/241101\_1\_5.0052233.pdf} {https://pubs.aip.org/aip/jap/article-pdf/doi/10.1063/5.0052233/19891745/241101\_1\_5.0052233.pdf} \BibitemShut {NoStop}%
\bibitem [{Gal(2025)}]{Gali2025Data}%
  \BibitemOpen
  \href@noop {} {\bibinfo {title} {{Data supporting “Additional data for the effective spin–orbit splitting of G4V centers in diamond”}}},\ \bibinfo {howpublished} {Available from the corresponding author upon reasonable request. Correspondence: \texttt{gali.adam@wigner.hun-ren.hu}} (\bibinfo {year} {2025})\BibitemShut {NoStop}%
\bibitem [{\citenamefont {Freysoldt}\ \emph {et~al.}(2009)\citenamefont {Freysoldt}, \citenamefont {Neugebauer},\ and\ \citenamefont {Van~de Walle}}]{Freysoldt2009}%
  \BibitemOpen
  \bibfield  {author} {\bibinfo {author} {\bibfnamefont {C.}~\bibnamefont {Freysoldt}}, \bibinfo {author} {\bibfnamefont {J.}~\bibnamefont {Neugebauer}},\ and\ \bibinfo {author} {\bibfnamefont {C.~G.}\ \bibnamefont {Van~de Walle}},\ }\bibfield  {title} {\bibinfo {title} {Fully ab initio finite-size corrections for charged-defect supercell calculations},\ }\href {https://doi.org/10.1103/PhysRevLett.102.016402} {\bibfield  {journal} {\bibinfo  {journal} {Phys. Rev. Lett.}\ }\textbf {\bibinfo {volume} {102}},\ \bibinfo {pages} {016402} (\bibinfo {year} {2009})}\BibitemShut {NoStop}%
\bibitem [{\citenamefont {Freysoldt}\ \emph {et~al.}(2011)\citenamefont {Freysoldt}, \citenamefont {Neugebauer},\ and\ \citenamefont {Van~de Walle}}]{Freysoldt2011}%
  \BibitemOpen
  \bibfield  {author} {\bibinfo {author} {\bibfnamefont {C.}~\bibnamefont {Freysoldt}}, \bibinfo {author} {\bibfnamefont {J.}~\bibnamefont {Neugebauer}},\ and\ \bibinfo {author} {\bibfnamefont {C.~G.}\ \bibnamefont {Van~de Walle}},\ }\bibfield  {title} {\bibinfo {title} {Electrostatic interactions between charged defects in supercells},\ }\href {https://doi.org/https://doi.org/10.1002/pssb.201046289} {\bibfield  {journal} {\bibinfo  {journal} {physica status solidi (b)}\ }\textbf {\bibinfo {volume} {248}},\ \bibinfo {pages} {1067} (\bibinfo {year} {2011})},\ \Eprint {https://arxiv.org/abs/https://onlinelibrary.wiley.com/doi/pdf/10.1002/pssb.201046289} {https://onlinelibrary.wiley.com/doi/pdf/10.1002/pssb.201046289} \BibitemShut {NoStop}%
\bibitem [{\citenamefont {Gajdo\ifmmode~\check{s}\else \v{s}\fi{}}\ \emph {et~al.}(2006)\citenamefont {Gajdo\ifmmode~\check{s}\else \v{s}\fi{}}, \citenamefont {Hummer}, \citenamefont {Kresse}, \citenamefont {Furthm\"uller},\ and\ \citenamefont {Bechstedt}}]{Gajdos2006}%
  \BibitemOpen
  \bibfield  {author} {\bibinfo {author} {\bibfnamefont {M.}~\bibnamefont {Gajdo\ifmmode~\check{s}\else \v{s}\fi{}}}, \bibinfo {author} {\bibfnamefont {K.}~\bibnamefont {Hummer}}, \bibinfo {author} {\bibfnamefont {G.}~\bibnamefont {Kresse}}, \bibinfo {author} {\bibfnamefont {J.}~\bibnamefont {Furthm\"uller}},\ and\ \bibinfo {author} {\bibfnamefont {F.}~\bibnamefont {Bechstedt}},\ }\bibfield  {title} {\bibinfo {title} {Linear optical properties in the projector-augmented wave methodology},\ }\href {https://doi.org/10.1103/PhysRevB.73.045112} {\bibfield  {journal} {\bibinfo  {journal} {Phys. Rev. B}\ }\textbf {\bibinfo {volume} {73}},\ \bibinfo {pages} {045112} (\bibinfo {year} {2006})}\BibitemShut {NoStop}%
\bibitem [{\citenamefont {Baroni}\ and\ \citenamefont {Resta}(1986)}]{Baroni1986}%
  \BibitemOpen
  \bibfield  {author} {\bibinfo {author} {\bibfnamefont {S.}~\bibnamefont {Baroni}}\ and\ \bibinfo {author} {\bibfnamefont {R.}~\bibnamefont {Resta}},\ }\bibfield  {title} {\bibinfo {title} {Ab initio calculation of the macroscopic dielectric constant in silicon},\ }\href {https://doi.org/10.1103/PhysRevB.33.7017} {\bibfield  {journal} {\bibinfo  {journal} {Phys. Rev. B}\ }\textbf {\bibinfo {volume} {33}},\ \bibinfo {pages} {7017} (\bibinfo {year} {1986})}\BibitemShut {NoStop}%
\bibitem [{\citenamefont {Elliott}\ and\ \citenamefont {Dawber}(1979)}]{elliott1979b}%
  \BibitemOpen
  \bibfield  {author} {\bibinfo {author} {\bibfnamefont {J.~P.}\ \bibnamefont {Elliott}}\ and\ \bibinfo {author} {\bibfnamefont {P.~G.}\ \bibnamefont {Dawber}},\ }\href {https://doi.org/10.1007/978-1-349-07635-2} {\emph {\bibinfo {title} {Symmetry in {{Physics Vol1}}}}}\ (\bibinfo  {publisher} {Macmillan Education UK},\ \bibinfo {address} {London},\ \bibinfo {year} {1979})\BibitemShut {NoStop}%
\end{thebibliography}%
\end{document}